\documentclass[12pt,english]{article}
\usepackage{libertine}
\usepackage[T1]{fontenc}
\usepackage[latin9]{inputenc}
\usepackage{color}
\definecolor{lyxnotefontcolor}{rgb}{0.792969, 0.105469, 0.253906}
\usepackage{colortbl}
\usepackage{babel}
\usepackage{array}
\usepackage{mathrsfs}
\usepackage{amsmath}
\usepackage{amsthm}
\usepackage{amssymb}
\usepackage{geometry}
\usepackage{setspace}
\usepackage[authoryear]{natbib}
\usepackage[pdfusetitle,
 bookmarks=true,bookmarksnumbered=false,bookmarksopen=false,
 breaklinks=false,pdfborder={0 0 1},backref=page,colorlinks=false]
 {hyperref}
\hypersetup{
 colorlinks,citecolor=darkblue,linkcolor=maroon,urlcolor=maroon,bookmarks=true,backref=page}

\makeatletter

\providecommand{\tabularnewline}{\\}

\usepackage{tikz}
\usetikzlibrary{calc}
\usetikzlibrary{arrows, arrows.meta,calc} 
\usetikzlibrary{patterns}

\usepackage{pgfplots}

\usepackage{color}
\definecolor{darkblue}{rgb}{0.0,0,.6}
\definecolor{maroon}{rgb}{0.68,0,0}
\definecolor{darkgreen}{rgb}{0,0.369,0.086}
\definecolor{gray}{rgb}{.5,.5,.5}

\usepackage{titletoc}
\usepackage[page,header]{appendix}

\setdisplayskipstretch{1}
\AtBeginDocument{%
  \setlength{\abovedisplayskip}{6pt plus 2pt minus 2pt}%
  \setlength{\belowdisplayskip}{6pt plus 2pt minus 2pt}%
  \setlength{\abovedisplayshortskip}{3pt plus 2pt minus 2pt}%
  \setlength{\belowdisplayshortskip}{6pt plus 2pt minus 2pt}%
}

\makeatletter
\newcommand{\tightdisplayskips}{%
  \setlength{\abovedisplayskip}{6pt plus 2pt minus 2pt}%
  \setlength{\belowdisplayskip}{6pt plus 2pt minus 2pt}%
  \setlength{\abovedisplayshortskip}{3pt plus 2pt minus 1pt}%
  \setlength{\belowdisplayshortskip}{4pt plus 2pt minus 2pt}%
}
\g@addto@macro\normalsize{\tightdisplayskips}
\g@addto@macro\small{\tightdisplayskips}
\g@addto@macro\footnotesize{\tightdisplayskips}
\makeatother

\AtBeginDocument{\allowdisplaybreaks[4]}

\ifdefined\showcaptionsetup
 \PassOptionsToPackage{caption=false}{subfig}
\fi
\usepackage{subfig}
\makeatother

\theoremstyle{plain}
\newtheorem{assumption}{\protect\assumptionname}
\newtheorem{prop}{\protect\propositionname}
\newtheorem{thm}{\protect\theoremname}
\theoremstyle{definition}
\newtheorem{defn}{\protect\definitionname}
\newtheorem{example}{\protect\examplename}
\theoremstyle{plain}
\newtheorem{lem}{\protect\lemmaname}
\providecommand{\assumptionname}{Assumption}
\providecommand{\definitionname}{Definition}
\providecommand{\examplename}{Example}
\providecommand{\lemmaname}{Lemma}
\providecommand{\propositionname}{Proposition}
\providecommand{\theoremname}{Theorem}

\begin{document}
\title{\textbf{Optimal Rating Design under Moral Hazard}\thanks{This paper was previously circulated under the title \textquotedblleft Optimal
Rating Design\textquotedblright . We thank Nageeb Ali, Ian Ball, James
Best, Aislinn Bohren, Odilon Camara, Joyee Deb, Rahul Deb, Maryam
Farboodi, Nima Haghpanah, Hugo Hopenhayn, Emir Kamenica, Navin Kartik,
Alexey Kushnir, Stephen Morris, Aniko Öry, Paula Onuchic, Jacopo Perego,
Ilya Segal, Vasiliki Skreta and Alex Wolitzky as well as numerous
seminar and conference participants for their helpful comments. We
have used Claude.ai similarly to an RA.}}
\author{Maryam Saeedi \thanks{Carnegie Mellon University, Email: \textcolor{blue}{msaeedi@andrew.cmu.edu}}
\and Ali Shourideh \thanks{Carnegie Mellon University, Email: \textcolor{blue}{ashourid@andrew.cmu.edu}}}
\maketitle
\begin{abstract}
We study optimal rating design under moral hazard and strategic manipulation.
An intermediary observes a noisy indicator of effort and commits to
a rating policy that shapes market beliefs and pay. Whether optimal
ratings reveal or censor information depends on how effort shifts
the outcome distribution: when effort increases tail risk, optimal
ratings use lower censorship, pooling poor outcomes to encourage risk-taking;
when effort reduces tail risk, upper censorship discourages negligence.
In multi-task settings with window dressing, a monotone relative informativeness
property again delivers upper- or lower-censorship ratings. In an
application to redistributive test design, optimal tests can feature
mid-censorship. These results follow from a general characterization
of optimal ratings via concavification of a gain function, accommodating
violations of monotone likelihood ratios and distributional concerns.

\end{abstract}
\newpage{}

\section{Introduction}

\noindent Many markets rely on information disclosure or ratings to
facilitate trade and incentivize quality provision. Environmental,
Social, and Governance (ESG) rating agencies aim to incentivize companies
to improve their environmental and social impact. Online platforms
such as Amazon, Airbnb, Upwork, and eBay design reputation systems
to incentivize and signal providers' quality. Standardized tests communicate
student ability to universities. In each case, an intermediary observes
signals about agent behavior and must decide how to convey this information
to a market that rewards agents based on perceived quality. A fundamental
challenge arises: agents strategically respond to rating policies
by potentially using window dressing and other manipulation strategies,
and the information disclosed shapes their incentives.

Despite the ubiquity of these systems in markets suffering from moral
hazard, several core theoretical questions are yet to be answered:
What are the fundamental trade-offs in designing rating systems when
participants can anticipate and react to them? How does the effort
technology shape optimal ratings? How should an intermediary design
information structures to account for multi-tasking and window-dressing
incentives? This paper answers these questions by incorporating information
design in a variant of the career concerns model introduced by \citet{ho99}.

Our model features an agent (e.g., a company seeking an ESG rating
or a seller on eBay) who takes costly actions that create value for
a competitive market. These actions also generate a noisy indicator
observed by an intermediary (e.g., an ESG rating firm or an online
platform) which must then decide how to convey this information via
a rating to the market. The market, in turn, pays the agent its expected
value based on the signal and its belief about the agent's action.
Our primary objective is to find the optimal information structure
to maximize a flexible welfare function for the intermediary. This
function can target a particular action or social welfare, capturing
environments where market values do not fully internalize the social
value of actions---such as the positive externalities of ESG activities---or
where a platform has concerns for fairness or redistribution.\footnote{Since \citet{ho99}, it is well known that the \emph{implicit incentives
}provided by career concerns do not necessarily lead to efficient
effort levels because they fail to fully internalize the social benefit
of the agent's action. This externality is also present in our model
and the intermediary's objective can be thought of as addressing such
externalities.}

Our main results shed light on optimal provision of incentives for
one- and multi-dimensional effort. First, we show how interactions
between the distribution of the indicators and effort affect the optimal
ratings when effort is a single variable. Namely, we show that when
effort increases tail risk, lower--censorship ratings maximize effort
while the opposite is true when effort decreases tail risk. Second,
in a multi-task model with window dressing as in \citet{holmstrom1991multitask}
and \citet{dewatripont1999economics}, we identify a condition on
the information technology, the monotone relative informativeness
property (MRIP), under which optimal ratings are upper or lower censorship,
with the censored tail determined by the relative informativeness
of the indicator about the two efforts.

The main technical challenge in formulating the optimal rating problem
is the interplay between information structures and the agent incentive
constraints. The main object of interest is the interim payoff of
the agent who knows the indicator (state) and does not know the outcome
of the possible random signal. We introduce the concept of \emph{interim
prices}---the agent's interim expectation of market's posteriors---as
the sufficient statistic that determines incentives from the agent's
perspective. In general, the set of implementable interim prices is
not well characterized: being a mean-preserving contraction of market
values is necessary but not sufficient.\footnote{In a standard sender--receiver setting, \citet{doval2024persuasion}
are also interested in the interim payoff of a sender. They use a
duality approach to characterize the frontier of interim payoff profiles
of the senders -- as a function of the state. Our comonotonicity
restriction allows us to simplify the problem and restrict interim
prices to be a mean preserving contraction of market values.} This object, which can be interpreted as the agent's second-order
belief, allows us to transform the problem of choosing an information
structure into a tractable constrained information design problem.
Our key theoretical result, Proposition \ref{thm:Suppose-that-T1},
shows that when these interim prices are comonotone with market values,
then a price schedule can be implemented by some rating if and only
if it is a mean-preserving contraction of the market values. This
implies that we can cast the problem of rating design as a standard
moral hazard problem with transfers subject to a majorization constraint.
Throughout the paper we therefore restrict attention to monotone ratings---those
whose interim prices are comonotone with market values---a restriction
we motivate by the agent's ability to manipulate the indicator downward
ex post. With this result in hand, in Theorem \ref{thm: 1}, we show
that optimal ratings can be characterized through concavification
of a\emph{ gain function} in the quantile space. As a result, the
optimal information structure is a deterministic monotone partition
of the indicator space: the intermediary either fully reveals the
indicator on some regions or pools contiguous intervals---a sharp
foundation for the prevalence of coarse, threshold-based rating systems.

Our first set of results derives general properties of optimal ratings
in classic moral hazard where the agent has a single dimensional effort.
When the intermediary\textquoteright s objective is to maximize effort
(absent distributional motives), the design problem reduces to finding
the rating that achieves the highest level of effort. Under the canonical
Monotone Likelihood Ratio Property (MLRP) assumption in the moral
hazard literature, we show that the gain function in this case is
concave and as a result full information disclosure is optimal.

Despite the prominence of the MLRP assumption, many economically relevant
activities violate MLRP in systematic ways. Innovative activities
often increase both upside potential and downside risk, i.e., R\&D
effort can lead to breakthroughs or failures. Conversely, maintenance
activities typically reduce variance through more consistent outcomes.
To capture these patterns, we introduce two new distributional properties:
the \emph{expanding likelihood ratio property} (ELRP), where increased
effort expands the distribution's tails, and the \emph{compressing
likelihood ratio property} (CLRP), where increased effort compresses
outcomes toward the center. Under ELRP, optimal ratings take the form
of lower censorship providing insurance against downside risk to encourage
risk-taking. In contrast and under CLRP, optimal ratings are upper
censorship, pooling high realizations while revealing low ones, which
punishes poor outcomes and encourages variance-reducing effort.

To see the intuition, first note that the marginal benefit of effort
is given by the covariance between the interim price and the marginal
percentage change in the probability of different outcomes induced
by a marginal increase in effort. In the MLRP case, higher effort
shifts the distribution toward higher realizations in a uniform, monotone
way. Full revelation is therefore optimal for incentives: because
interim prices must be mean-preserving contractions of the full-information
market values, no rating can make the price schedule \textquotedblleft steeper\textquotedblright{}
than under full revelation, and any pooling only flattens it.

Under ELRP, by contrast, higher effort expands tail events: it increases
the probability of both very high and very low outcomes, so full revelation
exposes the agent to downside risk that effort itself makes more likely.
Pooling low realizations of the indicator provides insurance against
these outcomes---not in the risk--sharing sense, since the agent
is risk neutral, but by withholding the penalty on realizations that
effort itself makes more likely---and thereby strengthens the covariance
between interim prices and likelihood changes, increasing incentives
to exert effort. Under CLRP, higher effort compresses the distribution
toward the center, so extreme high outcomes are not especially \textquotedblleft earned\textquotedblright{}
by effort. Pooling the top realizations while revealing low ones similarly
improves the alignment between interim prices and likelihood changes,
strengthening incentives to exert effort.

Our second set of results, in Section \ref{sec: Multitasking-1} pertains
to a multi-task moral hazard model with window dressing à la \citet{holmstrom1991multitask},
in the career--concerns variant of \citet{dewatripont1999economics}.
In the multi-task model, the agent allocates efforts across productive
activity and a window-dressing one (an action that boosts the indicator
without affecting market values) which differentially impact the observed
indicator and market value. An important determinant of optimal ratings
is whether the technology is reducible, i.e., the distribution of
the indicator is governed by a single function of effort.\footnote{Mathematically, this is equivalent to the set of equilibrium actions
for arbitrary ratings having dimension one.} When this is the case, the indicator is not differentially informative
of the two actions. The multi--task model, then, is \textit{reducible}
to the single task model and results from Section \ref{sec:Application-1:-Rating}
apply.

When the multi--task model is \textit{irreducible}, different values
of the indicator have different informational content about the two
actions undertaken by the agent. Thus, our analysis identifies a condition
on the information technology, what we refer to as the \textit{monotone
relative informativeness property} (MRIP). Under MRIP, the informativeness
of the indicator about productive effort relative to window dressing
is monotone across realizations; that is, high realizations of the
indicator are either always more informative of window dressing or
of productive activity. Our main result is that with MRIP, welfare-maximizing
ratings are either upper or lower censorship. In a class of location--scale
technologies, the direction of censorship is determined by whether
window dressing contributes more to the level or to the dispersion
of the indicator and how the marginal costs of effort are related
to their impact on the mean of indicators. Notably, censorship requires
no deviation from MLRP: even when the indicator is informative about
each effort in the standard way, the optimal rating censors, since
the designer attaches weights of opposite signs to the two incentive
constraints.

In the final section of our paper and to illustrate the range of problems
that can be addressed by our formulation, we apply our framework to
the problem of redistributive test design with heterogeneous students,
showing that optimal tests may involve ``mid-censorship'' to balance
incentive provision with redistributive motives towards disadvantaged
groups.

Our analysis thus offers practical guidance for regulators and rating
system design. As data collection has intensified, several institutions
have formed around using data to incentivize behavior which in turn
has created incentives for manipulation and window dressing (see for
example \citet{mayzlin2014promotional}). Our results provide guidance
on how ratings should be designed in such environments and rationalize
the heterogeneity observed across rating systems. For example, platforms
have adopted a variety of ratings: some platforms (such as Shipt or
Instacart) allow for low rating forgiveness and fresh start which
could be interpreted as lower censorship; others such as Airbnb have
too many high ratings (see for example \citet{zervas2021first}) that
can be interpreted as upper censorship. Our results suggest these
differences may reflect optimal responses to underlying differences
in how effort affects outcome distributions as we discussed above;
Table \ref{tab:taxonomy} in Section \ref{sec:Application-1:-Rating}
summarizes the resulting taxonomy. More broadly, the paper provides
a toolkit for evaluating rating policies across domains---from ESG
certification to educational testing to online marketplaces---by
connecting observable features of agent technology to the optimal
structure of information disclosure.

\subsection{Related Literature \label{subsec:Related-Literature}}

Our paper is related to a few strands of the literature in information
economics and mechanism design. It is closely related to a recent
literature that studies information design when strategic behavior
affects the state by the choice of the information structure (e.g.,
\citet{frankel2019muddled}, \citet{ball2019scoring}, and \citet{perez2022test}).
In contrast with \citet{ball2019scoring} and \citet{frankel2019muddled},
our mathematical result on second-order expectations allows us to
study a larger class of objectives and state spaces, albeit under
the restriction that ratings are monotone in the indicator. Our exercise
substantively differs from \citet{ball2019scoring} where the agent's
actions purely manipulate several indicators, preventing a receiver
from correctly predicting a correlated quality. In contrast, our setting
completely assumes away adverse selection and screening, and pooling
of indicators occurs solely in order to motivate the agent. Our analysis,
thus, identifies both the precise shape of the optimal information
structure and when it is optimal to use uncertain rating systems.
In our model, the presence of window-dressing incentives is similar
to the falsification model in \citet{perez2022test}. The main difference
with our setting is the existence of noise in the ability of the agent
to manipulate the signal observed by the intermediary.

A related paper to ours is \citet{boleslavsky2018bayesian}. They
study a model of Bayesian persuasion with moral hazard, similar to
ours, in which an agent chooses an effort level that affects the distribution
of the state, and a sender affects a receiver's action using an information
structure. The papers differ in terms of focus and technique: We focus
on a career concern model where the information structure only affects
the agent's incentive. Additionally, we use majorization tools which
allow us to work with larger state spaces. Beyond technique, the economic
difference is that our characterization ties the shape of the optimal
rating---full disclosure, lower or upper censorship---to how effort
shifts the distribution of the indicator, a question that is difficult
to answer in their setting. In a related dynamic career-concerns model,
\citet{horner2016motivational} elegantly show that optimal ratings
that aggregate the agent's past performance over time are linear in
histories. The two papers are different in scope and techniques. They
are interested in the dynamics of ratings while in our model all incentive
provision happens instantaneously and our focus is on how properties
of single and multiple task technologies translate into properties
of optimal ratings. Moreover, in their setting, the process for the
indicator is a Brownian process whose trend is controlled by the agent
and as a result instantaneous shocks are small. It remains to be seen
how large shocks, for example as in jump processes, can impact dynamics
of optimal ratings.\footnote{Relatedly, a recent paper by \citet{madsen2025performance} studies
a moral hazard model with non--monetary incentives. Our paper is
related to their work to the extent that our agent is incentivized
using ratings; a non--monetary instrument.}

From a technical perspective, our results are related to the new literature
in information economics that uses optimization under majorization
constraints -- see \citet{kleiner2020extreme}. Their solution method
uses the characterization of extreme points of the set of monotone
functions that majorize a certain function. Similarly, \citet{bergemann2022screening}
and \citet{bergemann2022optimal} use the same strategy as our work
to cast the problem in terms of quantiles and use concavification
to derive optimal mechanisms---see also \citet{rayo2013monopolistic}.
While their focus is on screening models with hidden information,
ours is closer to classic moral hazard. Our focus on monotone ratings
also connects our work to the literature on monotone persuasion---see
\citet{mensch2021monotone} and \citet{kolotilin2024monotone}---which
provides conditions under which optimal monotone signals are deterministic
and take censorship forms. In that literature the state is exogenous,
and the conditions for censorship are placed on the sender's payoff
over states. Here the distribution of the state is chosen by the agent,
so the analogous conditions fall on the effort technology---MLRP,
ELRP, and CLRP with a single task, MRIP with multiple---rather than
on the sender's preferences.

\noindent Our paper is also related to the literature concerned with
the problem of certification and its interactions with moral hazard:
\citet{albano2001strategic}, \citet{zubrickas2015optimal}, \citet{onuchic2021conveying},
and \citet{zapechelnyuk2020optimal}. A notable contribution is that
of \citet{albano2001strategic}, where the key assumption that the
intermediary can charge an arbitrary fee schedule leads to an indeterminacy
between using transfers and ratings to implement desired outcomes.
\citet{zubrickas2015optimal}, \citet{zapechelnyuk2020optimal}, and
\citet{onuchic2021conveying} also study related problems, but they
focus on \emph{deterministic} technologies where the agent's effort
deterministically translates into values for the market. In contrast
and in our model, the presence of noise allows us to disentangle the
indicator from market values which in turn leads to an inefficient
level of effort under full information and enables us to study window
dressing and manipulation incentives. Other related contributions
to this literature include \citet{rodina2020inducing} on inducing
effort through grades and \citet{xiao2025incentivizing} on incentivizing
agents through ratings. Relatedly and in the context of team production,
\citet{halac2021rank} show that uncertainty about a worker's compensation
ranking in a team can remove low effort equilibria. Our result on
how censorship for some technologies can improve incentives can be
regarded as the single agent version of their result.\footnote{\citet{ali2022sell} study a model with adverse selection (i.e., exogenous
state), where optimal disclosure involves uncertainty, but it is a
way of uniquely implementing an intermediary's desirable outcome.}

\noindent Finally, our paper complements the empirical literature
on certification and disclosure in markets with asymmetric information,
such as online platforms (e.g., \citet{huisaspata} and \citet{nosko2015limits}),
health insurance markets (\citet{vatter2021quality}), food labeling
(\citet{barahona2022equilibrium}), and ESG investing (\citet{berg2022aggregate}).
We contribute to this literature by developing theoretical methods
and general lessons for the optimal design of rating systems.

The remainder of the paper proceeds as follows. Section \ref{sec:General-Model-and-1}
presents the model and provides our key technical results. Section
\ref{sec:Application-1:-Rating} establishes general properties of
optimal ratings with a single task under alternative distributional
assumptions, including MLRP, ELRP, and CLRP. Section \ref{sec: Multitasking-1}
focuses on multi-task moral hazard and window dressing. Section 5
applies our results to redistributive test design. Proofs are relegated
to the Appendix.

\section{A Model of Moral Hazard and Optimal Ratings\label{sec:General-Model-and-1}}

In this section, we describe our basic model of rating design and
provide some preliminary analysis of the restrictions implied by the
fact that incentives are provided through ratings.

We are interested in settings in which an intermediary observes some
information about an agent's chosen actions and decides how to convey
this information to a competitive market, henceforth ``the market,''
who then pays its posterior mean as a price to the agent.

More specifically, the agent exerts an effort vector $a\in A\subset\mathbb{R}^{N}$
at a cost $c(a)$. This action generates a random outcome $\left(v,y\right)\in\mathbb{R}^{2}$,
where $v$ represents the value of the output to the market and $y$
is a noisy \emph{indicator} observed by the intermediary. We denote
the cumulative distribution function of the indicator $y$ conditional
on action $a$ by $G(y|a)$.

The market consists of competitive buyers who value the agent's output
at $v$. However, the market observes neither the true value $v$
nor the agent's action $a$ directly. Instead, it forms expectations
based on information provided by the intermediary. If the market observes
the indicator $y$ and believes the agent's action to be $\hat{a}$,
the expected value of the output is given by:
\[
\overline{v}(y;\hat{a})=\mathbb{E}[v\mid y,\hat{a}]
\]
 We refer to $\overline{v}\left(y;\hat{a}\right)$ as \textit{market
values}, i.e., the most informative assessment of the valuation of
the market.\footnote{Throughout the paper, we focus on equilibria in which the agent plays
a pure effort strategy. Our main characterization results, Proposition
\ref{prop: simple} and Theorem \ref{thm: 1}, hold when allowing
for mixed effort strategy by the agent.} Throughout the paper we impose the following monotonicity assumption:
\begin{assumption}
\label{assu:The-market-value} For all (deterministic) market beliefs
$\hat{a}$, the market value $\overline{v}\left(y;\hat{a}\right)$
is increasing in the indicator $y$.
\end{assumption}
This assumption states that market values are ranked based on the
values of the indicator. Without any other assumption on the distribution
function $G\left(y|a\right)$, e.g., first-order stochastic dominance
or MLRP, this assumption is innocuous as one can always relabel the
values of the indicator according to the market values. While our
main characterization results -- Proposition \ref{thm:Suppose-that-T1}
and Theorem \ref{thm: 1} -- hold without Assumption \ref{assu:The-market-value},
we maintain this assumption for tractability and convenience.

The intermediary observes the indicator $y$ (at no cost) and controls
the information observed by the market. Specifically, the intermediary
commits to an information structure $\left(S,\pi\left(\cdot|y\right)\right)$,
where $S$ is a set of signal realizations and $\pi\left(\cdot|y\right)\in\Delta\left(S\right)$
is the distribution over signals conditional on realization of $y$.
Having observed $s$, the market pays its expected value $\mathbb{E}\left[v|s,\hat{a}\right]$
to the agent when it believes the agent chooses $\hat{a}$.\footnote{We assume that the buyers are on the long side of the market, thus
willing to pay their expected value. Our analysis remains unchanged
if the market keeps a constant fraction of their expected value.} This expectation is calculated using the information available, $s$,
and the common belief about equilibrium play.\footnote{An information structure is a family of probability spaces $\left\{ \left(S,\mathscr{S},\pi\left(\cdot|y\right)\right)\right\} _{y\in Y}$,
where $S$ is the space of signal realizations and $\mathscr{S}$
is a $\sigma$-algebra. Throughout the paper, we work with $S$ as
a compact subset of some Euclidean space, and $\mathscr{S}$ as the
Borel $\sigma$-algebra associated with topology induced by the Euclidean
norm $S$. Henceforth, we drop references to $\sigma$-algebra in
our analysis. Additionally, when describing subsets, we refer to Borel
subsets.} The timing is summarized in Figure \ref{fig:Timing}: the intermediary
commits to an information structure, the agent chooses her effort,
the indicator realizes, a rating is drawn and sent to the market,
and the market pays the agent its posterior expectation of the value.

\begin{figure}
\centering{}\begin{tikzpicture} 
\coordinate (Origin) at (0,0);
\filldraw [fill=blue!90!yellow!20!, thick] ($(0,4)$)  coordinate (GeneralStart) -- ++(2.5,0) -- ++(0,-1) -- ++(-2.5,0) node[right=35pt, above=5pt]{Agent: $a\in A$} --cycle;
\draw[-stealth, very thick] (2.75,3.5) node[above=8pt,right=20pt]{\color{red}{$y\in \mathbb{R}$}} -- ++ (3,0);
\filldraw [fill=red!90!yellow!20!, thick] ($(6,4)$)  coordinate (GeneralStart) -- ++(3.5,0) -- ++(0,-1) -- ++(-3.5,0) node[right=50pt, above=3pt]{Int.: $\pi(\cdot|y)\in\Delta(S)$} --cycle;
\draw[-stealth, very thick] (7.75,2.75) node[below=13pt, right=2pt]{\color{red}{$s\in S$}} -- ++ (0,-1);
\filldraw [fill=green!90!yellow!5!, thick] ($(6,1.5)$)  coordinate (GeneralStart) -- ++(3.5,0) -- ++(0,-1) -- ++(-3.5,0) node[right=50pt, above=3pt]{Market: $v-\hat{p}$} --cycle;
\draw[-stealth, very thick] (5.5,1) node[left=38pt, above=20pt] {\color{blue}{pay $\hat{p}=\mathbf{E}\left[v|s,a\right]$}} .. controls (3.375,1.3) ..  (1.25,2.75); 
\end{tikzpicture}\caption{General structure of the model\label{fig:Timing}}
\end{figure}

Given an information structure $\left(S,\pi\left(\cdot|y\right)\right)$
and market belief $\hat{a}$, the agent's expected payoff is:
\begin{equation}
\int_{Y}\int_{S}\mathbb{E}\left[v|s,\hat{a}\right]d\pi\left(s|y\right)dG\left(y|a\right)-c\left(a\right).\label{eq: eqn1}
\end{equation}
In equilibrium, the agent chooses $a$ to maximize (\ref{eq: eqn1}).

The ex-post market price $\mathbb{E}\left[v|s,a\right]$ depends on
the information structure $\left(S,\pi\right)$ and also on the market's
prior about the distribution of $\left(a,y\right)$, which depends
on the agent's equilibrium strategy. More specifically, the market
uses its beliefs about the equilibrium strategy of the agent $a$
to form a prior $G\left(y|a\right)$ and uses Bayes' rule to form
the posterior expectation $\mathbb{E}\left[v|s,a\right]$ satisfying
\begin{equation}
\int_{Y}\int_{S'}\mathbb{E}\left[v|s,a\right]d\pi\left(s|y\right)dG\left(y|a\right)=\int_{Y}\overline{v}\left(y;a\right)\pi\left(S'|y\right)dG\left(y|a\right),\forall S'\subset S.\label{eq:Bayes}
\end{equation}

The above defines a Bayesian Nash equilibrium given the information
structure $\left(S,\pi\right)$. More specifically, given an information
structure $\left(S,\pi\right)$, an equilibrium is an effort $a$
together with market's belief $\hat{a}=a$ and $\mathbb{E}\left[v|s,\hat{a}\right]$
such that $a$ maximizes expression (\ref{eq: eqn1}), and given $a$,
the market beliefs satisfy Bayesian updating as defined in equation
(\ref{eq:Bayes}).

We now turn to the intermediary's problem. The remainder of this section
reduces optimal rating design to a tractable problem in two steps:
we first characterize the set of interim price schedules that some
rating can implement, and we then use this characterization to solve
for the optimal rating.

The intermediary chooses an information structure to maximize an objective
that may differ from total surplus. We consider a class of objectives
in the form 
\begin{equation}
W\left(a\right)+\int p\left(y\right)\alpha\left(y\right)dG\left(y|a\right),\label{eq: obj}
\end{equation}
where $W(a)$ captures externalities or direct preferences over effort,
and $\alpha\left(y\right)\geq0$ represents distributional weights
on agent payoffs. This class of objective functions fits various examples:
the term $W\left(a\right)$ captures direct externalities---as in
the ESG and career-concerns examples---so that $W\left(a\right)-V\left(a\right)$,
with $V\left(a\right)=\mathbb{E}\left[v|a\right]-c\left(a\right)$
the total surplus, represents the external effects that are not captured
by the market; the weights $\alpha\left(y\right)$ capture distributional
concerns, such as a platform aiming to guarantee a minimum payoff
level for its sellers or a college reweighting test outcomes for its
admission policies.

The environment fits a number of applications. Online platforms observe
rich performance data about their providers and convey it to the market
through certification policies such as Airbnb's Superhost or eBay's
Top Rated Seller; such policies correspond to the information structure
in our model, see \citet{hui2016reputation}. Ratings also invite
manipulation and window dressing: the agent may take costly actions
that inflate the indicator without enhancing fundamental value, as
when sellers pay for positive reviews, see \citet{he2022market};
we develop this application in Section \ref{sec: Multitasking-1}.
Finally, when market values vary with the agent's action, as in models
of career concerns, equilibrium effort is inefficient and rating design
shapes this inefficiency. We characterize welfare maximizing ratings
in settings with a single or multiple tasks.

\subsection{Interim Prices: Definition and Characterization\label{subsec:Interim-Prices:-Definition}}

In this section, we introduce a mathematical object, \emph{interim
prices}, that allows us to simplify the problem of rating design in
the environment described above. The notion of interim price is simple.
This mathematical object determines the agent's incentives in choice
of effort and will be present in the incentive constraints for the
agent. Specifically, we define \emph{interim prices} as
\begin{equation}
p\left(y\right)=\int\mathbb{E}\left[v|s\right]d\pi\left(s|y\right).\label{eq: interim}
\end{equation}
In words, $p$ is the expected payment to the agent conditional on
the indicator $y$, integrating over possible signals $s$ given the
rating system. Additionally, it is an equilibrium object as it depends
on $\mathbb{E}\left[v|s\right]$ which depends on the market's beliefs
about the agent's action profile.\footnote{Throughout this section and the next one, we suppress the market belief
$\hat{a}$ in the notation in order to avoid clutter. This is without
loss since the payoff of the agent depends on market beliefs only
through the interim prices.} It can also be interpreted as the agent's ``second-order belief'':
their belief about the belief of the market on values.

Critically, interim prices are a sufficient statistic for the information
structure from the agent's perspective. Specifically, for any choice
of $a$, the agent's payoff is given by
\[
\int p\left(y\right)dG\left(y|a\right)-c\left(a\right).
\]

Thus, the problem of designing an optimal rating system is isomorphic
to the problem of choosing an interim price schedule $p$, subject
to the constraint that $p$ must be implementable via some information
structure $(S,\pi)$. Thus, we need to characterize the set of feasible
interim prices.

Generally, there are no simple conditions to characterize the set
of interim price profiles that result from a particular information
structure and action profiles. However, as we will show next, under
some restriction on information structures, a simple characterization
exists.

To see the restrictions that ratings place on interim prices, note
that market values are the interim prices associated with the fully
revealing information structure, and that any interim price schedule
is an iterated conditional expectation of market values. Viewed as
random variables, we must therefore have $p\succeq_{\text{cv}}\overline{v}$,
i.e., $p$ is a mean preserving contraction of $\overline{v}$.\footnote{The relation $p\succeq_{\text{cv}}\overline{v}$ represents the concave
order which implies that for all concave functions $\phi:\mathbb{R}\rightarrow\mathbb{R}$,
$\mathbb{E}\left[\phi\left(p\right)\right]\geq\mathbb{E}\left[\phi\left(\overline{v}\right)\right]$.
Since Bayes plausibility implies $\mathbb{E}\left[p\right]=\mathbb{E}\left[\overline{v}\right]$,
this definition is equivalent to majorization, second order stochastic
dominance and increasing concave order -- see for example \citet{shaked2007stochastic}
Section 4.A.} Intuitively, a rating can only average market values across realizations
of the indicator, and averaging reduces dispersion.

Given the results in the literature -- see for example \citet{rothschild1970increasing}
or \citet{Gentzkow} -- it is tempting to suggest that the reverse
of the above observation is also true: that mean preserving contraction
is a sufficient condition for existence of ratings. Below we show
that this is indeed true when interim prices and market values are
comonotone.\footnote{In Appendix \ref{sec:Importance-of-Co=002013Monotonicity}, we provide
an example that illustrates that without comonotonicity mean preserving
contraction is no longer sufficient and additional conditions are
needed. We also discuss its relationship with similar results in the
literature.} Formally, we say that $p$ and $\overline{v}$ are comonotone if:
\[
p\left(y\right)>p\left(y'\right)\Rightarrow\overline{v}\left(y;\hat{a}\right)>\overline{v}\left(y';\hat{a}\right).
\]

In words, higher prices are associated with higher market values,
so the two random variables never move in opposite directions.
\begin{prop}
\label{thm:Suppose-that-T1}Suppose that $p$ is a function that maps
values of $y$ into $\mathbb{R}$ such that
\begin{enumerate}
\item $p$ is comonotone with $\overline{v}$, and
\item $p\succeq_{\text{cv}}\overline{v}$.
\end{enumerate}
Then, there exists an information structure $\left(S,\pi\right)$
such that $p\left(y\right)=\int\mathbb{E}\left[v|s\right]d\pi\left(s|y\right)$.
\end{prop}
Proposition \ref{thm:Suppose-that-T1} is the main technical tool
that allows us to significantly simplify the problem of rating design
into the constrained concavification in Theorem \ref{thm: 1}. The
main challenge is the fact that interim prices are second order expectations
-- expectation of the market's expectation conditional on the realization
of the state. In our proof, we use the comonotonicity to construct
a sequence of garblings each of which is a randomization between full
revelation and pooling of consecutive states.\footnote{In our proof, we first establish the result when there are finitely
many market values and then approximate arbitrary distribution of
market values by discretizing the values and use compactness arguments.}

This proposition implies that the comonotone equilibria of the game
are characterized by three conditions: the effort is incentive compatible
given the interim prices,
\begin{equation}
a\in\arg\max_{\hat{a}\in A}\int p\left(y\right)dG\left(y|\hat{a}\right)-c\left(\hat{a}\right)\label{eq: IC}
\end{equation}
the interim prices dominate market values in the concave order, and
interim prices and market values are comonotone.

This reduction allows us to transform the optimal rating design problem
into a standard mechanism design problem with transfers, where the
``transfers'' are the interim prices constrained by the concave
order.

\paragraph*{Remark on Comonotonicity}

In the rest of the paper, we focus on ratings that satisfy this constraint.
One can justify this decision using ex-post manipulation: if the agent
can manipulate the indicator downwards but not upwards (hide some
positive outcomes), then the effective interim prices are monotone.\footnote{The classic moral hazard literature has imposed such constraints.
The seminal paper by \citet{innes1990limited} is an example.}

\subsection{The Main Characterization}

Given this class of objectives and the comonotonicity restriction,
the problem of optimal rating design can be stated as maximizing the
objective in (\ref{eq: obj}) subject to incentive compatibility (\ref{eq: IC}),
comonotonicity and majorization.

We solve this problem in two steps. First, we transform prices and
values into quantile space (for the indicator), where the concave
order becomes a majorization constraint and the value of any objective
can be computed by concavification. Second, we incorporate the first
order version of the incentive constraints through a Lagrangian argument,
which leads to our main characterization.

Let $v_{Q}(i)$ denote the market value associated with the $i$-th
quantile of the indicator distribution. Formally,
\begin{equation}
v_{Q}(i)=\overline{v}(G^{-1}(i|a);a).\label{eq: quant}
\end{equation}
For ease of notation, we have dropped $a$ from the expression of
quantile value. Similarly, let $p_{Q}(i)$ be the quantile representation
of the interim price. Given the comonotonicity assumption of $p$
and $\overline{v}$, $p_{Q}\left(i\right)$ is the interim price associated
with market value $v_{Q}\left(i\right)$.

The characterization of implementable interim prices allows us to
evaluate any objective by concavification: as we show in Appendix
\ref{subsec:Proof-of-Proposition}, maximizing an unconstrained objective
over implementable interim prices amounts to concavifying an integrated
version of its weighting function in the quantile space. Intuitively,
pooling an interval of realizations replaces the objective's value
on that interval with its average, and the best monotone pooling is
described by a concave envelope.

In order to incorporate the incentive compatibility, we first define
\[
F\left(i|\hat{a};a\right)=G\left(G^{-1}\left(i|a\right)|\hat{a}\right).
\]
In words, $F$ is the distribution over quantiles when the agent chooses
$\hat{a}$ but quantiles are defined according to $a$. In quantile
space, incentive compatibility (\ref{eq: IC}) requires that $a$
maximizes $-\int^{1}_{0}F\left(i|\hat{a};a\right)dp_{Q}\left(i\right)-c\left(\hat{a}\right)$
over $\hat{a}\in A$.

\paragraph{The First Order Approach.}

In order to sidestep many of the complications that typically arise
in moral hazard problems, we will use the first order approach (FOA)
throughout the paper. That is, we replace the incentive constraint
(\ref{eq: IC}) with its first order condition. Since our characterization
identifies a candidate rating---in the leading cases, an upper- or
lower-censorship of the indicator---it is sufficient to verify that
the agent's payoff is quasi--concave in effort given this candidate,
so that the local incentive constraint implies global incentive compatibility.
In Appendix \ref{sec:Validity-of-the}, we will use an approach similar
to \citet{chade2020no} to provide sufficient conditions on the distribution
functions for the validity of the first order approach. The conditions
often reduce to comparing the curvature of the c.d.f. function $G\left(y|a\right)$
with respect to $a$ to that of the cost function $c\left(a\right)$
and are typically satisfied provided that $c\left(a\right)$'s are
sufficiently convex.\footnote{The literature on verification of first order approach in moral hazard
is quite extensive. A few examples are \citet{mirrlees1999theory},
\citet{rogerson1985first}, \citet{jewitt1988justifying}, and more
recently \citet{conlon2009two} and \citet{jung2015information}.}

The following provides a full characterization of optimal ratings
under FOA:
\begin{thm}
\label{thm: 1} If $w^{*}$ is the highest value of the objective
(\ref{eq: obj}) and under the validity of FOA, there exists a real
vector $\lambda\in\mathbb{R}^{N}$ such that 
\begin{align}
w^{*}= & \max_{a}W\left(a\right)+\int\text{cav}\Gamma\left(i;\lambda,a\right)dv_{Q}\left(i\right)\tag{D}\label{eq: D}\\
 & \text{ s.t. }i=G\left(\left\{ y:\overline{v}\left(y;a\right)\leq v_{Q}\left(i\right)|a\right\} \right)\nonumber 
\end{align}
where 
\begin{align*}
\Gamma\left(i;\lambda,a\right)= & \int_{\left\{ y:\overline{v}\left(y;a\right)>v_{Q}\left(i\right)\right\} }\alpha\left(y\right)dG-\sum^{N}_{n=1}\lambda_{n}\left[\left.\frac{\partial}{\partial\hat{a}_{n}}F\left(i|\hat{a};a\right)\right|_{\hat{a}=a}+\frac{\partial}{\partial a_{n}}c\left(a\right)\right]
\end{align*}
\end{thm}
\begin{figure}
\begin{centering}
\begin{tikzpicture}[scale=0.5]
\coordinate (Origin) at (0,0);

\draw[-stealth] (12-.3,0)  -- (22.,0) node[anchor=north west,right=5pt, below=-5pt] {$i$}; 
\draw[-stealth] (12,0) -- (12,10);
\draw[very thick, color=blue] (12,8) node[right = 16pt,below=8pt] {$\Gamma(i;\lambda,a)$}..  controls (13.5,6) and (19,10.5) .. (21,0);
\draw  (12.0,.3) -- (12,-.3) node[below]{$0$};
\draw  (21.0,.3) -- (21,-.3) node[below]{$1$};
\draw[ultra thick, red]  (12,8) -- (15.8,7.3) node[above=4pt]{$\text{cav}\Gamma(i;\lambda,a)$} .. controls (18,7.) and (20.5,4.5) .. (21,0);
\draw[thick, dashed,darkgreen]  (15.8,0) node[below]{\footnotesize $\hat{i}$}-- (15.8,7.3);
\draw[very thick, darkgreen, <->]  (12.2,3) -- (15.6,3) node[midway, left=3.5pt, below]{\footnotesize $p = \mathbb{E}[\bar{v}|y<\hat{y}]$};
\draw[very thick, darkgreen, <->]  (16.,3) -- (21,3) node[midway, below]{\footnotesize $p=\bar{v}$};
\draw[ thick, dashed,darkgreen]  (21,0) -- (21,3);
\end{tikzpicture}
\par\end{centering}
\caption{An illustration of the concavification of the gain function $\Gamma$
and construction of optimal ratings \label{fig:Concavification-of-}}
\end{figure}

We refer to $\Gamma$ as the gain function. It summarizes the weight
that the intermediary puts on a particular realization of the indicator
and the associated IC conditions. As it is demonstrated in Figure
\ref{fig:Concavification-of-}, the concave envelope of the gain function
determines the optimal ratings. When it coincides with $\Gamma$,
optimal rating should be fully revealing while over intervals where
the concave envelope is linear, the optimal rating is pooling the
realizations within that interval. Theorem \ref{thm: 1} implies that
the rating design problem can be solved by solving a one-dimensional
concavification problem of the gain function and then finding optimal
values of effort and the multipliers associated with the incentive
compatibility constraints (\ref{eq: IC}).

\textbf{Optimality of Threshold Based Ratings}

The first implication of Theorem \ref{thm: 1} is that optimal rating
systems are simple. In fact, since the function to be\textit{ concavified}
is only a function of the quantile $i$, by Caratheodory theorem any
convex combination of values of $\Gamma\left(i;\lambda,a\right)$
can be achieved by using at most two points. This logic establishes
that the optimum in (\ref{eq: D}) is always achieved by a deterministic
monotone partition: the intermediary either fully reveals the indicator
on some regions or pools contiguous intervals into coarse categories.

While the optimum value in (\ref{eq: D}) is always achieved by an
interim price associated with a deterministic monotone partitional
signal, such a rating should also satisfy the incentive compatibility.
To ensure that this is indeed possible, we make the following assumption
on the distribution function $G\left(y|a\right)$:
\begin{assumption}
\textbf{\label{assu:Independence.}Independence. }For all $a\in A$:
\begin{enumerate}
\item $G\left(y|a\right)$ has full support over a convex subset of $\mathbb{R}$.
\item For any interval $I\subset\text{Supp}G\left(y|a\right)$, the function
$\alpha\left(y\right)g\left(y|a\right)$ cannot be written as a non--zero
linear combination of $g\left(y|a\right),\left\{ \frac{\partial g\left(y|a\right)}{\partial a_{n}}\right\} ^{N}_{n=1}$
for all values of $y\in I$.
\end{enumerate}
\end{assumption}
The independence assumption ensures that there is enough variation
in $y$ conditional on effort $a$ so that only coarse moments of
$p\left(y\right)$ matter for incentives.\footnote{An example that violates Assumption \ref{assu:Independence.} is one
in which $y=\overline{y}\left(a\right)$, an increasing function of
$a$. In this case, any change in the interim price function affects
the incentives of the agent. In a previous version of this paper,
we have established that if Assumption \ref{assu:Independence.} is
violated, optimal ratings can involve randomization.} An example of a family that satisfies the above is when $\alpha=0$
and $G$ is the scale--location family $G\left(y|a\right)=H\left(\frac{y-\mu\left(a\right)}{\sigma\left(a\right)}\right)$
with $H$ having a support of the entire real line with density $h\left(\varepsilon\right)$
such that $h\left(\varepsilon\right),h'\left(\varepsilon\right),$
and $\varepsilon h'\left(\varepsilon\right)$ are linearly independent.
Given Assumption \ref{assu:Independence.}, Theorem \ref{thm: 1},
and Lemma \ref{lem:Let--be} (in the Appendix), we have the following
proposition:
\begin{prop}
\label{prop: simple} Suppose that Assumption \ref{assu:Independence.}
holds. Then the optimal interim price in (\ref{eq: D}) is always
associated with a deterministic monotone partitional rating. Moreover,
whenever $\text{cav}\Gamma\left(i;\lambda,a\right)=\Gamma\left(i;\lambda,a\right)$,
optimal rating reveals the value $\overline{v}=v_{Q}\left(i\right)$
to the market. When $\text{cav}\Gamma\left(i;\lambda,a\right)>\Gamma\left(i;\lambda,a\right)$,
then there exists an interval $i\in\left[i_{1},i_{2}\right]$ such
that optimal rating reveals that $\overline{v}\in\left[v_{Q}\left(i_{1}\right),v_{Q}\left(i_{2}\right)\right]$.
\end{prop}
Theorem \ref{thm: 1} and Proposition \ref{prop: simple} together
provide a full characterization of optimal ratings. In what follows,
we provide general properties of optimal ratings with single and multiple
efforts/tasks.

\section{Optimal Single Task Ratings\label{sec:Application-1:-Rating}}

In this section, we provide general properties of optimal ratings
and how they depend on the joint distribution of the indicator function
and market values. As we show, whether optimal ratings involve full
disclosure or censorship---and whether censorship occurs at the bottom
or the top of the distribution---is determined by how effort shifts
the distribution of the indicator. Table \ref{tab:taxonomy} at the
end of this section summarizes the resulting taxonomy. In this part,
we focus on problems in which effort is one dimensional. In Section
\ref{sec: Multitasking-1}, we study multiple tasks.

\subsection*{Targeting An Action}

When does an intermediary that wishes to maximize effort benefit from
concealing information? To answer this question, we start our analysis
by considering objectives that only target an action, i.e., $\alpha\left(y\right)=0$
in (\ref{eq: obj}). In this case, the problem of solving optimal
rating design boils down to a characterization of the set of implementable
efforts. As we have shown, an effort $a\in A$ is implementable when
there exists an interim price function $p\left(y\right)$ such that
$p\left(y\right)$ is a mean-preserving contraction of market values
$\overline{v}\left(y;a\right)$ and $a$ is incentive compatible given
$p\left(y\right)$.

To make the model tractable, let us assume the following:
\begin{assumption}
\label{assu: targetting} The action space $A$ and the distribution
function $g\left(y|a\right)$ satisfy the following
\begin{enumerate}
\item The action space is $A=\left[0,\overline{a}\right]\subset\mathbb{R}$.
\item For all $a>0$, the support of $g\left(y|a\right)$ is a (potentially
unbounded) interval $I=\left[\underline{y},\overline{y}\right]$ and
$g\left(y|a\right)$ is twice differentiable.
\item Cost function $c\left(a\right)$ is non-negative, strictly convex,
increasing and twice differentiable for all $a\in A$.
\item For any effort, $a\in A$, $\overline{v}\left(y;a\right)$ is increasing
in $y$.
\end{enumerate}
\end{assumption}
The first three parts of Assumption \ref{assu: targetting} are fairly
common in the moral hazard literature. The last assumption notably
implies that $y$ is an indicator that is positively correlated with
market values. This assumption on its own is innocuous since the indicator
$y$ itself is not payoff relevant.

By Theorem \ref{thm: 1}, under FOA, the optimal rating is found by
a concavification of the function $\Gamma\left(i;\lambda,a\right)=-\lambda\left[\left.\frac{\partial F\left(i|\hat{a};a\right)}{\partial\hat{a}}\right|_{\hat{a}=a}+c'(a)\right]$
where $\lambda$ is the Lagrange multiplier on local incentive-compatibility
constraint, and $F\left(i|\hat{a};a\right)$ is the induced distribution
of the quantiles of the indicator when the agent chooses effort $\hat{a}$
while the market believes it to be $a$. The following calculation
ties the object to be concavified to properties of the distribution
function $G\left(y|a\right)$:
\begin{align*}
-\lambda\frac{\partial^{2}}{\partial i^{2}}\left.\frac{\partial F\left(i|\hat{a}\right)}{\partial\hat{a}}\right|_{\hat{a}=a} & =-\lambda\frac{1}{g\left(y|a\right)}\left.\frac{\partial^{2}}{\partial y\partial a}\log g\left(y|a\right)\right|_{y=G^{-1}\left(i|a\right)}
\end{align*}
In other words, the concavity of $\Gamma\left(i;\lambda,a\right)$
at a particular quantile $i$ is determined by the sign of the cross
partial of the log-likelihood function $\log g\left(y|a\right)$.
This implies that optimal ratings are directly tied to the supermodularity
of the log-likelihood function $\log g\left(y|a\right)$. In what
follows, we discuss a few cases and their economic interpretation
and implication for optimal rating.

Let us start from the canonical assumption made in the moral hazard
literature, the so-called \textit{MLRP }assumption:
\begin{defn}
\label{assu:MLRP}\textbf{ }A distribution function $g\left(y|a\right)$
is said to satisfy Monotone Likelihood Ratio Property (\textbf{MLRP})
when $g\left(y|a\right)$ is log--supermodular. That is $\frac{\partial^{2}}{\partial a\partial y}\log g\left(y|a\right)=\frac{\partial}{\partial y}\frac{g_{a}\left(y|a\right)}{g\left(y|a\right)}\geq0,\forall y\in I,a\in A$.
\end{defn}
MLRP implies that an increase in effort leads to a rightward shift
of the distribution of indicator realizations. Moreover, it also implies
that the same is true for the conditional distribution of the indicator
when restricted to an interval of values of $y$.\footnote{Formally, MLRP is equivalent to the statement that an increase in
$a$ increases the distributions over the indicator $y$ according
to the likelihood ratio order. See \citet{shaked2007stochastic},
section 1C.}

Our first result establishes that in the presence of MLRP, the highest
implementable effort is indeed associated with full information:
\begin{prop}[Full Disclosure under MLRP]
\label{prop: MLRP}Suppose Assumptions \ref{assu:Independence.}
and \ref{assu: targetting} hold, FOA is valid, and $g\left(y|a\right)$
satisfies MLRP. Then the highest implementable effort is associated
with interim price $p\left(y\right)=\overline{v}\left(y;a\right)$.
That is, it is the highest effort level that satisfies
\[
a_{FI}\in\arg\max_{a\in A}\int\overline{v}\left(y;a_{FI}\right)g\left(y|a\right)dy-c\left(a\right).
\]
\end{prop}
The above result states that the often assumed MLRP has strong implications
for what can be achieved via ratings. Specifically, it states that
using ratings, it is not possible to increase the level of effort
beyond what the market can achieve by fully observing the indicators.

We should note that the definition of highest implementable effort
$a_{FI}$ involves calculation of a fixed point. This is because market
values $\overline{v}\left(y;a\right)$ should be calculated under
the belief of the market that the action taken is $a_{FI}$ while
the agent is able to deviate from it. In Proposition \ref{prop: MLRP},
$a_{FI}$ is defined as the highest such fixed point.

The proof of Proposition \ref{prop: MLRP} follows straight from Theorem
\ref{thm: 1}. Specifically, under MLRP, the function $\Gamma\left(i;\lambda,a\right)$
is either concave or convex for all values of $i$ depending on the
sign of $\lambda$. This means that when $\lambda>0$, $\Gamma$ is
concave and coincides with its concavification. Thus given our construction
of optimal ratings in Section \ref{sec:General-Model-and-1}, optimal
rating becomes fully revealing. In turn, if $\lambda<0$, $\Gamma$
is convex in $i$ and thus, its concavification is simply the 0 function
-- since $\Gamma\left(0;\lambda,a\right)=\Gamma\left(1;\lambda,a\right)=0$.
In other words, optimal rating involves providing no information which
results in $a=0$ which cannot be optimal, so $\lambda$ cannot be
negative.

\subsection{Expanding and Compressing Likelihood Ratios}

Many economic activities violate MLRP in systematic ways. For example,
activities where greater effort affects not just the mean outcome
but also its variance. Innovative activities often increase both upside
potential and downside risk---greater R\&D effort can lead to breakthroughs
or failures. On the other hand, activities such as maintenance typically
reduce variance---more careful attention produces more consistent
outcomes. These patterns correspond to distributions where the cross-derivative
of the log-likelihood changes sign.

In what follows, we define two classes of distributions and characterize
the optimal ratings.
\begin{defn}
A distribution function $g\left(y|a\right)$ is said to satisfy:
\begin{enumerate}
\item Expanding likelihood ratio property \textbf{(ELRP) }if for any $a\in A$,
there exists $\hat{y}$ such that $\frac{\partial^{2}}{\partial a\partial y}\log g\left(y|a\right)\geq0$
when $y\geq\hat{y}$ and $\frac{\partial^{2}}{\partial a\partial y}\log g\left(y|a\right)\leq0$
when $y\leq\hat{y}$,
\item Compressing likelihood ratio property \textbf{(CLRP) }if for any $a\in A$,
there exists $\hat{y}$ such that $\frac{\partial^{2}}{\partial a\partial y}\log g\left(y|a\right)\leq0$
when $y\geq\hat{y}$ and $\frac{\partial^{2}}{\partial a\partial y}\log g\left(y|a\right)\geq0$
when $y\leq\hat{y}$.
\end{enumerate}
\end{defn}
The terminology reflects how effort affects the signal distribution's
tails. Under ELRP, increased effort expands the tails, while under
CLRP, increased effort compresses the distribution toward the center.

For example, consider $\log y\sim\mathcal{N}\left(\log a,a^{\gamma}\right)$,
that is, $\log y$ has a normal distribution with mean $\log a$ and
variance $a^{\gamma}$. In this case, we can use the definition of
the density of the normal distribution to show that
\[
\frac{\partial^{2}}{\partial a\partial y}\log g\left(y|a\right)=\frac{1+\gamma\log y/a}{a^{1+\gamma}y}.
\]
When $\gamma>0$, $g$ satisfies ELRP since the above is positive
if and only if $y/a\geq e^{-1/\gamma}$. In contrast, when $\gamma<0$,
$g$ satisfies CLRP since the above is positive if and only if $y/a\leq e^{-1/\gamma}$.

A version of these examples, with mean parameter $a$ in place of
$\log a$, is depicted in Figure \ref{fig:Distributions-with-ELRP}.
As can be seen, in case of ELRP, the tail densities increase as effort
$a$ increases while the densities for mid--realizations decline.
In contrast, under CLRP, tail densities decline while the densities
for mid--realizations increase. In both cases, the two densities
intersect exactly twice which is in contrast with single crossing
of MLRP.

Given these definitions, we can state our result on optimal ratings
for these classes of distributions:
\begin{prop}[Censorship under ELRP and CLRP]
\label{prop: ELRPCLRP} Suppose Assumptions \ref{assu:Independence.}
and \ref{assu: targetting} hold and FOA is valid.
\begin{enumerate}
\item If $g\left(y|a\right)$ satisfies ELRP, then the highest implementable
effort $a_{LC}$ is the highest value of effort that is incentive
compatible for an interim price associated with lower-censorship ratings,
i.e., ratings that pool values of $y$ below a threshold and reveal
higher values.
\item If $g\left(y|a\right)$ satisfies CLRP, then the highest implementable
effort $a_{UC}$ is the highest value of effort that is incentive
compatible for an interim price associated with upper-censorship ratings,
i.e., ratings that pool values of $y$ above a threshold and reveal
lower values.
\end{enumerate}
\end{prop}
\begin{center}
\begin{figure}[h]
\subfloat[ELRP]{\begin{centering}
\begin{tikzpicture}[scale=.9]
\begin{axis}[
     domain=0.01:10,
    samples=200,
    axis lines=middle,
    xlabel={$y$},
    ylabel={$g(y|a)$},
    xlabel style={at={(axis description cs:1.05,0)}, anchor=west},
    ylabel style={at={(axis description cs:0,1.05)}, anchor=east},
    xtick=\empty,
    ytick=\empty]

\addplot[blue, thick] 
    {1/(x*1*sqrt(2*pi)) * exp(-((ln(x)-1)^2)/(2*1))};

\addplot[red, dashed, thick] 
    {1/(x*sqrt(1.75)*sqrt(2*pi)) * exp(-((ln(x)-1.2)^2)/(2*1.75))};

\end{axis}
\draw (5,5) node{$\log y\sim \mathcal{N}(a,a^3)$};
\draw[blue] (2.9,3) node{$a=1$};
\draw[red] (1.6,2) node{$a=1.2$};
\end{tikzpicture}
\par\end{centering}
}\subfloat[CLRP]{\begin{centering}
\begin{tikzpicture}[scale=.9]
\begin{axis}[
     domain=0.01:17,
    samples=200,
    axis lines=middle,
    xlabel={$y$},
    ylabel={$g(y|a)$},
    xlabel style={at={(axis description cs:1.05,0)}, anchor=west},
    ylabel style={at={(axis description cs:0,1.05)}, anchor=east},
    xtick=\empty,
    ytick=\empty]

\addplot[blue, thick] 
    {1/(x*1*sqrt(2*pi)) * exp(-((ln(x)-1)^2)/(2*1))};

\addplot[red, dashed, thick] 
    {1/(x*sqrt(.6)*sqrt(2*pi)) * exp(-((ln(x)-1.2)^2)/(2*.6))};

\end{axis}
\draw (5,5) node{$\log y\sim \mathcal{N}(a,a^{-3})$};
\draw[blue] (.9,2.5) node{$a=1$};
\draw[red] (2.7,2.7) node{$a=1.2$};
\end{tikzpicture}
\par\end{centering}
}

\caption{Distributions with ELRP (left) and CLRP (right)\label{fig:Distributions-with-ELRP}}
\end{figure}
\par\end{center}

As the above proposition establishes, optimal ratings for ELRP and
CLRP distributions are fairly simple. They involve either lower censorship
(in case of ELRP) or upper censorship (in case of CLRP). In what follows
we provide an example and discuss its implications for various tasks
and technologies.

Suppose that $y\sim\mathcal{N}\left(a,\left(ka\right)^{2}\right)$,
that $y$ determines market values, i.e., $\overline{v}\left(y;a\right)=y$,
and that cost is $c\left(a\right)=a^{2}/2$. Under a full information
rating, $p\left(y\right)=y$, profit of the agent is $a-a^{2}/2$
which is maximized at $a_{FI}=1$. It can be easily checked that in
this case $G\left(y|a\right)$ satisfies ELRP. For any value of $i\in\left[0,1\right]$,
we can find the highest level of effort that is a best response to
pooling of $i$ lowest realizations of the indicator $y$. This is
depicted in Figure \ref{fig:Optimal-efforts-for} (left panel) for
values of $k=1,2,3,4$. At the lowest value, $i=0$, optimal effort
is $a_{FI}=1$. Optimal effort peaks at some threshold, 0.56, 0.69,
0.73, 0.75 respectively, and falls to zero as $i$ tends to 1. It
should be noted that as the variance of $y$ becomes more responsive
to $a$, the highest possible value of effort increases. Figure \ref{fig:Optimal-efforts-for}
(right panel) depicts the marginal change in the quantile as a result
of an increase in $a$, $-F_{a}\left(i|a;a\right)$, and its concavification
(in the case of $k=1$). The threshold for pooling on the right coincides
with the peak of the left plot since optimal ratings take the form
of lower censorship.
\begin{center}
\begin{figure}
\begin{centering}
\subfloat[Optimal effort for pooling the lowest $i$ realizations of $y$]{\begin{centering}
\begin{tikzpicture}[scale=1]
\node[inner sep=0pt] (0,0)
    {\includegraphics[width=.5\textwidth]{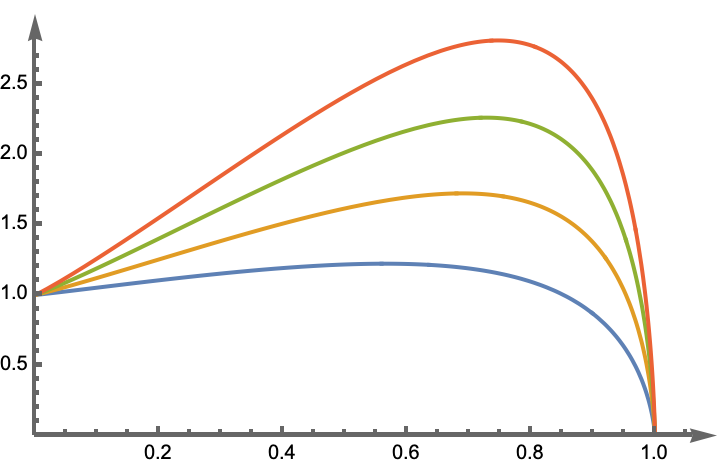}};
\draw[black] (3.9,-2.3) node[above] { $i$};
\draw[black] (-4.3,2.3) node[right] { $a$};
\draw[blue] (-1.5,-.6) node[right] {$k=1$};
\draw[orange] (-.6,0.) node[right] {$k=2$};
\draw[darkgreen] (.5,.8) node[right] {$k=3$};
\draw[red] (.9,1.6) node[right] {$k=4$};
\end{tikzpicture}
\par\end{centering}
}\subfloat[Concavification of the marginal change in quantiles]{\begin{centering}
\begin{tikzpicture}[scale=1]
\node[inner sep=0pt] (0,0)
    {\includegraphics[width=.5\textwidth]{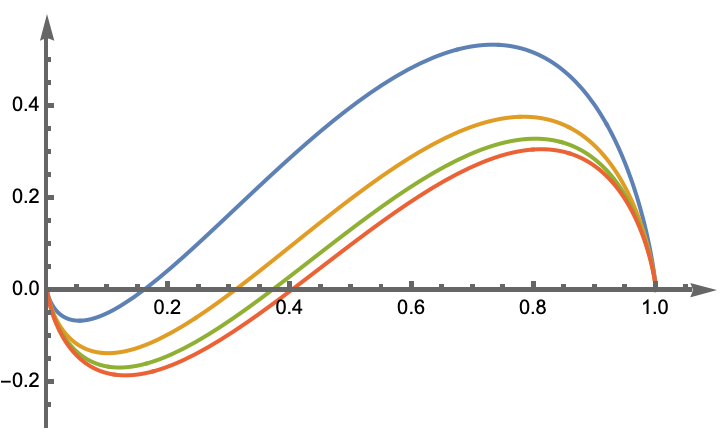}};
\draw[black] (4,-.7) node[above] { $i$};
\draw[black] (-4.2,2.6) node[right] {$-F_a(i|a;a)$};
\draw [very thick,dashed,blue] (-3.52,-.84) -- ++ (3.72,2.33);
\draw[blue, thick] (-1.2,2.1-.72) node[above] {"pooling"} .. controls (-1.2,1.1) .. (-1,1.0);
\end{tikzpicture}
\par\end{centering}
}
\par\end{centering}
\caption{Optimal efforts for lower-censorship policies \label{fig:Optimal-efforts-for}}
\end{figure}
\par\end{center}

We should note that the notions of ELRP and CLRP are tied to whether
an increase in effort $a$ leads to a higher or lower variance of
the indicator $y$. Specifically, suppose that $y=f\left(\sigma\left(a\right)\varepsilon+m\left(a\right)\right)$
with $m$ and $f$ increasing, and $\varepsilon$ has density $e^{h\left(\varepsilon\right)}$
such that $h\left(\varepsilon\right)$ is concave and $\varepsilon+h'\left(\varepsilon\right)/h''\left(\varepsilon\right)$
is increasing in $\varepsilon$. In this case, $y$ exhibits ELRP
(CLRP) if $\sigma\left(a\right)$ is increasing (decreasing) in $a$.
Several classes of distributions satisfy these properties for $h$:
Normal, Gumbel, Generalized Normal, Logistic, etc. For this class
of distributions, a constant $\sigma\left(a\right)$ leads to MLRP.

The above findings point to a practical property of optimal ratings
in targeting an action. Namely, how the variance of the indicator
interacts with the desired action determines the best way to incentivize
it. Specifically, for an activity where more effort leads to a more
precise outcome (CLRP), such as maintenance activities, full revelation
at low values and pooling at higher values encourages higher effort
to avoid punishments for low outcomes. On the other hand, if higher
effort leads to a riskier outcome (ELRP), such as innovative activities,
then pooling of low realizations via lower-censorship ratings provides
insurance against possible downsides and encourages risk taking (increasing
effort). When effort does not affect variance (MLRP), optimal rating
is full revelation and no pooling is needed.

We should end this section by emphasizing that while we have focused
on the highest possible effort that is implementable, any lower value
of effort can also be targeted by rating systems. The analysis in
this section is specifically useful in identifying values of effort
that are higher than those achieved by a fully revealing rating system.
Especially in markets with positive externalities where fully informative
ratings lead to inefficiently low levels of effort, one can use ratings
(absent MLRP) to improve market efficiency.

It is useful to conclude this section with a summary of the results.
Table \ref{tab:taxonomy} collects the taxonomy that emerges from
our analysis: the properties of the technology $g\left(y|a\right)$
and the objective of the intermediary determine both whether information
should be censored and where in the distribution censorship occurs.

\begin{table}
\begin{centering}
\begin{tabular}{>{\raggedright}p{3cm}>{\raggedright}p{3.6cm}>{\raggedright}p{3.4cm}>{\raggedright}p{3.6cm}}
\hline 
\textbf{Technology} & \textbf{Economic interpretation} & \textbf{Optimal rating} & \textbf{Examples}\tabularnewline
\hline 
MLRP & Effort shifts outcomes uniformly upward & Full disclosure (Proposition \ref{prop: MLRP}) & Routine tasks\tabularnewline
ELRP & Effort increases tail risk; innovative activities, R\&D & Lower censorship (Proposition \ref{prop: ELRPCLRP}) & Low-rating forgiveness (Shipt, Instacart)\tabularnewline
CLRP & Effort reduces tail risk; maintenance activities & Upper censorship (Proposition \ref{prop: ELRPCLRP}) & Compressed top ratings (Airbnb)\tabularnewline
\hline 
\end{tabular}
\par\end{centering}
\caption{A taxonomy of optimal ratings\label{tab:taxonomy}}
\end{table}

\section{Multi-task Moral Hazard and Window Dressing \label{sec: Multitasking-1}}

In this section, we illustrate the value of our characterization results
above by applying them to the classic multi-task moral hazard model.
Since the seminal work of \citet{holmstrom1991multitask}, the multi-task
principal-agent models have become the workhorse of analyzing incentives
in settings where agents can use several actions to affect the observed
outcomes -- see also \citet{baker1992incentive} and \citet{dewatripont1999economics}.\footnote{Since \citet{dewatripont1999economics} analyze a variant of \citet{ho99}'s
career concern model, their model is closer to ours where the agent's
compensation is determined by market expectations as opposed to an
endogenous contract chosen by the principal as in \citet{holmstrom1991multitask}
and \citet{baker1992incentive}.

Several empirical studies have looked at variants of the multi-task
moral hazard model. A partial list includes \citet{dumont2008physicians}
and \citet{alexander2020doctors} for compensation of doctors, \citet{acemoglu2020perils}
for incentives in military and security forces, \citet{de2023subjective}
for incentives of government employees, \citet{andrabi2022subjective}
for incentives of teachers, and \citet{mayzlin2014promotional} and
\citet{hui2025designing} for incentives on online platforms.} Despite this prevalence of multi--task moral hazard in applied settings
and the importance of disclosure/rating policies in it, a unified
theoretical analysis of rating policies is essentially non--existent.
In this section, we consider a variant of the model in \citet{dewatripont1999economics}
to understand how rating design can be used to mitigate multi-task
incentive problems. As we will see, a few general lessons will arise
and in some special cases, lessons from Section \ref{sec:Application-1:-Rating}
go through.

As before, the agent chooses a vector of efforts $a=\left(a_{1},a_{2}\right)\in\left[0,\overline{a}\right]^{2}$
with $a_{1}$ being the \textit{productive }effort and $a_{2}$ being
a \textit{window dressing} effort. The difference between productive
and window dressing effort is defined in Assumption \ref{assu:byIn-words--the}
below:
\begin{assumption}[Productive Effort vs. Window Dressing]
\label{assu:byIn-words--the}  The distribution of $\left(v,y\right)$
conditional on $a$ satisfies:
\begin{enumerate}
\item market value $\overline{v}\left(y;a\right)=\mathbb{E}\left[v|y;a\right]$
is strictly increasing in $a_{1}$ and strictly decreasing in $a_{2}$.
\item both efforts increase the indicator, i.e., $\mathbb{E}\left[y|a\right]$,
is increasing in $a_{n}$ for $n=1,2$.
\item welfare is decreasing in $a_{2}$, i.e., $V\left(a\right)=\mathbb{E}\left[v|a\right]-c\left(a\right)$
is decreasing in $a_{2}$ and quasi--concave in $a_{1}$ given $a_{2}$.
\end{enumerate}
\end{assumption}
The above are standard properties of multi--tasking models. They
imply that efficiency leads to $a^{*}_{2}=0$ and $a^{*}_{1}\in\arg\max_{a_{1}}V\left(a_{1},0\right)$.

An example of this is the standard model of \citet{dewatripont1999economics}
where
\begin{align*}
\left(\begin{array}{c}
v\\
y
\end{array}\right) & =\left(\begin{array}{c}
b_{v}\cdot a\\
b_{y}\cdot a
\end{array}\right)+\varepsilon, & \varepsilon\sim\mathcal{N}\left(0,\left(\begin{array}{cc}
\sigma^{2}_{y} & \sigma_{vy}\\
\sigma_{vy} & \sigma^{2}_{v}
\end{array}\right)\right),\sigma_{vy}>0
\end{align*}
where $b_{v},b_{y}\in\mathbb{R}^{2}_{+}$ capture the effect of $a$
on the market value and indicator. Using properties of the normal
distribution, we can show that market values conditional on $y$ and
belief $\hat{a}$ are:
\[
\overline{v}\left(y;\hat{a}\right)=\mathbb{E}\left[v|y;\hat{a}\right]=\beta y+\left(b_{v}-\beta b_{y}\right)\cdot\hat{a}
\]
 where $\beta=\sigma_{vy}/\sigma^{2}_{y}>0$. Since $a_{2}$ must
be a window dressing effort, we should have $b_{v,2}=0,b_{y,2}>0$.

Throughout the section, our focus will be on welfare maximizing rating
design. The key trade-off is that since window--dressing is a completely
wasteful activity but has private benefits for the agent, an intermediary
that cares about welfare might want to hide some information to differentially
affect the incentives for the two activities. In Section \ref{subsec:Reducible-Multi=002013Tasking},
we discuss cases in which it is impossible to differentially affect
incentives for productive and window dressing efforts; \textit{reducible
}multi--tasking. Section \ref{subsec:A-Nonreducible-Two=002013Task}
then discusses \textit{irreducible }multi--tasking technologies and
establishes new general insights in this case.

\subsection{Reducible Multi--Tasking \label{subsec:Reducible-Multi=002013Tasking}}

We say that a multi-task technology is\emph{ reducible} if the effort
vector $a$ affects the distribution of the indicator solely through
a one--dimensional index $m\left(a\right)$, i.e., $G\left(y|a\right)=\hat{G}\left(y|m\left(a\right)\right)$
for some family of distributions $\hat{G}$, such as in the above
example. In this case, the multi-task problem collapses to a single--action
problem: given any level of the index, the agent chooses the cost-minimizing
vector of efforts, and the rating design problem is reducible to a
choice of an interim price function $p\left(\cdot\right)$ and the
index $m$. As a result, the analysis of Section \ref{sec:Application-1:-Rating}
applies directly and the MLRP principles derived there drive optimal
ratings.

Formally, with a reducible technology, one can simply contract on
$m$. This is because the cost of effort is separable from its expected
benefit and we can define the following indirect cost function:
\begin{align*}
C\left(\hat{m}\right)=\min_{a\in\left[0,\overline{a}\right]^{2},m\left(a\right)=\hat{m}} & c\left(a\right)
\end{align*}
and a selection of its solution be $\tilde{a}\left(m\right)$. Given
this, the problem becomes essentially the same as that in Section
\ref{sec:Application-1:-Rating} in which the intermediary is choosing
the index $m$ and interim prices $p\left(y\right)$. As a result,
we can simply apply the same ideas around MLRP (ELRP and CLRP), in
this case applied to $\hat{G}$.

\subsection{Irreducible Multi--Tasking\label{subsec:A-Nonreducible-Two=002013Task}}

The key benefit of the setup above was that it was reducible to the
single effort setup characterized in Section \ref{sec:Application-1:-Rating}.
Here we discuss the case in which the technology is not reducible
to a single effort and its implication on optimal rating design. The
setup is the same as before, i.e., Assumption \ref{assu:byIn-words--the}
holds, except for the fact that $G\left(y|a\right)$ is no longer
reducible. Specifically, let us denote the \textit{score} of task
$n$ by
\[
\ell_{n}\left(y|a\right)=\frac{\partial g\left(y|a\right)}{\partial a_{n}}/g\left(y|a\right).
\]
Then the multi--tasking technology is \textit{irreducible} when $\ell_{1}\left(y|a\right)/\ell_{2}\left(y|a\right)$
is \emph{not} independent of $y$. In this case, different parts of
the indicator distribution carry different information about the two
tasks, and censorship can be used to discriminate between them.

Before proceeding, it is worth pointing out an important difference
between the censorship results below and those of Section \ref{sec:Application-1:-Rating}.
In the single effort model, censorship was optimal only when the technology
deviated from MLRP, that is, when effort expanded (ELRP) or compressed
(CLRP) the distribution of the indicator. In the multi-task model,
no such deviation is needed: even when the indicator is informative
about each effort in the standard way, the welfare-maximizing rating
censors, since the designer attaches weights of opposite signs to
the two incentive constraints.

To make this precise, let us apply Theorem \ref{thm: 1}. Under welfare
maximization, the objective of rating design places no distributional
weights on the realization of the indicator, i.e., $\alpha\left(y\right)=0$
in (\ref{eq: obj}), and is thus independent of $p\left(y\right)$.
Moreover, the shape of the optimal rating is determined by a weighted
value of the marginal change in the quantiles:
\[
\Gamma\left(i|a\right)=-\lambda_{1}\left.\frac{\partial G\left(G^{-1}\left(i|\hat{a}\right)|a\right)}{\partial a_{1}}\right|_{\hat{a}=a}-\lambda_{2}\left.\frac{\partial G\left(G^{-1}\left(i|\hat{a}\right)|a\right)}{\partial a_{2}}\right|_{\hat{a}=a}
\]
 and its concavification. A simple calculation implies that the derivative
of $\Gamma$ with respect to $i$ evaluated at $i=G\left(y|a\right)$
is:
\begin{equation}
\left.\Gamma'\left(i|a\right)\right|_{i=G\left(y|a\right)}=-\lambda_{1}\ell_{1}\left(y|a\right)-\lambda_{2}\ell_{2}\left(y|a\right)\label{eq: GammaMult}
\end{equation}
and thus the local concavity of $\Gamma$ is determined by whether
the above is increasing (convex) or decreasing (concave) in $y$.

Suppose that $\lambda_{1}>0>\lambda_{2}$ which means that it is optimal
to discourage $a_{2}$ and encourage $a_{1}$, which is the natural
case to consider. Note that under MLRP, which means that the $\ell_{n}\left(y\right)$'s
are increasing in $y$, the above is the difference of two increasing
functions and thus the logic from Section \ref{sec:Application-1:-Rating}
does not hold. Indeed, under MLRP where both $\ell_{n}\left(y|a\right)$'s
are increasing, the local concavity of $\Gamma$ changes exactly where
the ratio
\[
r\left(y|a\right)=\frac{\partial\ell_{1}\left(y|a\right)/\partial y}{\partial\ell_{2}\left(y|a\right)/\partial y}
\]
 crosses $\left|\lambda_{2}\right|/\lambda_{1}$. Whether this happens
once, and on which side it begins, is what we take up next. To understand
this better, consider the following example.
\begin{example}
\label{ex: irreducible} Assume:
\[
v=a_{1}\left(\varepsilon_{1}+1\right),y=ba_{1}\left(\varepsilon_{1}+1\right)+a_{2}+\varepsilon_{2}
\]
 with $\varepsilon_{1},\varepsilon_{2}$ being two standard normal
and independent random variables. In this case, two increasing functions
$\mu\left(a\right)$, $\sigma\left(a\right)$ exist such that $G\left(y|a\right)=\Phi\left(\frac{y-\mu\left(a\right)}{\sigma\left(a\right)}\right)$
with $\Phi$ the c.d.f. of standard normal.\footnote{For this example, $\mu\left(a\right)=ba_{1}+a_{2},\sigma\left(a\right)=\sqrt{\left(ba_{1}\right)^{2}+1}$.}
Simple calculations show that in this case, the expression in (\ref{eq: GammaMult})
is the sum of two quadratic functions in $y$ which means that its
derivative is linear and thus changes sign only once. In other words,
$\Gamma\left(i|a\right)$ is either convex--concave or concave--convex.
This implies that optimal rating is either upper or lower censorship.
This can be seen in terms of 
\[
r\left(y|a\right)=b+2\frac{y-\mu\left(a\right)}{\sigma\left(a\right)^{2}}b^{2}a_{1},
\]
which is linear and monotone in $y$. This together with $\ell_{2}$
being increasing in $y$, implies that $\Gamma\left(i|a\right)$ is
either convex--concave or concave--convex.
\end{example}
The technology in this example satisfies MLRP with respect to both
efforts, and involves no expansion or compression of the tails in
the sense of Section \ref{sec:Application-1:-Rating}. In the single
effort model such a technology calls for full disclosure.

This example is a special case of a more general class of distributions
that we define below. The ratio $r\left(y|a\right)$ introduced above
can be interpreted as the relative informativeness of an increase
in $y$ about $a_{1}$ relative to $a_{2}$. In fact, we can define
$I_{n}\left(i|a\right)=-\left.\frac{\partial G\left(G^{-1}\left(i|\hat{a}\right)|a\right)}{\partial a_{n}}\right|_{\hat{a}=a}$
and this is the local Lehmann informativeness, \citet{lehmann1988comparing}
and \citet{persico2000information}, of an increase in $a_{n}$. Simple
application of chain rule says that 
\[
I_{n}'\left(i|a\right)=-\ell_{n}\left(G^{-1}\left(i|a\right)|a\right)
\]
which implies that $r$ coincides with $I_{1}''/I_{2}''$.

For $r\left(y|a\right)$ to be an appropriate notion of relative informativeness
of $y$ for $a_{1}$ and $a_{2}$, MLRP has to hold. For general irreducible
technologies, the following defines the key property used in this
section:
\begin{defn}
\label{def: MRIP} A multi--tasking technology $G\left(y|a\right)$
exhibits \textit{Monotone Relative Informativeness Property (MRIP)
}if for all $y_{2}>y_{1}$ and $a\in\left[0,\overline{a}\right]^{2}$:
\[
D\left(y_{1},y_{2}|a\right)=\frac{\partial\ell_{1}\left(y_{1}|a\right)}{\partial y}\frac{\partial\ell_{2}\left(y_{2}|a\right)}{\partial y}-\frac{\partial\ell_{1}\left(y_{2}|a\right)}{\partial y}\frac{\partial\ell_{2}\left(y_{1}|a\right)}{\partial y}
\]
always has the same sign, i.e., it is non--negative or non--positive
for all such $y_{1},y_{2}$ and $a$.
\end{defn}
Using the Lehmann informativeness analogy, one can interpret MRIP
as stating that a higher value of $y$ provides uniformly more, or
uniformly less information about $a_{1}$ relative to $a_{2}$. When
$G\left(y|a\right)$ exhibits MLRP with respect to $a_{2}$, i.e.,
$\ell_{2}\left(y|a\right)$ is increasing in $y$, $G$ exhibits MRIP
when $r\left(y|a\right)$ is monotone, which is the case we verified
in the example above. Formally, $D\left(y_{1},y_{2}|a\right)$ having
the same sign is equivalent to $\Gamma''\left(i|a\right)$ changing
sign at most once for any $\lambda_{1},\lambda_{2}$. Hence, we have
the following theorem:
\begin{thm}
\label{thm: MultiTask}Suppose that $G\left(y|a\right)$ is an irreducible
technology that exhibits MRIP and FOA is valid. Then welfare maximizing
ratings are either upper or lower censorship.
\end{thm}
The significance of Theorem \ref{thm: MultiTask} is that it ties
a natural notion of informativeness, relative Lehmann informativeness
between the productive and the window-dressing efforts, to general
properties of welfare--maximizing ratings. We should also note that
MRIP is not merely a convenience. When $r\left(y|a\right)$ crosses
$\left|\lambda_{2}\right|/\lambda_{1}$ more than once, the local
concavity of $\Gamma$ alternates and its concavification pools more
than one interval, which leads to the mid-censorship ratings we study
in Section \ref{subsec:Redistributive-Test-Design}.

Theorem \ref{thm: MultiTask} does not determine the direction of
the censorship. While one expects the sign of $D\left(y_{1},y_{2}|a\right)$
to be indicative of that direction, this ultimately depends on one
important factor: how changes in incentives affect market beliefs
which in turn affect the incentives through their effect on $\overline{v}\left(y,a\right)$.
This is in addition to the complementarity between the two efforts
in terms of cost. To this end, we make two assumptions: First, we
assume that $\overline{v}\left(y;a\right)$ is affine in $y$; Second,
we assume that $G\left(y|a\right)$ belongs to the location--scale
class of distributions. Namely, we consider the location--scale family
$G\left(y|a\right)=H\left(\frac{y-\mu\left(a\right)}{\sigma\left(a\right)}\right)$
such that $a$ controls the mean and variance of $y$, with $H$ the
c.d.f. of a distribution with a log--concave density. Several classes
of distributions satisfy MRIP within this family. These include Normal,
Logistic, Gumbel, Generalized Normal, Gamma (with shape parameter
$\geq1$), and Weibull (with shape parameter $\geq1$). We have the
following proposition:\footnote{MRIP fails for $H$ being Cauchy or Frechet but neither of these have
log--concave densities. While MRIP and log--concavity of $h$ are
not equivalent and it is possible to come up with examples that are
log--concave but not MRIP, among the textbook classes of distributions
this is the case.}
\begin{prop}
\label{prop: direction} Suppose that $G$ is a location--scale technology
that exhibits MRIP with a log--concave density and that $\overline{v}\left(y;a\right)$
is a linear function of $y$. Moreover, suppose that FOA is valid.
Then the welfare--maximizing rating is upper censorship if and only
if
\begin{equation}
\frac{\frac{\partial\mu}{\partial a_{2}}\frac{\partial c}{\partial a_{1}}-\frac{\partial\mu}{\partial a_{1}}\frac{\partial c}{\partial a_{2}}}{\frac{\partial\mu}{\partial a_{1}}\frac{\partial\sigma}{\partial a_{2}}-\frac{\partial\sigma}{\partial a_{1}}\frac{\partial\mu}{\partial a_{2}}}\geq0\label{eq: impact-1}
\end{equation}
 and it is lower censorship otherwise.
\end{prop}
In words, if window dressing is the cheaper way to raise the mean
of the indicator -- so that the numerator of (\ref{eq: impact-1})
is positive -- then when window dressing has a larger impact on the
dispersion of the indicator relative to its mean, optimal rating is
upper censorship; lower censorship arises when the impact of window
dressing is more pronounced on the level of the indicator relative
to its dispersion. When instead productive effort is the cheaper source
of the mean of the indicator, the numerator of (\ref{eq: impact-1})
is negative and the direction of censorship is reversed. We provide
the derivations behind Proposition \ref{prop: direction} in Appendix
\ref{sec: OnlineDirection}.

The logic behind this result is one of implementation. Under welfare
maximization, the objective depends on the rating only through the
action that it implements, and thus the role of the rating is to make
the implemented action incentive compatible. Within the location--scale
class, the rating affects incentives only through two statistics:
the marginal reward that it attaches to the level of the indicator
and the marginal reward that it attaches to its dispersion. The agent's
first order conditions determine both at the implemented action. When
window dressing is the cheaper source of the level of the indicator,
a rating that rewards the level alone induces too much window dressing.
The only remaining margin is dispersion and since window dressing
is the more dispersive effort, incentive compatibility requires that
dispersion is punished. Since pooling high realizations of the indicator
removes the reward for upside risk, upper censorship is the rating
that punishes dispersion. When either of the two comparisons in (\ref{eq: impact-1})
is reversed, the same logic delivers lower censorship.

The above insight is the analogue of the results of Section \ref{sec:Application-1:-Rating}.
While there the direction of the censorship was determined by whether
more effort expanded (ELRP) or compressed (CLRP) tail events, here
something similar holds. The difference, however, is that in this
multi--task model what matters is the relative impact of window dressing
and productive efforts on tail events as illustrated by (\ref{eq: impact-1}),
weighed against the relative cost of the two efforts in moving the
level of the indicator. When window dressing has a larger impact on
tail events, high realizations of the indicator should be censored,
and when productive effort has a higher impact on tail events, low
realizations should be censored. One can thus view MRIP as the multi-task
counterpart of the MLRP, ELRP, and CLRP conditions: it plays the same
role for the question of which task the indicator speaks about that
those conditions play for the question of how much it speaks. This
is also the sense in which our results go beyond the logic of low-powered
incentives in \citet{holmstrom1991multitask}: since prices are determined
by the market, the power of incentives cannot be lowered uniformly,
but a rating can remove incentives locally, and it is optimal to do
so exactly where the indicator mainly reflects manipulation.

\section{Application: Redistributive Test Design \label{subsec:Redistributive-Test-Design}}

Recent public discourse in the education realm has highlighted the
biases of standardized testing (such as the SAT) and testing of difficult
subjects (such as math) against students with socioeconomic disadvantages.\footnote{In 2023, the California Board of Education passed the controversial
California Mathematics Framework which sets guidelines for mathematics
education in California public schools. Citing the students' socioeconomic
disadvantages, the framework calls for some relaxations in testing
standards and education of mathematics. For more information, see
the article in \href{https://www.newyorker.com/science/elements/california-students-are-struggling-in-math-will-reforms-make-the-problem-worse}{the New Yorker on the California Mathematics Framework}.} These observations have prompted a movement to relax the requirement
that students take such tests. This includes several universities'
policies to make the SAT optional (see \citet{dessein2025test}) and
attempts at making mathematics education more accessible and easier.
Inspired by this debate, in this section, we provide an alternative
answer to this debate in the form of optimal test design. Namely,
we will show that our main characterization results in Section \ref{sec:General-Model-and-1}
can be used to shed light on this question.

To see this, consider a student who could be of type $\theta\in\left\{ R,P\right\} $
with probabilities $f_{R},f_{P}$. Suppose that both types can exert
effort $a_{\theta}$ which leads to a distribution of an indicator
$y$ which is distributed according to $g\left(y|a_{\theta}\right)$
with support given by $I=\left[\underline{y},\overline{y}\right]$
-- with the possibility that $\underline{y}=-\infty$ and $\overline{y}=\infty$.
The cost of effort for each type is $k_{\theta}c\left(a\right)$ where
$c\left(a\right)$ is a convex and increasing function where $0<k_{R}<k_{P}$.
Finally, let $v=y$ so that the market values are simply the value
of the indicators and let $p\left(y\right)$ be the interim price
function that is increasing.

Consider a rating designer who wishes to maximize the following objective
\begin{align}
\alpha_{P}f_{P}\left[\int_{I}p\left(y\right)dG\left(y|a_{P}\right)-k_{P}c\left(a_{P}\right)\right]+ & \alpha_{R}f_{R}\left[\int_{I}p\left(y\right)dG\left(y|a_{R}\right)-k_{R}c\left(a_{R}\right)\right]\label{eq: ObjRed}
\end{align}
where $\alpha_{P}f_{P}+\alpha_{R}f_{R}=1$ and $\alpha_{P}>1>\alpha_{R}$.
Let us assume that an increase in $a$ leads to an increase in $g\left(y|a\right)$
in the sense of first order stochastic dominance.

The distributional weights $\alpha_{\theta}$ are attached to the
student's type---that is, to the socioeconomic disadvantage captured
by the cost parameter $k_{\theta}$---and not to the realization
of the test score itself. The formulation with weights $\alpha\left(y\right)$
on realizations of the indicator in Section \ref{sec:Application-1:-Rating}
can thus be viewed as the reduced form of the problem studied here
when the designer cannot condition on types.

The problem of optimal rating design is then to find $p\left(y\right)$
and $a_{R},a_{P}$ to maximize the objective in (\ref{eq: ObjRed})
subject to incentive compatibility for both types and that $p\left(y\right)$
is a mean preserving contraction of $y$ which is distributed according
to $f_{P}G\left(y|a_{P}\right)+f_{R}G\left(y|a_{R}\right)=\overline{G}\left(y|a_{R},a_{P}\right)$.
In this case, a similar proof to that of Theorem \ref{thm: 1} implies
that under the validity of FOA, the optimal rating can be found by
concavification of the gain function 
\begin{align*}
\Gamma\left(i\right)= & -\alpha_{P}f_{P}G\left(\overline{G}^{-1}\left(i\right)|a_{P}\right)-\alpha_{R}f_{R}G\left(\overline{G}^{-1}\left(i\right)|a_{R}\right)\\
 & -\lambda_{P}G_{a}\left(\overline{G}^{-1}\left(i\right)|a_{P}\right)-\lambda_{R}G_{a}\left(\overline{G}^{-1}\left(i\right)|a_{R}\right)
\end{align*}
where in the above $i$ is the quantile of $y$ according to $\overline{G}$
and $\lambda_{\theta}$'s are the multipliers on the associated incentive
compatibility constraint. As in the previous sections, we maintain
Assumption \ref{assu:Independence.} and the validity of FOA throughout
this section.

The following proposition guarantees that for a class of distribution
functions, the second derivative of the above object switches sign
at most three times. This would imply that optimal rating is always
switching between at most four regions of pooling and separation:
\begin{prop}
\label{prop:midcensor} Suppose that $\log g\left(y|a\right)=f\left(y\right)+r\left(y\right)m\left(a\right)-b\left(a\right)$
where $r\left(y\right),m\left(a\right),b\left(a\right)$ are increasing
functions and $m\left(a\right)>0$. Then the optimal rating that maximizes
(\ref{eq: ObjRed}) always has at most four alternating intervals
of pooling and revelation.
\end{prop}
The class of distributions considered in Proposition \ref{prop:midcensor}
includes some of the fairly common ones that are used in applied work
including 1. a normally distributed $y$ where either the mean or
the variance is controlled by the action $a$, 2. a log--normally
distributed $y$ such that $a$ controls the mean of $\log y$, 3.
when $y$ is distributed according to an extreme-value distribution
of types 1 and 2 (Gumbel and Fréchet) where the scale parameter is
controlled by $a$, among others. Moreover, the assumption implies
that $\log g$ is supermodular in $\left(y,a\right)$, i.e., it satisfies
MLRP.

Interestingly, one might expect redistributive motives to always favor
pooling low realizations, since pooling at the bottom shields the
disadvantaged type. However, the difference here is that there are
two incentive constraints. Under certain conditions on the likelihood
function $g_{a}/g$ for low realizations of $y$ -- specifically
as it becomes arbitrarily large as $y\rightarrow\underline{y}$, the
incentive effect dominates the redistributive motives for low realizations
and optimal ratings become mid-censorship. To see this, let us consider
the following example.
\begin{example}
\label{exa:Let-us-illustrate}Suppose that $y=a+\varepsilon$ where
$\varepsilon\sim\mathcal{N}\left(0,1\right)$ and that $c\left(a\right)=a^{2}/2$.
Let us also assume that $\alpha_{P}=1/f_{P},\alpha_{R}=0$ so that
objective is to maximize the payoff of the high cost type. As mentioned
before, this example satisfies the requirement of Proposition \ref{prop:midcensor}
and hence optimal ratings switch at most three times between pooling
and revelation. Our calculations illustrate that optimal ratings are
indeed mid-censorship: those that pool middle observations of $y$
and separate the extreme realizations. We further assume that $k_{R}=1/2$
and that $f_{P}=f_{R}=1/2$. Figure \ref{fig:Determining-the-optimal}
depicts the optimal rating and the concavification of the gain function
described above for different values of costs for type $P$.\footnote{In order for the difference between the concavification and the function
to be more visible, we subtract $\left(\alpha_{R}f_{R}+\alpha_{P}f_{P}\right)\left(1-i\right)$
from the gain function $\Gamma$ in the plot. Since this subtraction
is linear, it does not affect the resulting concavification in terms
of the pooling and separating intervals.} In the left panel, the cost of type $P$ is closer to that of type
$R$, $k_{P}=3/4$. In this case, optimal rating pools observations
approximately between the 6th and 71st percentiles of the $y$ distribution.
When the cost for type--$P$ increases to 5/4, the optimal rating
pools observations between the 1st and the 78th percentiles. As can
be seen, as the difference between the two cost types increases, the
rating policy becomes less informative in order to redistribute more
across the types.
\end{example}
\begin{center}
\begin{figure}
\begin{centering}
\subfloat[Low cost for type $P$, $k_{P}=3/4$]{\begin{tikzpicture}[scale=.9]
\node[inner sep=0pt] (0,0)
    {\includegraphics[width=.35\textwidth]{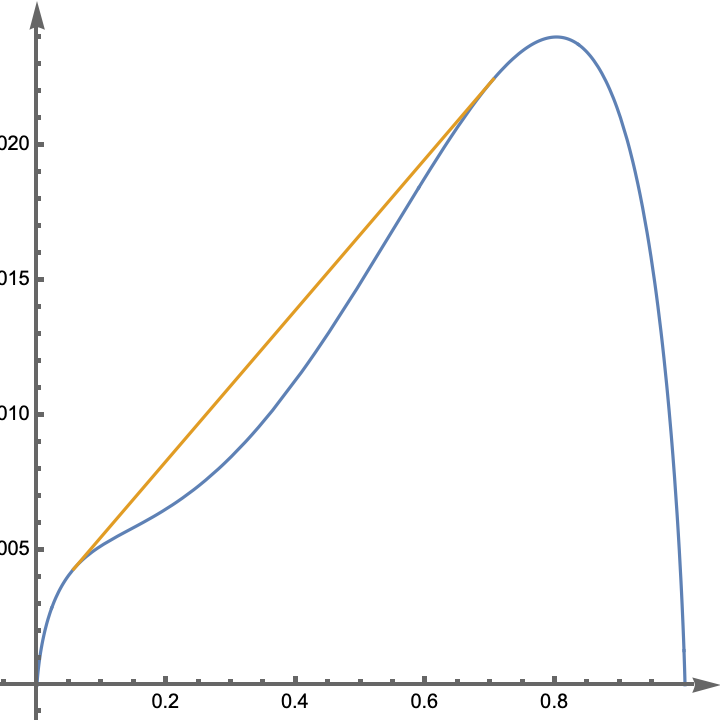}};
\draw[black] (3.4,-3.5) node[above] { $i$};
\draw[red] (-1.3,1.7) node[right] { $\text{cav}\Gamma(i)$};
\draw[blue] (1.3,1.7) node[right] { $\Gamma(i)$};
\end{tikzpicture}

}\subfloat[High cost for type $P$, $k_{P}=5/4$]{\begin{tikzpicture}[scale=.9]
\node[inner sep=0pt] (0,0)
    {\includegraphics[width=.35\textwidth]{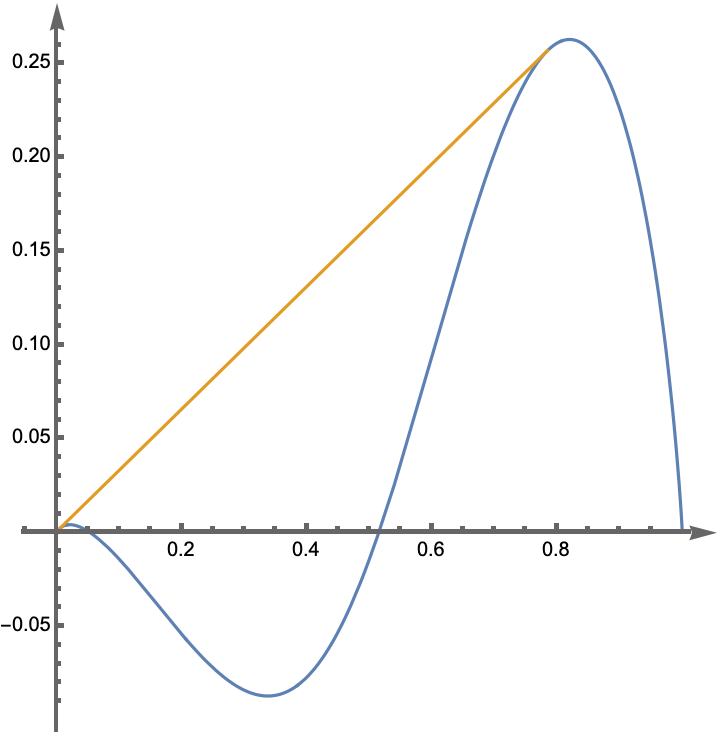}};
\draw[black] (3.4,-2.) node[above] { $i$};
\draw[red] (-1.3,1.7) node[right] { $\text{cav}\Gamma(i)$};
\draw[blue] (1.3,1.7) node[right] { $\Gamma(i)$};
\end{tikzpicture}

}
\par\end{centering}
\caption{Determining the optimal rating for Example \ref{exa:Let-us-illustrate}\label{fig:Determining-the-optimal}}
\end{figure}
\par\end{center}

Viewed through the lens of the current debate on standardized testing,
our analysis suggests an alternative to making tests optional: rather
than discarding the information altogether which is done by making
the tests optional, a designer with redistributive motives can pool
intermediate scores while preserving the incentive effects of the
extremes.

\section{Conclusion}

In this paper, we have developed a general framework for the design
of rating systems in the presence of moral hazard and strategic manipulation.
Our approach makes this problem tractable by identifying ``interim
prices'' as the sufficient statistic for the agent's incentives.
Under a natural monotonicity restriction, implementability is equivalent
to a majorization restriction: feasible interim prices are mean-preserving
contractions of the full-information market value. This converts rating
design into the classic moral-hazard problem with a majorization constraint.

Building on this characterization, we provide a general solution method.
Under a first-order approach to incentive compatibility, optimal rating
design reduces to concavifying a gain function in quantile space.
This formulation yields two broad takeaways. First, it delivers a
set of sufficient statistics for optimal transparency: the technology
matters through how effort shifts the quantile distribution of the
indicator and, in turn, the distribution of market values. Second,
it implies structure on optimal information policies. Under mild conditions,
optimal ratings are simple deterministic monotone partitions of the
indicator space: the intermediary either fully reveals the indicator
on some regions or pools contiguous regions into coarse categories.

The economic insights that emerge from our analysis connect the statistical
properties of the task to the structure of optimal disclosure. In
the canonical benchmark with monotone likelihood ratios, full information
disclosure achieves the highest implementable effort. Departures from
this benchmark generate systematic and testable patterns of censorship.
For innovative activities where greater effort expands outcome variance
(ELRP), \emph{lower-censorship} ratings that pool poor realizations
provide insurance against downside risk, encouraging risk-taking.
For maintenance activities where effort compresses variance (CLRP),
\emph{upper-censorship} ratings that pool high realizations punish
poor outcomes by deterring negligence and encouraging consistency.
Strong redistributive motives create a tension between incentive provision
and protecting disadvantaged agents, which our test-design application
resolves with mid-censorship. More broadly, the framework clarifies
why ``more transparency'' is not a universal remedy: whether additional
information strengthens incentives depends on which parts of the outcome
distribution effort affects.

Our second set of results concerns multi-task environments with window
dressing, where more informative ratings can intensify incentives
for manipulable activities that improve measured performance without
an increase in underlying value. Under the monotone relative informativeness
property (MRIP)---the multi-task counterpart of the single-task taxonomy---welfare-maximizing
ratings are again upper or lower censorship, even when the technology
satisfies MLRP and the single-task logic would call for full disclosure.
Finally, in our application to redistributive test design, we show
how the same concavification logic rationalizes \emph{mid-censorship}
rules that pool intermediate outcomes while separating extremes. These
results formalize a common policy intuition: the optimal granularity
of evaluation depends jointly on incentive provision, manipulability,
and distributional objectives.

\bibliographystyle{ecta}
\bibliography{econbib}

\appendix

\section{Proofs}

\subsection{Proof of Proposition \ref{thm:Suppose-that-T1}\label{subsec:Proof-of-Proposition}}
\begin{proof}
The if side of the statement, that is if $p$ is constructed from
some information structure $\left(\pi,S\right)$ then $p\succcurlyeq_{\text{cv}}\overline{v}$
is immediate from the text. For the reverse, we will first prove the
proposition when the support of the indicator $Y$ is finite. We will
then show that the proposition can be extended to the case when $Y$
is a compact Euclidean space.

\textbf{1. When $Y$ is finite.} Let $Y=\left\{ y_{1},y_{2},\cdots,y_{n}\right\} $
such that $\overline{v}\left(y_{1}\right)\leq\overline{v}\left(y_{2}\right)\leq\cdots\leq\overline{v}\left(y_{n}\right)$.
When $p\left(y\right)$ is co-monotone with $\overline{v}\left(y\right)$,
we must have that $p\left(y_{1}\right)\leq p\left(y_{2}\right)\leq\cdots\leq p\left(y_{n}\right)$.
In this case, $\mathcal{S}\subset\mathbb{R}^{n}$. For simplicity,
we also let $f_{i}=\mu_{y}\left(\left\{ y_{i}\right\} \right)$ and
$\overline{v}_{i}=\overline{v}\left(y_{i}\right)$ and $p_{i}=p\left(y_{i}\right)$.

We prove the claim by induction on $n$.

\emph{First step: The claim holds for $n=2$.}

If $n=2$, then majorization implies that
\begin{align*}
f_{1}\overline{v}_{1}+f_{2}\overline{v}_{2} & =f_{1}p_{1}+f_{2}p_{2},\\
0\leq p_{2}-p_{1} & \leq\overline{v}_{2}-\overline{v}_{1}.
\end{align*}
Consider a signal structure that only sends a low signal when the
state is $y_{1}$. When the state is $y_{2}$, it sends the low signal
with probability $\alpha$ and otherwise a high signal. Let 
\[
0\leq\alpha=\frac{p_{1}-\overline{v}_{1}}{\overline{v}_{2}-p_{1}}\frac{f_{1}}{f_{2}}\leq1.
\]
Then, the ex post price upon observing the low signal is 
\[
\frac{\overline{v}_{1}f_{1}+\alpha\overline{v}_{2}f_{2}}{f_{1}+\alpha f_{2}}=\frac{\overline{v}_{1}f_{1}+\frac{p_{1}-\overline{v}_{1}}{\overline{v}_{2}-p_{1}}f_{1}\overline{v}_{2}}{f_{1}+\frac{p_{1}-\overline{v}_{1}}{\overline{v}_{2}-p_{1}}f_{1}}=\frac{\overline{v}_{1}f_{1}\left(\overline{v}_{2}-p_{1}\right)+\left(p_{1}-\overline{v}_{1}\right)f_{1}\overline{v}_{2}}{f_{1}\left(\overline{v}_{2}-p_{1}\right)+\left(p_{1}-\overline{v}_{1}\right)f_{1}}=p_{1}.
\]
Thus the interim price when the state $y_{1}$ is $p_{1}$. The mean
of $p_{1}$ and $p_{2}$ is the same as that of $\overline{v}_{1},\overline{v}_{2}$
which means that $p_{2}$ is the second order expectation of $\overline{v}$
when the state is $y_{2}$. This proves the claim.

\emph{Second Step: If the claim is true for $n-1$, then it is true
for $n$.}

Consider an interim price function $\left\{ p_{i}\right\} $. If for
some $i$, $p_{i}=p_{i+1}$, we can reduce the number of states by
considering the interim price function $p_{1}\leq\cdots\leq p_{i}\leq p_{i+2}\leq\cdots\leq p_{n}$
distributed according to $f_{1},\cdots,f_{i-1},f_{i}+f_{i+1},f_{i+2},\cdots,f_{n}$
and market values of $\overline{v}_{1}\leq\cdots\leq\overline{v}_{i-1}\leq\frac{f_{i}\overline{v}_{i}+f_{i+1}\overline{v}_{i+1}}{f_{i}+f_{i+1}}\leq\overline{v}_{i+2}\leq\cdots\leq\overline{v}_{n}$.
Since by the induction assumption, an information structure $\pi$
exists that generates this interim price function, simply pooling
$y_{i}$ and $y_{i+1}$ and using the information structure $\pi$
generates the original interim price function.

Suppose, on the other hand, that $p_{i}<p_{i+1}$ for all $i$. Let
$\lambda=\min_{i\leq n-1}\frac{p_{i+1}-p_{i}}{\overline{v}_{i+1}-\overline{v}_{i}}$.
Let $\hat{p}_{j}$ be defined by
\[
\hat{p}_{j}=\frac{p_{j}-\lambda\overline{v}_{j}}{1-\lambda}
\]
Then $\left(1-\lambda\right)\left(\hat{p}_{j+1}-\hat{p}_{j}\right)=p_{j+1}-p_{j}-\lambda\left(\overline{v}_{j+1}-\overline{v}_{j}\right)\geq0$.
Moreover,
\[
\sum^{k}_{j=1}f_{j}\left(\hat{p}_{j}-\overline{v}_{j}\right)=\sum^{k}_{j=1}f_{j}\frac{p_{j}-\overline{v}_{j}}{1-\lambda}\geq0,
\]
 and, finally, if $\lambda=\left(p_{i+1}-p_{i}\right)/\left(\overline{v}_{i+1}-\overline{v}_{i}\right)$,
then $\hat{p}_{i}=\hat{p}_{i+1}$. This implies that an argument similar
to the above shows that an information structure $\left(\hat{\pi},\hat{S}\right)$
exists that generates $\hat{p}$ as its interim price. Now consider
an information structure that reveals the state with probability $\lambda$
and otherwise it is the same as $\hat{\pi}$. That is
\[
\pi\left(s|y_{j}\right)=\begin{cases}
\lambda & s=y_{j}\\
\left(1-\lambda\right)\hat{\pi}\left(s|y_{j}\right) & s\in\hat{S}.
\end{cases}
\]
Then, since the set of signals that reveal the state does not overlap
with $\hat{S}$, we must have that
\begin{align*}
\sum_{s}\pi\left(s|y_{j}\right)\frac{\sum_{i}\pi\left(s|y_{i}\right)f_{i}\overline{v}_{i}}{\sum_{i}\pi\left(s|y_{i}\right)f_{i}} & =\\
\sum_{s\in\hat{S}}\left(1-\lambda\right)\hat{\pi}\left(s|y_{j}\right)\frac{\sum_{i}\hat{\pi}\left(s|y_{i}\right)f_{i}\overline{v}_{i}}{\sum_{i}\hat{\pi}\left(s|y_{i}\right)f_{i}}+\lambda\overline{v}_{j} & =\\
\left(1-\lambda\right)\hat{p}_{j}+\lambda\overline{v}_{j} & =p_{j}.
\end{align*}
This concludes the proof.

\textbf{2. When $Y$ is a subset of $\mathbb{R}$. }Let $Y$ be the
support of the indicator $y$ and a subset of $\mathbb{R}$. Consider
a sequence of partitions $Y^{n}=\left\{ \min Y=y^{n}_{0}<y^{n}_{1}<\cdots<y^{n}_{n}=\max Y\right\} $
for $n=1,2,\cdots$ with $\max_{0\leq i\leq n-1}y^{n}_{i+1}-y^{n}_{i}\rightarrow0$
and $Y^{n}\subset Y^{n+1}$. We define
\begin{align*}
g^{n}_{i}=\Pr\left(y\in\left[y^{n}_{i-1},y^{n}_{i}\right)\right),1\leq i\leq n
\end{align*}
\begin{align*}
\overline{v}^{n}_{i}=\begin{cases}
\frac{\int\overline{v}\left(y\right)\mathbf{1}\left[y\in\left[y^{n}_{i-1},y^{n}_{i}\right)\right]dG}{g^{n}_{i}} & g^{n}_{i}>0,i\leq n\\
\frac{\overline{v}\left(y^{n}_{i-1}\right)+\overline{v}\left(y^{n}_{i}\right)}{2} & g^{n}_{i}=0,i\geq1
\end{cases} & \ \ \ \ p^{n}_{i}=\begin{cases}
\frac{\int p\left(y\right)\mathbf{1}\left[y\in\left[y^{n}_{i-1},y^{n}_{i}\right)\right]dG}{g^{n}_{i}} & g^{n}_{i}>0,i\leq n\\
\frac{p\left(y^{n}_{i-1}\right)+p\left(y^{n}_{i}\right)}{2} & g^{n}_{i}=0,i\geq1.
\end{cases}
\end{align*}
In words, the above constructs a discretization of the buyer values
$v$ and the intermediary's interim prices $p$. Since $p\left(y\right)$
is comonotone with $\overline{v}\left(y\right)$, so are $\overline{v}^{n}$
and $p^{n}$. Moreover, $p^{n}\succcurlyeq_{\text{SOSD}}\overline{v}^{n}$
and $Y^{n}$ is finite, we can apply the result from the first part.
That is, an information structure $\left(S^{n},\pi^{n}\right)$ exists
where $\pi^{n}:Y^{n}\rightarrow\Delta\left(S^{n}\right)$ such that
$p^{n}\left(y^{n}_{i}\right)=\sum_{s\in S^{n}}\pi^{n}\left(\left\{ s\right\} |y^{n}_{i}\right)\mathbb{E}\left[\overline{v}|s\right].$

Note that each $\left(S^{n},\pi^{n}\right)$ induces a distribution
over posterior beliefs of the buyers given by $\tau^{n}\in\Delta\left(\Delta\left(Y^{n}\right)\right)$,
since any probability measure in $\Delta\left(Y^{n}\right)$ can be
embedded in $\Delta\left(Y\right)$. This is because for any $\mu\in\Delta\left(Y^{n}\right)$,
we can construct $\hat{\mu}\in\Delta\left(Y\right)$ defined by $\hat{\mu}\left(A\right)=\sum^{n}_{i=1}\mu^{n}_{i}\mathbf{1}\left[y^{n}_{i}\in A\right]$,
where $A$ is an arbitrary Borel subset of $Y$. Similarly, we can
find $\hat{\tau}^{n}\in\Delta\left(\Delta\left(Y\right)\right)$,
which is equivalent to $\tau^{n}$.\\
Now consider the probability measure $\zeta^{n}$ representing the
joint distribution of $y^{n}$ and posterior mean $\mathbb{E}_{\mu}\left[v\right]=\int vd\mu$
for any $\mu\in\text{Supp}\left(\hat{\tau}^{n}\right)$ induced by
$\hat{\tau}^{n}$. Note that $\zeta^{n}\in\Delta\left(Y\times Y\right)$.
By an application of Riesz representation theorem (see Theorem 14.12
in \citet{charalambos2013infinite}), $\Delta\left(Y\times Y\right)$
is compact according to the weak-{*} topology. This implies that the
sequence $\left\{ \zeta^{n}\right\} $ must have a convergent subsequence
whose limit is given by $\zeta\in\Delta\left(Y\times Y\right)$. Let
$\mathcal{G}^{n}$ be the $\sigma$-field generated by the sets $\left\{ \left[y^{n}_{i},y^{n}_{i+1}\right)\right\} _{i\leq n}$
and let $\mathcal{F}^{n}=\mathcal{G}^{n}\times\left\{ \emptyset,\Delta\left(Y\right)\right\} $.
In words, $\mathcal{F}^{n}$ conveys the information that $y\in\left[y^{n}_{i},y^{n}_{i+1}\right)$.
Note that $\mathcal{F}^{n}\subset\mathcal{F}^{n+1}$ because $Y^{n}\subset Y^{n+1}$.
Moreover,
\[
\mathbb{E}\left[\zeta^{n}|\mathcal{F}^{n}\right]=\left(\overline{v}^{n},p^{n}\right),
\]
where $\left(\overline{v}^{n},p^{n}\right)$ is the random variable
with values $\left(\overline{v}^{n}_{i},p^{n}_{i}\right)$ with probability
$g^{n}_{i}$. Note that the above holds by the construction of $\tau^{n}$
and $\zeta^{n}$. As a result
\begin{align*}
\mathbb{E}\left[\zeta^{n+1}|\mathcal{F}^{n}\right] & =\mathbb{E}\left[\mathbb{E}\left[\zeta^{n+1}|\mathcal{F}^{n+1}\right]|\mathcal{F}^{n}\right]=\mathbb{E}\left[\left(\overline{v}^{n+1},p^{n+1}\right)|\mathcal{F}^{n}\right]=\left(\overline{v}^{n},p^{n}\right),
\end{align*}
where the last equality follows because $\mathbb{E}\left[p\left(y\right)|\mathcal{F}^{n}\right]=p^{n},\mathbb{E}\left[\overline{v}\left(y\right)|\mathcal{F}^{n}\right]=\overline{v}^{n}$
given the definition of $p^{n}$ and $\overline{v}^{n}$ above. All
of this implies that $\mathcal{F}^{n}$ is a filtration and $\left(\zeta^{n},\mathcal{F}^{n}\right)$
forms a bounded martingale (for a definition see \citet{doob2012measure}).
Hence by Doob's martingale convergence theorem (see Theorem XI.14
in \citet{doob2012measure}), we must have that
\[
\lim_{n\rightarrow\infty}\mathbb{E}\left[\zeta^{n}|\mathcal{F}^{n}\right]=\mathbb{E}\left[\zeta|\mathcal{F}\right].
\]
Therefore, $\mathbb{E}_{\zeta}\left[\int\overline{v}\left(y'\right)d\mu|y\right]=p\left(y\right)$.
This concludes the proof.
\end{proof}

\subsection{Proof of Theorem \ref{thm: 1}}

To prove this theorem we first need to prove the following lemma.
\begin{lem}
\label{lem:Let--be}Let $h\left(y\right)$ be an integrable function.
Suppose that $H\left(i\right)=\int_{\left\{ y:\overline{v}\left(y\right)>v_{Q}\left(i\right)\right\} }h\left(y\right)dG$
and let $\text{cav}H$ be the concave envelope of $H$, i.e., the
lowest concave function that is higher than $H\left(i\right)$. Then
\begin{equation}
\max_{\begin{array}{c}
p:p\succcurlyeq_{\text{cv}}\overline{v},\\
p,\overline{v}:\text{comonotone}
\end{array}}\int h\left(y\right)p\left(y\right)dG=\int^{1}_{0}\text{cav}H\left(i\right)dv_{Q}\left(i\right)\label{eq: equality}
\end{equation}
Moreover, an optimal $p$ that achieves the above value satisfies
$p\left(y\right)=\overline{v}\left(y;a\right)$ when $H\left(G\left(y|a\right)\right)=\text{cav}H\left(G\left(y|a\right)\right)$.
Moreover, if for some maximal interval $I\subset\left[0,1\right]$,
$\text{cav}H\left(i\right)<H\left(i\right),i=G\left(y|a\right)\in\text{int}I$,
$p\left(y\right)=\mathbb{E}\left[\overline{v}|G\left(y|a\right)\in I\right]$.
\end{lem}
\begin{proof}
In this proof, we assume that $-\infty<v_{Q}\left(0\right)<v_{Q}\left(1\right)<\infty$.
The cases with $v_{Q}\left(1\right)=\infty$ or $v_{Q}\left(0\right)=-\infty$
can be proved using a limiting argument. Before proceeding, we prove
the following lemma:
\begin{lem}
\label{lem: invert} Let $p,\overline{v}$ be comonotone and $p_{Q}\left(i\right),v_{Q}\left(i\right)$
be their associated quantile representation as defined in (\ref{eq: quant}).
If $F_{p}\left(v\right),F_{v}\left(v\right)$ are the cumulative distribution
functions of $p,\overline{v}$ respectively, then $p\succcurlyeq_{\text{cv}}\overline{v}$
if and only if $F_{v}\left(v\right)\succcurlyeq_{\text{cv}}F_{p}\left(v\right)$
where $v$ is uniformly distributed over $V=\left[v_{Q}\left(0\right),v_{Q}\left(1\right)\right]$.
In other words,
\begin{align*}
\int^{1}_{0}\phi\left(p_{Q}\left(i\right)\right)di & \geq\int^{1}_{0}\phi\left(v_{Q}\left(i\right)\right)di,\forall\phi:V\rightarrow\mathbb{R}:\text{concave}\Leftrightarrow\\
 & \int_{V}\psi\left(F_{v}\left(v\right)\right)dv\geq\int_{V}\psi\left(F_{p}\left(v\right)\right)dv,\forall\psi:\left[0,1\right]\rightarrow\mathbb{R}:\text{concave}
\end{align*}
\end{lem}
An example that illustrates Lemma \ref{lem: invert} is depicted in
Figure \ref{fig:interim-prices.-market}. On the left, we have an
example of $p,\overline{v}$ (not their quantile version) where $p$
is a mean--preserving contraction of $\overline{v}$ while this is
reversed for their c.d.f.'s. The idea behind Lemma \ref{lem: invert}
is simple. Since c.d.f.'s are inverses of the quantile functions mean
preserving contraction for one implies mean preserving spread for
the other. We provide its proof in the Online Appendix, Section \ref{sec:Proof-of-Lemma}.
In the above, we can view $i=F_{v}\left(v\right)$ as having a distribution
according to $\frac{v_{Q}\left(i\right)-v_{Q}\left(0\right)}{v_{Q}\left(1\right)-v_{Q}\left(0\right)}$
and $j=F_{p}\left(v\right)$ as having a distribution $\frac{p_{Q}\left(j\right)-v_{Q}\left(0\right)}{v_{Q}\left(1\right)-v_{Q}\left(0\right)}$
with probability $\frac{v_{Q}\left(1\right)-p_{Q}\left(1\right)}{v_{Q}\left(1\right)-v_{Q}\left(0\right)}$
on $j=1$.

\begin{figure}
\begin{centering}
\begin{tikzpicture}[scale=0.4]
\coordinate (Origin) at (0,0);
\draw[-stealth] (0,0) node[left] {$v_{\min}$} -- (10.,0) node[anchor=north west,right=5pt, below=-5pt] {$y$}; 
\draw[-stealth] (0.3,0) -- (0.3,10);
\draw[very thick, color=blue] (0.3,0) ..  controls (2.5,7.5) and (3.5,7) .. (9,9) node[above = 6pt,left=6pt] {$\overline{v}(y) = \mathbb{E}\left[v|y\right]$};
\draw[very thick, color=red] (0.3,3)  ..  controls (.8,4)  ..  (4.5,5) ..  controls (6,6) .. (9,7) node[below=10pt] {$p(y)$};
\draw  (0.3,.3) -- (0.3,-.3) node[below]{$\underline{y}$};
\draw  (9.3,.3) -- (9.3,-.3) node[below]{$\overline{y}$};
\draw [dashed] (0,9) node[left=1pt]{$v_{\max}$} -- (9,9);
\draw[-stealth] (12,0) node[left] {$0$} -- (22.,0) node[anchor=north west,right=5pt, below=-5pt] {$v$}; 
\draw[-stealth] (12.3,0) -- (12.3,10);
\draw[very thick, color=blue] (12.3,0) ..  controls (18.5,3.5) and (20,5) .. (21,9) node[below = 13pt,right=5pt] {$F_v(v)$};
\draw[very thick, color=red] (12.3,0) -- (15.3,0) ..  controls (16.,1) .. (17,6) .. controls (18.5,7.5) .. (19,9) node[left=15pt,above=2pt] {$F_p(v)$} -- (21,9);
\draw  (12.3,.3) -- (12.3,-.3) node[below]{$v_{\min}$};
\draw  (21.3,.3) -- (21.3,-.3) node[below]{$v_{\max}$};
\draw[dashed] (12,9) node[left=1pt]{$1$} -- (21,9);
\end{tikzpicture}
\par\end{centering}
\caption{Example of $p\succcurlyeq_{\text{cv}}\overline{v}$ (left) and their
associated CDF's (right) that satisfy $F_{v}\succcurlyeq_{\text{cv}}F_{p}$\label{fig:interim-prices.-market}}
\end{figure}

Now, if $p\succcurlyeq_{\text{cv}}\overline{v}$, Lemma \ref{lem: invert}
implies that $F_{v}\succcurlyeq_{\text{cv}}F_{p}$. By Blackwell's
theorem (see also \citet{kolotilin2018optimal} and \citet{Gentzkow}),
there must exist a signal structure (or a garbling) where $F_{v}\left(v\right)=i$
is the conditional mean of $F_{p}=j$ upon realization of the signal
with $i$ having distribution $dv_{Q}\left(i\right)/\left(v_{Q}\left(i\right)-v_{Q}\left(0\right)\right)$
and similarly for $j$. Let $\mu\left(\cdot|i\right)\in\Delta\left[0,1\right]$
be the posterior associated with $F_{v}\left(v\right)=i$. Since the
distribution of $i$ is given by $dv_{Q}\left(i\right)/\left(v_{Q}\left(i\right)-v_{Q}\left(0\right)\right)$
we can write Bayes plausibility as
\begin{equation}
\int\mu\left(A|i\right)\frac{dv_{Q}\left(i\right)}{\left|V\right|}=\frac{\int_{A}dp_{Q}\left(j\right)}{\left|V\right|},A\subset\left[0,1\right]\label{eq: BPinvert}
\end{equation}
such that
\begin{equation}
i=\int jd\mu\left(j|i\right),\forall i\in\left[0,1\right]\label{eq: posteriorm}
\end{equation}
where in the above $\left|A\right|$ is the Lebesgue measure of $A$
and $\left|V\right|=v_{Q}\left(1\right)-v_{Q}\left(0\right)$. Equation
(\ref{eq: posteriorm}) simply states that $i=F_{v}\left(v\right)$
is the posterior mean of $j=F_{p}\left(v\right)$ according to the
distribution $\mu\left(\cdot|i\right)$.

We can then write
\begin{align*}
\int_{Y}p\left(y\right)h\left(y\right)dG & =-\int^{1}_{0}p_{Q}\left(j\right)dH\left(j\right)\\
 & =-\int^{1}_{0}\mu\left(\left[0,j\right]|i\right)dv_{Q}\left(i\right)dH\left(j\right)\text{ by setting \ensuremath{A}=\ensuremath{\left[0,j\right]} in (\ref{eq: BPinvert})}\\
 & =-\int^{1}_{0}\int^{1}_{0}\mu\left(\left[0,j\right]|i\right)dH\left(j\right)dv_{Q}\left(i\right)\text{ Fubini's Theorem}\\
 & =-\int^{1}_{0}\left[H\left(1\right)-\int^{1}_{0}H\left(j\right)d\mu\left(j|i\right)\right]dv_{Q}\left(i\right)\text{ Int. by parts for inside integral}\\
 & =\int^{1}_{0}\int^{1}_{0}H\left(j\right)d\mu\left(j|i\right)dv_{Q}\left(i\right)\text{ since by construction \ensuremath{H\left(1\right)=0}}
\end{align*}
Now, consider the expression $\int^{1}_{0}H\left(j\right)d\mu\left(j|i\right)$.
In this expression $\mu\left(\cdot|i\right)$ is a probability distribution
over $j$ whose mean value is $i$. In other words, this is a convex
combination of values of $H\left(j\right)$ with average of $i$.
Given the definition of the concave envelope, we have that $\int^{1}_{0}H\left(j\right)d\mu\left(j|i\right)\leq\text{cav}H\left(i\right)$.
Moreover, since the space of measures over $\left[0,1\right]$ is
compact according to weak--{*} topology (Banach-Alaoglu theorem)
for each $i$ there must exist $\mu\left(\cdot|i\right)\in\Delta\left[0,1\right]$
such that $\int^{1}_{0}H\left(j\right)d\mu\left(j|i\right)=\text{cav}H\left(i\right)$.
We can then use the above procedure to construct an interim price
function that delivers $\int\text{cav}H\left(i\right)dv_{Q}\left(i\right)$.
This proves the equality (\ref{eq: equality}).

To prove the second part, by Caratheodory theorem, for any $i$ either
$\text{cav}H\left(i\right)=H\left(i\right)$ or that there exists
$i_{1}<i<i_{2}$ such that
\[
\text{cav}H\left(i\right)=\frac{i_{2}-i}{i_{2}-i_{1}}H\left(i_{1}\right)+\frac{i-i_{1}}{i_{2}-i_{1}}H\left(i_{2}\right)
\]
In the first case, $\mu\left(\left\{ i\right\} |i\right)=1$ and in
the second case $\mu\left(\left\{ i_{1}\right\} |i\right)=\frac{i_{2}-i}{i_{2}-i_{1}}=1-\mu\left(\left\{ i_{2}\right\} |i\right)$.
In other words, concavification of $H$ partitions the range of $\left[0,1\right]$
into subintervals $\left(i_{1,\alpha},i_{2,\alpha}\right)$, $\left[i_{1,\beta},i_{2,\beta}\right]$
where for any $\alpha$ there exists $\beta$ such that $i_{2,\alpha}=i_{1,\beta}$
and another $\beta$ for which $i_{1,\alpha}=i_{2,\beta}$ where for
all $i\in\left[i_{1,\beta},i_{2,\beta}\right]$, $\mu\left(\left\{ i\right\} |i\right)=1$
and for all $i\in\left(i_{1,\alpha},i_{2,\alpha}\right)$, $\mu\left(\left\{ i_{1,\alpha}\right\} |i\right)=\frac{i_{2,\alpha}-i}{i_{2,\alpha}-i_{1,\alpha}}=1-\mu\left(\left\{ i_{2,\alpha}\right\} |i\right)$.
From above we know that 
\[
p_{Q}\left(j\right)=\int^{1}_{0}\mu\left(\left[0,j\right]|i\right)dv_{Q}\left(i\right)
\]
Given the construction of $\mu$, we have
\[
\mu\left(\left[0,j\right]|i\right)=\begin{cases}
\mathbf{1}\left[j\geq i\right] & \text{cav}H\left(i\right)=H\left(i\right)\\
\frac{i_{2,\alpha}-i}{i_{2,\alpha}-i_{1,\alpha}}\mathbf{1}\left[i_{2,\alpha}>j\geq i_{1,\alpha}\right]+\mathbf{1}\left[j\geq i_{2,\alpha}\right] & \text{otherwise}
\end{cases}
\]
Now, suppose that $j\in\left[i_{1,\beta},i_{2,\beta}\right]$ for
some $\beta$. This means that if $i>j$, $\mu\left(\left[0,j\right]|i\right)=0$
since all higher quantiles either put weights only on $i$ or on values
of the form $i_{1,\alpha},i_{2,\alpha}$ which are higher than $j$.
Then we can write
\begin{align*}
p_{Q}\left(j\right) & =\int^{1}_{0}\mu\left(\left[0,j\right]|i\right)dv_{Q}\left(i\right)\\
 & =\int^{j}_{0}dv_{Q}\left(i\right)=v_{Q}\left(j\right)
\end{align*}
Moreover, if $j\in\left(i_{1,\alpha},i_{2,\alpha}\right)$ for some
$\alpha$, then it has to be that if $i>i_{2,\alpha}$, then $\mu\left(\left[0,j\right]|i\right)=0$.
So we can write
\begin{align*}
p_{Q}\left(j\right) & =\int^{i_{2,\alpha}}_{0}\mu\left(\left[0,j\right]|i\right)dv_{Q}\left(i\right)\\
 & =\int^{i_{1,\alpha}}_{0}dv_{Q}\left(i\right)+\int^{i_{2,\alpha}}_{i_{1,\alpha}}\mu\left(\left[0,j\right]|i\right)dv_{Q}\left(i\right)\\
 & =v_{Q}\left(i_{1,\alpha}\right)+\int^{i_{2,\alpha}}_{i_{1,\alpha}}\frac{i_{2,\alpha}-i}{i_{2,\alpha}-i_{1,\alpha}}dv_{Q}\left(i\right)\\
 & =v_{Q}\left(i_{1,\alpha}\right)-v_{Q}\left(i_{1,\alpha}\right)+\int^{i_{2,\alpha}}_{i_{1,\alpha}}\frac{v_{Q}\left(i\right)}{i_{2,\alpha}-i_{1,\alpha}}di\\
 & =\mathbb{E}\left[v_{Q}\left(i\right)|i\in\left(i_{1,\alpha},i_{2,\alpha}\right)\right]
\end{align*}
This establishes the claim.
\end{proof}

\subsubsection{Proof of the Theorem}
\begin{proof}
Consider the problem of finding the best interim price function in
quantile form for a given action $a$:
\[
\max_{p_{Q}}W\left(a\right)-\int p_{Q}\left(i\right)dH\left(i;a\right)=\max_{p_{Q}}W\left(a\right)+T_{H}p_{Q}
\]
 subject to $p_{Q}\succcurlyeq_{\text{cv}}v_{Q}$, monotonicity of
$p_{Q}$ and the first order IC constraints. We will show that for
any $a\in A$, Lagrange multipliers associated with the first order
IC constraints exist so that the constrained optimization gives the
same value as the unconstrained optimization of the Lagrangian over
the space of $p_{Q}$'s that are a mean preserving contraction of
$v_{Q}$ and are monotone. This combined with Lemma \ref{lem:Let--be}
implies the desired result.

Let us view $p_{Q}$ as a member of the space $L_{\infty}\left(\left[0,1\right]\right)$.
Let us refer to the first order IC constraints with respect to $a_{n}$
as $T_{n}p_{Q}=0$ where $T_{n}$ is an affine transformation that
maps $L_{\infty}\left(\left[0,1\right]\right)$ to $\mathbb{R}$ (same
is true for $T_{H}$) and we can define $Tp_{Q}=\left(T_{1}p_{Q},\cdots,T_{N}p_{Q}\right)$
which is an affine transformation from $L_{\infty}\left(\left[0,1\right]\right)$
to $\mathbb{R}^{N}$. Finally, let us refer to the set of $p_{Q}$'s
that satisfy $p_{Q}\succcurlyeq_{\text{cv}}v_{Q}$ and monotonicity
of $p_{Q}$ as $\mathcal{P}$ and the subset of $\mathcal{P}$ that
satisfies $Tp_{Q}=0$ as $\mathcal{Q}$.

For any subset $S\subset\left\{ 1,\cdots,N\right\} $, let $S^{c}=\left\{ 1,\cdots,N\right\} \backslash S$
and let us consider the following sets
\[
\mathcal{P}\left(S\right)=\left\{ p_{Q}\in L_{\infty}\left(\left[0,1\right]\right)|p_{Q}\succcurlyeq_{\text{cv}}v_{Q},p_{Q}\text{ increasing},T_{n}p_{Q}\geq0,n\in S,T_{n'}p_{Q}\leq0,n'\in S^{c}\right\} 
\]
\begin{lem}
\label{lem: There-exists-}There exists $S$ such that $\max_{p_{Q}\in\mathcal{P}\left(S\right)}T_{H}p_{Q}=\max_{p_{Q}\in\mathcal{Q}}T_{H}p_{Q}$.
\begin{proof}
Since all members of $\mathcal{Q}$ satisfy $T_{n}p_{Q}=0$, it must
be that $\mathcal{Q}\subset\mathcal{P}\left(S\right)$ for all $S$.
This implies that $\max_{p_{Q}\in\mathcal{P}\left(S\right)}T_{H}p_{Q}\geq\max_{p_{Q}\in\mathcal{Q}}T_{H}p_{Q}$.
Now, suppose to the contrary that for all $S$, the left hand side
is strictly higher than the right hand side. This implies that for
any $S\subset\left\{ 2,\cdots,N\right\} $, there exists $p\in\mathcal{P}\left(S\right),p'\in\mathcal{P}\left(S\cup\left\{ 1\right\} \right)$
such that $T_{H}p,T_{H}p'>\max_{p_{Q}\in\mathcal{Q}}T_{H}p_{Q}$.
Since $T_{1}p\leq0\leq T_{1}p'$ there must exist $\lambda$ such
that $T_{1}\left(\lambda p+\left(1-\lambda\right)p'\right)=0$. Let
$p^{\left(1\right),S}=\lambda p+\left(1-\lambda\right)p'$ and recall
that $p^{\left(1\right),S}\in\mathcal{P}\left(S\right)$. We must
also have that $T_{H}p^{\left(1\right),S}>\max_{p_{Q}\in\mathcal{Q}}T_{H}p_{Q}$.
Now, we know that $T_{2}p^{\left(1\right),S}\leq0\leq T_{2}p^{\left(1\right),S\cup\left\{ 2\right\} }$
for any $S\subset\left\{ 3,\cdots,N\right\} $. By using the same
argument, we can find $p^{\left(1,2\right),S}$ such that $T_{1}p^{\left(1,2\right),S}=T_{2}p^{\left(1,2\right),S}=0,$
$p^{\left(1,2\right),S}\in\mathcal{P}\left(S\right)$ and $T_{H}p^{\left(1,2\right),S}>\max_{p_{Q}\in\mathcal{Q}}T_{H}p_{Q}$.
By continuing this construction, we can find $p^{\left(1,2,\cdots,N\right)}$
such that $T_{1}p^{\left(1,\cdots,N\right)}=\cdots=T_{N}p^{\left(1,\cdots,N\right)}=0$
and that $T_{H}p^{\left(1,\cdots,N\right)}>\max_{p_{Q}\in\mathcal{Q}}T_{H}p_{Q}$
which is a contradiction since $p^{\left(1,\cdots,N\right)}\in\mathcal{Q}$.
\end{proof}

\end{lem}
Now, suppose that $\hat{S}\subset\left\{ 1,\cdots,N\right\} $ satisfies
the condition in Lemma \ref{lem: There-exists-}. Let us define $\mathcal{P}=\left\{ p_{Q}|p_{Q}\succcurlyeq_{\text{cv}}v_{Q},p_{Q}\text{ increasing}\right\} $.
Then, $T$ maps members of $\mathcal{P}$ into $\mathbb{R}^{N}$.
Moreover, $T$ maps members of $\mathcal{P}\left(\hat{S}\right)$
into a convex cone. Since the image of $\mathcal{P}\left(\hat{S}\right)$
under $T$ is convex in $\mathbb{R}^{N}$, it must have a non--empty
relative interior.\footnote{See for example Theorem 6.2 in \citet{rockafellar1970convex}.}
This implies that we can apply standard results for existence of Lagrange
multipliers (strong duality) -- see for example, Theorem 8.3.1. in
\citet{luenberger1997optimization}. Hence, it must be that $\lambda\neq0\in\mathbb{R}^{N}$
exists such that $\lambda_{n}\geq0$ for all $n\in\hat{S}$ and $\lambda_{n}\leq0$
for all $n\in\hat{S}^{c}$ such that
\begin{align*}
\max_{p_{Q}\in\mathcal{P}}W\left(a\right)-\int p_{Q}\left(i\right)dH\left(i;a\right) & =\\
\max_{p_{Q}\in\mathcal{P}}W\left(a\right)+\int\left[H\left(i;a\right)-\sum^{N}_{n=1}\lambda_{n}\left.\frac{\partial}{\partial\hat{a}_{n}}F\left(i|\hat{a};a\right)\right|_{\hat{a}=a}\right]dp_{Q}-\sum^{N}_{n=1}\lambda_{n}\frac{\partial c\left(a\right)}{\partial a_{n}}
\end{align*}
The rest of the claim follows from Lemma \ref{lem:Let--be}.
\end{proof}

\subsection{Proof of Proposition \ref{prop: simple}}
\begin{proof}
Given the statement of Theorem \ref{thm: 1}, we know that the unconstrained
objective in (\ref{eq: D}) can be achieved by a monotone partition.
Note that by \citet{kleiner2020extreme}, the extreme points of the
convex set $\mathcal{P}$ -- the set of $p_{Q}$'s that are mean
preserving contractions of $v_{Q}$ and are monotone -- are associated
with the monotone partitions. In what follows, we show that under
the Assumption \ref{assu:Independence.}, no two extreme points of
$\mathcal{P}$ can deliver the same value of the Lagrangian 
\[
L\left(p_{Q},\lambda;a\right)=W\left(a\right)-\int p_{Q}\left(i\right)d\left(H\left(i;a\right)-\sum^{N}_{n=1}\lambda_{n}\left.\frac{\partial}{\partial\hat{a}_{n}}F\left(i|\hat{a};a\right)\right|_{\hat{a}=a}\right)-\sum^{N}_{n=1}\lambda_{n}\frac{\partial c\left(a\right)}{\partial a_{n}}
\]
This would imply that $L\left(p_{Q},\lambda;a\right)$ has a unique
maximizer for any $\lambda$ which establishes the claim.

Suppose that there are two interim price functions $p_{1},p_{2}\in\mathcal{P}$
that are associated with monotone partitions. If $p_{1}\neq p_{2}$,
then there must exist an interval $I\subset\left[0,1\right]$ so that
all of its members satisfy $p_{1,Q}\left(i\right)=v_{Q}\left(i\right)$
and $p_{2,Q}\left(i\right)$ is constant for all $i\in I$. By Lemma
\ref{lem:Let--be}, if $p_{1,Q}\left(i\right)=v_{Q}\left(i\right)$
is optimal for an interval $I$, then $\Gamma\left(i;a,\lambda\right)=H\left(i;a\right)-\sum^{N}_{n=1}\lambda_{n}\left.\frac{\partial}{\partial\hat{a}_{n}}F\left(i|\hat{a};a\right)\right|_{\hat{a}=a}$
should coincide with its concave envelope. Moreover, suppose that
the maximal interval containing $I$ for which $p_{2,Q}$ is constant
is $\tilde{I}=\left(i_{1},i_{2}\right)$. Let the value of $p_{2,Q}$
be $\tilde{p}$ over $\tilde{I}$.

We show that this implies that $\Gamma\left(i;a,\lambda\right)$ is
linear over the interval $I$. Suppose to the contrary that for some
sub--interval $I'\subset I$, $\Gamma\left(i;a,\lambda\right)$ is
strictly concave. Then,
\[
-\int_{\tilde{I}}p_{2,Q}\left(i\right)d\Gamma\left(i;a,\lambda\right)=-\tilde{p}\left[\Gamma\left(i_{2};a,\lambda\right)-\Gamma\left(i_{1};a,\lambda\right)\right]
\]
We also have that 
\begin{align*}
v_{Q}\left(i_{1}\right) & <\tilde{p}=\frac{\int_{\tilde{I}}v_{Q}\left(i\right)di}{i_{2}-i_{1}}<v_{Q}\left(i_{2}\right)
\end{align*}
 Let $j$ be an interior point of $I'$. Let us define 
\[
\hat{p}_{Q}\left(i\right)=\begin{cases}
p_{2,Q}\left(i\right) & i\in\left[0,1\right]\backslash\tilde{I}\\
\underline{p} & i\in\left(i_{1},j\right)\\
\overline{p} & i\in\left(j,i_{2}\right)
\end{cases}
\]
where
\[
\underline{p}=\frac{\int^{j}_{i_{1}}v_{Q}\left(i\right)di}{j-i_{1}},\overline{p}=\frac{\int^{i_{2}}_{j}v_{Q}\left(i\right)di}{i_{2}-j}
\]
We have
\begin{align*}
-\int\hat{p}_{Q}\left(i\right)d\Gamma\left(i;\lambda,a\right)+\int p_{2,Q}\left(i\right)d\Gamma\left(i;\lambda,a\right) & =\\
-\frac{\int^{j}_{i_{1}}v_{Q}\left(i\right)di}{j-i_{1}}\left[\Gamma\left(j;a,\lambda\right)-\Gamma\left(i_{1};a,\lambda\right)\right]-\frac{\int^{i_{2}}_{j}v_{Q}\left(i\right)di}{i_{2}-j}\left[\Gamma\left(i_{2};a,\lambda\right)-\Gamma\left(j;a,\lambda\right)\right]\\
+\frac{\int^{i_{2}}_{i_{1}}v_{Q}\left(i\right)di}{i_{2}-i_{1}}\left[\Gamma\left(i_{2};a,\lambda\right)-\Gamma\left(i_{1};a,\lambda\right)\right]
\end{align*}
In the above, since $\Gamma\left(i;a,\lambda\right)$ is strictly
concave over parts of $I$, we must have that 
\[
\frac{\Gamma\left(j;a,\lambda\right)-\Gamma\left(i_{1};a,\lambda\right)}{j-i_{1}}>\frac{\Gamma\left(i_{2};a,\lambda\right)-\Gamma\left(j;a,\lambda\right)}{i_{2}-j}
\]
Let us also define $\pi_{1}=\frac{j-i_{1}}{i_{2}-i_{1}}=1-\pi_{2}$
. Since $\underline{p}<\overline{p},$we can write
\begin{align*}
\pi_{1}\underline{p}\frac{\Gamma\left(j;a,\lambda\right)-\Gamma\left(i_{1};a,\lambda\right)}{j-i_{1}}+\pi_{2}\overline{p}\frac{\Gamma\left(i_{2};a,\lambda\right)-\Gamma\left(j;a,\lambda\right)}{i_{2}-j} & <\\
\left(\pi_{1}\underline{p}+\pi_{2}\overline{p}\right)\left(\pi_{1}\frac{\Gamma\left(j;a,\lambda\right)-\Gamma\left(i_{1};a,\lambda\right)}{j-i_{1}}+\pi_{2}\frac{\Gamma\left(i_{2};a,\lambda\right)-\Gamma\left(j;a,\lambda\right)}{i_{2}-j}\right) & =\\
\frac{\int^{i_{2}}_{i_{1}}v_{Q}\left(i\right)di}{i_{2}-i_{1}}\frac{\Gamma\left(i_{2};a,\lambda\right)-\Gamma\left(i_{1};a,\lambda\right)}{i_{2}-i_{1}}
\end{align*}
which implies that 
\[
-\int\hat{p}_{Q}\left(i\right)d\Gamma\left(i;\lambda,a\right)+\int p_{2,Q}\left(i\right)d\Gamma\left(i;\lambda,a\right)>0
\]
and thus $p_{2,Q}$ cannot be optimal. Therefore, 
\[
\Gamma'\left(i;a,\lambda\right)=c,\forall i\in I
\]
 for some $c$. Using the definition of $H$ and $F$, we have
\[
\Gamma'\left(i;a,\lambda\right)=-\alpha\left(G^{-1}\left(i|a\right)\right)-\frac{1}{g\left(y|a\right)}\sum\lambda_{n}\frac{\partial g\left(G^{-1}\left(i|a\right)|a\right)}{\partial a_{n}}=c
\]
This is indeed in contradiction with the independence assumption which
establishes the claim.
\end{proof}

\subsection{Proof of Proposition \ref{prop: MLRP}}
\begin{proof}
In this case, the function $\Gamma\left(i;a,\lambda\right)$ satisfies
\[
\Gamma'\left(i;a,\lambda\right)=-\lambda\frac{\partial g\left(G^{-1}\left(i|a\right)|a\right)}{\partial a}\frac{1}{g\left(G^{-1}\left(i|a\right)|a\right)}
\]
Since $g$ exhibits MLRP, if $\lambda>0$ then, $\Gamma'$ is decreasing
in $i$ and so $\Gamma$ is concave. If $\lambda<0$, then $\Gamma'$
is increasing in $i$ and so $\Gamma$ is convex. Clearly $\lambda$
cannot be negative since convex $\Gamma$ leads to a fully uninformative
rating.
\end{proof}

\subsection{Proof of Proposition \ref{prop: ELRPCLRP}}
\begin{proof}
Recall that $\Gamma\left(i;\lambda,a\right)=-\lambda\left.\frac{\partial F\left(i|\hat{a};a\right)}{\partial\hat{a}}\right|_{\hat{a}=a}$,
where we drop the constant $-\lambda c'\left(a\right)$ term since
it shifts $\Gamma$ uniformly and does not affect its concavification.
As we have shown in Section \ref{sec:Application-1:-Rating},
\begin{align*}
\Gamma''\left(i;a,\lambda\right) & =-\lambda\left.\frac{1}{g\left(y|a\right)}\frac{\partial^{2}\log g\left(y|a\right)}{\partial a\partial y}\right|_{y=G^{-1}\left(i|a\right)}
\end{align*}
Given our definition of ELRP, when $\lambda$ is negative, the above
is concave--convex and when $\lambda$ is positive, the above is
convex--concave. We wish to show that under ELRP, $\lambda$ is positive
and thus optimal rating has to be lower censorship.

Suppose to the contrary that $\lambda$ is negative. In this case,
since $\Gamma\left(0;a,\lambda\right)=\Gamma\left(1;a,\lambda\right)=0$,
there are two possibilities: 1. $\Gamma'\left(0;a,\lambda\right)<0$
in which case $\Gamma\left(i;a,\lambda\right)$ is non--positive
for all values of $i$ and its concave envelope is the zero function
associated with no information; 2. $\Gamma'\left(0;a,\lambda\right)>0$
in which case for an interval $\left[0,i_{1}\right]$ the concave
envelope coincides with $\Gamma$ and for higher values $\text{cav}\Gamma$
is linear. This is associated with an upper-censorship optimal rating.
In the first case, the marginal return to effort is zero and effort
with positive marginal cost cannot be supported.

In the second case, let $y_{1}$ be the value of indicator associated
with $i_{1}$. In this case, the marginal benefit of effort is given
by
\begin{align*}
\int p\left(y\right)\frac{\partial g\left(y|a\right)}{\partial a}dy & =\\
\int^{y_{1}}_{\underline{y}}\overline{v}\left(y;a\right)\frac{\partial\log g\left(y|a\right)}{\partial a}dG+\overline{p}\int^{\overline{y}}_{y_{1}}\frac{\partial g\left(y|a\right)}{\partial a}dy
\end{align*}
where in the above $\overline{p}$ is the average value of $\overline{v}$
when $y\geq y_{1}$. Since $\Gamma$ is concave over $\left[0,i_{1}\right]$,
$\frac{\partial\log g\left(y|a\right)}{\partial a}$ is decreasing
in $y$ over the interval $\left[\underline{y},y_{1}\right]$ and
thus, using the fact that $\overline{v}$ is increasing in $y$, we
can use Chebyshev's sum inequality to write the above as
\begin{align*}
\int^{y_{1}}_{\underline{y}}\overline{v}\left(y;a\right)\frac{\partial\log g\left(y|a\right)}{\partial a}dG+\overline{p}\int^{\overline{y}}_{y_{1}}\frac{\partial g\left(y|a\right)}{\partial a}dy & \leq\\
\frac{\int^{y_{1}}_{\underline{y}}\overline{v}\left(y;a\right)dG}{G\left(y_{1}|a\right)}\int^{y_{1}}_{\underline{y}}\frac{\partial\log g\left(y|a\right)}{\partial a}dG+\overline{p}\int^{\overline{y}}_{y_{1}}\frac{\partial g\left(y|a\right)}{\partial a}dy & =\\
\frac{\int^{y_{1}}_{\underline{y}}\overline{v}\left(y;a\right)dG}{G\left(y_{1}|a\right)}\frac{\partial G\left(y_{1}|a\right)}{\partial a}-\overline{p}\frac{\partial G\left(y_{1}|a\right)}{\partial a}
\end{align*}
where in the above we have used the fact that $\frac{\partial G\left(\overline{y}|a\right)}{\partial a}=0$.
Since $\Gamma\left(i_{1}\right)=-\lambda\frac{\partial G\left(y_{1}|a\right)}{\partial a}>0$,
the above expression satisfies
\[
\left(\mathbb{E}\left[\overline{v}|y\leq y_{1}\right]-\mathbb{E}\left[\overline{v}|y\geq y_{1}\right]\right)\frac{\partial G\left(y_{1}|a\right)}{\partial a}\leq0
\]
 which cannot be the case since the cost of effort is increasing.
Hence, $\lambda\geq0$ and thus optimal rating is lower censorship.
When $g$ exhibits CLRP, the argument is the mirror of the current
argument.
\end{proof}

\subsection{Proof of Theorem \ref{thm: MultiTask}}
\begin{proof}
From the text, we know that in this case the second derivative of
the gain function is given by
\[
\frac{\left.\Gamma''\left(i|a\right)\right|_{i=G\left(y|a\right)}}{g\left(y|a\right)}=-\lambda_{1}\frac{\partial\ell_{1}\left(y|a\right)}{\partial y}-\lambda_{2}\frac{\partial\ell_{2}\left(y|a\right)}{\partial y}.
\]
In order to prove the result, we show that under MRIP the above changes
sign at most once. This means that $\Gamma$ is either concave--convex
or convex--concave and therefore, optimal rating is either lower
or upper censorship. To see this, consider three values $y_{1}<y_{2}<y_{3}$.
Note that we can assume without loss of generality that $D\geq0$
as when $D\leq0$ the reverse argument would work. Let us define the
vectors $\mathbf{z}_{i}=\left(\frac{\partial\ell_{1}\left(y_{i}|a\right)}{\partial y},\frac{\partial\ell_{2}\left(y_{i}|a\right)}{\partial y}\right),\boldsymbol{\lambda}=\left(\lambda_{1},\lambda_{2}\right)$
and for an arbitrary vector $x$ let $x^{+}=\left(x_{2},-x_{1}\right)$
be the perpendicular vector of $x$; we know that $x^{+}\cdot x=0$.
Then MRIP implies that
\[
\mathbf{z}_{2},\mathbf{z}_{3}\in\left\{ \mathbf{x}|\mathbf{z}^{+}_{1}\cdot\mathbf{x}\geq0\right\} ,\mathbf{z}_{3}\in\left\{ \mathbf{x}|\mathbf{z}^{+}_{2}\cdot\mathbf{x}\geq0\right\} 
\]
This means that $\mathbf{z}_{2}$ belongs to the half space above
the line through $\mathbf{z}_{1}$ and the same holds for $\mathbf{z}_{3}$
and $\mathbf{z}_{2}$. In other words, $\mathbf{z}_{1},\mathbf{z}_{2},\mathbf{z}_{3}$
should relate to each other as in Figure \ref{fig:The-vectors-of}
where $\mathbf{z}_{2}$ has a higher angle with the $x$--axis relative
to $\mathbf{z}_{1}$ and $\mathbf{z}_{3}$ has a higher angle than
$\mathbf{z}_{2}$. This means that $\mathbb{R}^{2}$ is partitioned
by the following sets
\[
\left\{ \mathbf{x}|\mathbf{z}^{+}_{1}\cdot\mathbf{x}\geq0>\mathbf{z}^{+}_{2}\cdot\mathbf{x}\right\} ,\left\{ \mathbf{x}|\mathbf{z}^{+}_{2}\cdot\mathbf{x}\geq0>\mathbf{z}^{+}_{3}\cdot\mathbf{x}\right\} ,\left\{ \mathbf{x}|\mathbf{z}^{+}_{3}\cdot\mathbf{x}\geq0>\mathbf{z}^{+}_{1}\cdot\mathbf{x}\right\} ,\left\{ \mathbf{x}|0>\mathbf{z}^{+}_{1}\cdot\mathbf{x}\right\} .
\]
This implies that $\boldsymbol{\lambda}^{+}$ has to belong in one
of the above sets. Suppose that $\boldsymbol{\lambda}^{+}$ belongs
to the first set (the situation depicted in Figure \ref{fig:The-vectors-of}),
then
\[
\mathbf{z}^{+}_{1}\cdot\boldsymbol{\lambda}^{+}\geq0>\mathbf{z}^{+}_{2}\cdot\boldsymbol{\lambda}^{+}\Rightarrow\begin{cases}
\frac{\partial\ell_{2}\left(y_{1}|a\right)}{\partial y}\lambda_{2}+\frac{\partial\ell_{1}\left(y_{1}|a\right)}{\partial y}\lambda_{1}\geq0\\
\frac{\partial\ell_{2}\left(y_{2}|a\right)}{\partial y}\lambda_{2}+\frac{\partial\ell_{1}\left(y_{2}|a\right)}{\partial y}\lambda_{1}<0
\end{cases}
\]
This means that if $\boldsymbol{\lambda}^{+}=\alpha_{1}\mathbf{z}_{1}+\alpha_{2}\mathbf{z}_{2}$,
then 
\begin{align*}
\mathbf{z}^{+}_{1}\cdot\boldsymbol{\lambda}^{+} & =\alpha_{2}\mathbf{z}^{+}_{1}\cdot\mathbf{z}_{2}\Rightarrow\alpha_{2}\geq0\\
\mathbf{z}^{+}_{2}\cdot\boldsymbol{\lambda}^{+} & =\alpha_{1}\mathbf{z}^{+}_{2}\cdot\mathbf{z}_{1}\Rightarrow\alpha_{1}\geq0
\end{align*}
This in turn means that 
\begin{align*}
\mathbf{z}^{+}_{3}\cdot\boldsymbol{\lambda}^{+} & =\mathbf{z}_{3}\cdot\boldsymbol{\lambda}=\mathbf{z}^{+}_{3}\cdot\left(\alpha_{1}\mathbf{z}_{1}+\alpha_{2}\mathbf{z}_{2}\right)\\
 & =\alpha_{1}\mathbf{z}^{+}_{3}\cdot\mathbf{z}_{1}+\alpha_{2}\mathbf{z}^{+}_{3}\cdot\mathbf{z}_{2}\leq0
\end{align*}
where the above holds because $\mathbf{z}^{+}_{1}\cdot\mathbf{z}_{3},\mathbf{z}^{+}_{2}\cdot\mathbf{z}_{3}\geq0$.
This means that $\mathbf{z}_{1}\cdot\boldsymbol{\lambda}\geq0\geq\mathbf{z}_{2}\cdot\boldsymbol{\lambda},\mathbf{z}_{3}\cdot\boldsymbol{\lambda}$
which is the desired result. The other cases can be shown the same
way. 
\begin{figure}
\begin{centering}
\begin{tikzpicture}[>=Stealth, scale=.8]

\path[use as bounding box] (-2.15,-1.25) rectangle (4.4,4.45);

\coordinate (O) at (0,0);
\coordinate (z1) at (20:3.6);
\coordinate (z2) at (45:3.0);
\coordinate (z3) at (100:3.6);

\draw[->, thick] (-1.9,0) -- (4.2,0) node[below left] {$x_1$};
\draw[->, thick] (0,-1.0) -- (0,4.2) node[below right] {$x_2$};

\draw[thick, densely dashed, blue!65!black] (-1.9,-0.95) -- (4.1,2.05) node[sloped, above, pos=0.72, blue!65!black] {$\lambda\cdot x=0$};

\draw[->, very thick, blue!65!black] (O) -- ($(O)+(116.57:1.7)$) node[above left] {$\lambda$};

\draw[blue!65!black] ($(O)+(26.57:0.42)$) -- ($(O)+(26.57:0.42)+(116.57:0.42)$) -- ($(O)+(116.57:0.42)$);

\foreach \p in {z1,z2,z3}{\draw[gray!55] (O) -- (\p);}

\fill (z1) circle (2.4pt) node[right=4pt] {$z_1$};
\fill (z2) circle (2.4pt) node[above right=2pt] {$z_2$};
\fill (z3) circle (2.4pt) node[above=4pt] {$z_3$};

\node[below left] at (O) {$0$};

\end{tikzpicture}
\par\end{centering}
\caption{The vectors of likelihood changes under MRIP.\label{fig:The-vectors-of}}
\end{figure}
\end{proof}

\subsection{Proof of Proposition \ref{prop:midcensor}}
\begin{proof}
It is immediate by using an argument similar to Theorem \ref{thm: 1}
that optimal ratings can be found by concavification of the function
\[
\Gamma\left(i\right)=-\alpha_{P}f_{P}G\left(\overline{G}^{-1}\left(i\right)|a_{P}\right)-\alpha_{R}f_{R}G\left(\overline{G}^{-1}\left(i\right)|a_{R}\right)-\lambda_{P}G_{a}\left(\overline{G}^{-1}\left(i\right)|a_{P}\right)-\lambda_{R}G_{a}\left(\overline{G}^{-1}\left(i\right)|a_{R}\right)
\]
for some Lagrange multipliers $\lambda_{P},\lambda_{R}$. We have
\[
\Gamma'\left(i\right)=-\alpha_{P}f_{P}\frac{g\left(\overline{G}^{-1}\left(i\right)|a_{P}\right)}{\overline{g}\left(\overline{G}^{-1}\left(i\right)\right)}-\alpha_{R}f_{R}\frac{g\left(\overline{G}^{-1}\left(i\right)|a_{R}\right)}{\overline{g}\left(\overline{G}^{-1}\left(i\right)\right)}-\lambda_{P}\frac{g_{a}\left(\overline{G}^{-1}\left(i\right)|a_{P}\right)}{\overline{g}\left(\overline{G}^{-1}\left(i\right)\right)}-\lambda_{R}\frac{g_{a}\left(\overline{G}^{-1}\left(i\right)|a_{R}\right)}{\overline{g}\left(\overline{G}^{-1}\left(i\right)\right)}
\]
where in the above $\overline{g}\left(y\right)=f_{P}g\left(y|a_{P}\right)+f_{R}g\left(y|a_{R}\right)$.
Given the functional form of $g$, we have
\begin{align*}
g\left(y|a\right) & =e^{f\left(y\right)+r\left(y\right)m\left(a\right)-b\left(a\right)}\\
\frac{g\left(y|a_{P}\right)}{g\left(y|a_{R}\right)} & =e^{r\left(y\right)\left(m\left(a_{P}\right)-m\left(a_{R}\right)\right)+b\left(a_{R}\right)-b\left(a_{P}\right)}\\
\frac{g_{a}\left(y|a\right)}{g\left(y|a\right)} & =m'\left(a\right)r\left(y\right)-b'\left(a\right)
\end{align*}
Replacing in the formula for $\Gamma'$ implies
\[
\Gamma'\left(\overline{G}\left(y\right)\right)=-\alpha_{P}f_{P}\frac{1}{f_{P}+f_{R}\frac{g\left(y|a_{R}\right)}{g\left(y|a_{P}\right)}}-\alpha_{R}f_{R}\frac{\frac{g\left(y|a_{R}\right)}{g\left(y|a_{P}\right)}}{f_{P}+f_{R}\frac{g\left(y|a_{R}\right)}{g\left(y|a_{P}\right)}}-\lambda_{P}\frac{\frac{g_{a}\left(y|a_{P}\right)}{g\left(y|a_{P}\right)}}{f_{P}+\frac{g\left(y|a_{R}\right)}{g\left(y|a_{P}\right)}}-\lambda_{R}\frac{\frac{g\left(y|a_{R}\right)}{g\left(y|a_{P}\right)}\frac{g_{a}\left(y|a_{R}\right)}{g\left(y|a_{R}\right)}}{f_{P}+f_{R}\frac{g\left(y|a_{R}\right)}{g\left(y|a_{P}\right)}}
\]
If we refer to $m\left(a_{R}\right)-m\left(a_{P}\right)$ as $\Delta m$
and similarly for $b\left(a_{R}\right)-b\left(a_{P}\right)$, we can
write the above as
\[
\Gamma'\left(\overline{G}\left(y\right)\right)=-\frac{\alpha_{P}f_{P}+\alpha_{R}f_{R}e^{r\Delta m-\Delta b}+\lambda_{P}\left(m_{P}'r-b_{P}'\right)+\lambda_{R}\left(m_{R}'r-b_{R}'\right)e^{r\Delta m-\Delta b}}{f_{P}+f_{R}e^{r\Delta m-\Delta b}}
\]
If we define $x=e^{r\Delta m}$, then the above has the form
\[
-\frac{A_{1}+A_{2}x+A_{3}\log x+A_{4}x\log x}{B_{1}+B_{2}x}
\]
We can argue that $\lambda_{R}>0$. This is because if we consider
the problem by replacing the IC for the $R$ type with its inequality
version imposing that the marginal return to $a_{R}$ be higher than
its cost. In this problem, if this constraint remains slack, one can
simply increase $a_{R}$ and shifts all $p\left(y\right)$'s upwards
by the same amount and improve the payoffs. Hence, $\lambda_{R}\geq0$.
Since $m_{R}'\geq0$ by assumption, we must have that $A_{4}>0$.

We then have
\begin{align*}
\left(B_{1}+B_{2}x\right)^{2}\frac{\Gamma''\left(\overline{G}\left(y\right)\right)\overline{g}\left(y\right)}{\frac{dx}{dy}}= & \left(B_{2}A_{3}-B_{1}A_{4}\right)\log x-B_{1}A_{3}/x-B_{2}A_{4}x\\
 & +B_{2}\left(A_{1}-A_{3}\right)-B_{1}\left(A_{1}+A_{4}\right)\\
= & B_{2}A_{4}\left(\alpha_{1}\log x+\alpha_{2}/x-x+\alpha_{3}\right)
\end{align*}
Note that in the above $A_{4}B_{2}>0$. The derivative of the above
with respect to $x$ is given by $\frac{\alpha_{1}}{x}-\alpha_{2}/x^{2}-1=\frac{\alpha_{1}x-\alpha_{2}-x^{2}}{x}$.
Suppose that $\alpha_{2}>0$. Since the numerator is a quadratic function,
it has at most two roots and this means that $\Gamma''$ switches
sign at most three times. This establishes the claim.
\end{proof}

\newpage{}

\part*{Online Appendix}

\section{Multi--Tasking with Location--Scale Technologies\label{sec: OnlineDirection}}

In this section, we describe optimal ratings when $G\left(y|a\right)=H\left(\frac{y-\mu\left(a\right)}{\sigma\left(a\right)}\right)$,
i.e., $a$ controls the location and scale of the indicator $y$,
and provide a proof of Proposition \ref{prop: direction}. Throughout,
we write $\mu_{n}=\partial\mu\left(a\right)/\partial a_{n}$, $\sigma_{n}=\partial\sigma\left(a\right)/\partial a_{n}$,
$c_{n}=\partial c\left(a\right)/\partial a_{n}$, $\varepsilon=\left(y-\mu\left(a\right)\right)/\sigma\left(a\right)$
and $\Psi\left(a\right)=\mu_{1}\sigma_{2}-\mu_{2}\sigma_{1}$, all
evaluated at the implemented action.

Before providing the analysis, let us describe what MRIP implies within
the location--scale class. Consider the family $G\left(y|a\right)=H\left(\frac{y-\mu\left(a\right)}{\sigma\left(a\right)}\right)=H\left(\varepsilon\right)$
such that $a$ controls the mean and variance of $y$ with $H$ the
c.d.f. of a distribution with a log--concave density $H'\left(\varepsilon\right)=h\left(\varepsilon\right)=e^{\eta\left(\varepsilon\right)}$.
In this case,
\begin{equation}
D\left(y_{1},y_{2}|a\right)=\eta''\left(\varepsilon_{1}\right)\eta''\left(\varepsilon_{2}\right)\Psi\left(a\right)\left[\varepsilon_{2}+\frac{\eta'\left(\varepsilon_{2}\right)}{\eta''\left(\varepsilon_{2}\right)}-\varepsilon_{1}-\frac{\eta'\left(\varepsilon_{1}\right)}{\eta''\left(\varepsilon_{1}\right)}\right]\label{eq: Dsl-1}
\end{equation}
 where $\varepsilon_{n}=\frac{y_{n}-\mu\left(a\right)}{\sigma\left(a\right)}$
and $\Psi\left(a\right)=\frac{\partial\mu}{\partial a_{1}}\frac{\partial\sigma}{\partial a_{2}}-\frac{\partial\sigma}{\partial a_{1}}\frac{\partial\mu}{\partial a_{2}}$.
Since $h$ is log--concave, i.e., $\eta''\leq0$, the leading product
is non--negative, and MRIP for this class is equivalent to monotonicity
of $\varepsilon+\frac{\eta'\left(\varepsilon\right)}{\eta''\left(\varepsilon\right)}$.
Monotonicity of $\varepsilon+\eta'\left(\varepsilon\right)/\eta''\left(\varepsilon\right)$
holds for the Normal (where it equals $2\varepsilon$), Logistic,
Gumbel, Generalized Normal, Gamma (with shape parameter $\geq1$),
and Weibull (with shape parameter $\geq1$) families, so all of these
exhibit MRIP with a log--concave density.

Fix the implemented action $a$ and let $\pi\left(\varepsilon\right)=p\left(\mu\left(a\right)+\sigma\left(a\right)\varepsilon\right)$
be the interim price expressed in units of the standardized indicator.
The payoff of the agent from effort $a'$ is $\int p\left(y\right)dG\left(y|a'\right)-c\left(a'\right)$
and differentiating with respect to $a_{n}'$ at $a'=a$ yields
\[
B_{n}=\mu_{n}P_{1}+\sigma_{n}P_{2},\qquad P_{1}=-\frac{1}{\sigma\left(a\right)}\int\pi\left(\varepsilon\right)h'\left(\varepsilon\right)d\varepsilon,\quad P_{2}=-\frac{1}{\sigma\left(a\right)}\int\pi\left(\varepsilon\right)\left(\varepsilon h\left(\varepsilon\right)\right)'d\varepsilon
\]
 Integrating by parts, $P_{1}=\frac{1}{\sigma\left(a\right)}\int h\left(\varepsilon\right)d\pi\left(\varepsilon\right)$
and $P_{2}=\frac{1}{\sigma\left(a\right)}\int\varepsilon h\left(\varepsilon\right)d\pi\left(\varepsilon\right)$,
where $d\pi$ is the non--negative measure generated by the non--decreasing
schedule $\pi$. Thus the rating affects the incentives of the agent
only through two statistics: $P_{1}$, the marginal reward that the
rating attaches to the level of the indicator, and $P_{2}$, the marginal
reward that it attaches to its dispersion. Moreover, $P_{1}>0$ unless
the rating is uninformative.
\begin{lem}
\label{lem: P1P2} Suppose that the welfare-maximizing rating implements
an interior action $a$ and that $\Psi\left(a\right)\neq0$. Then
\[
P_{1}=\frac{\sigma_{2}c_{1}-\sigma_{1}c_{2}}{\Psi},\qquad P_{2}=\frac{\mu_{1}c_{2}-\mu_{2}c_{1}}{\Psi}
\]
\end{lem}
\begin{proof}
The first order conditions of the agent are $\mu_{n}P_{1}+\sigma_{n}P_{2}=c_{n}$
for $n=1,2$, two linear equations in $\left(P_{1},P_{2}\right)$
whose determinant is $\Psi\left(a\right)$. Solving establishes the
claim.
\end{proof}

Note that $P_{1}>0$ requires that $\sigma_{2}c_{1}-\sigma_{1}c_{2}$
and $\Psi\left(a\right)$ share a sign, which holds automatically
in the leading case in which productive effort compresses the dispersion
of the indicator while window dressing expands it, i.e., $\sigma_{1}\leq0\leq\sigma_{2}$.
\begin{lem}
\label{lem: signP2} Suppose that $\overline{v}\left(y;a\right)$
is a linear function of $y$ with a positive slope and that $h$ is
log--concave. Then any upper-censorship rating satisfies $P_{2}<0$,
any lower-censorship rating satisfies $P_{2}>0$, and the fully revealing
rating satisfies $P_{2}=0$.
\end{lem}
\begin{proof}
Let $\beta\left(a\right)>0$ be the slope of $\overline{v}$ and consider
an upper-censorship rating with threshold $\hat{y}=\mu\left(a\right)+\sigma\left(a\right)\hat{\varepsilon}$.
The measure $d\pi$ has density $\beta\left(a\right)\sigma\left(a\right)$
below $\hat{\varepsilon}$ and an atom of size $\beta\left(a\right)\sigma\left(a\right)m\left(\hat{\varepsilon}\right)$
at $\hat{\varepsilon}$, where $m\left(x\right)=\mathbb{E}\left[\varepsilon-x|\varepsilon\geq x\right]$
is the mean residual life of $\varepsilon$. Therefore, 
\[
P_{2}=\beta\left(a\right)\left[\Theta\left(\hat{\varepsilon}\right)+m\left(\hat{\varepsilon}\right)\hat{\varepsilon}h\left(\hat{\varepsilon}\right)\right],\qquad\Theta\left(x\right)=\int^{x}_{-\infty}\varepsilon h\left(\varepsilon\right)d\varepsilon<0
\]
 When $\hat{\varepsilon}\leq0$, both terms are non--positive and
the first is strictly negative. When $\hat{\varepsilon}>0$, since
$\varepsilon$ has mean zero, $-\Theta\left(\hat{\varepsilon}\right)=\left(m\left(\hat{\varepsilon}\right)+\hat{\varepsilon}\right)\left(1-H\left(\hat{\varepsilon}\right)\right)$
and thus $P_{2}<0$ is equivalent to 
\[
\hat{\varepsilon}m\left(\hat{\varepsilon}\right)h\left(\hat{\varepsilon}\right)<\left(m\left(\hat{\varepsilon}\right)+\hat{\varepsilon}\right)\left(1-H\left(\hat{\varepsilon}\right)\right)
\]
 Log--concavity of $h$ implies log--concavity of $1-H$, which
is equivalent to $m\left(x\right)\leq\left(1-H\left(x\right)\right)/h\left(x\right)$;
the left hand side is thus at most $\hat{\varepsilon}\left(1-H\left(\hat{\varepsilon}\right)\right)$,
which is strictly below the right hand side since $m\left(\hat{\varepsilon}\right)>0$.
The argument for lower-censorship ratings is the mirror of this argument
and uses $\mathbb{E}\left[x-\varepsilon|\varepsilon\leq x\right]\leq H\left(x\right)/h\left(x\right)$,
implied by log--concavity of $H$. Finally, under full disclosure,
$d\pi=\beta\sigma\left(a\right)d\varepsilon$ and $P_{2}=\beta\int\varepsilon h\left(\varepsilon\right)d\varepsilon=0$.
\end{proof}

We can now prove Proposition \ref{prop: direction}.
\begin{proof}
An uninformative rating satisfies $B_{n}=0$ and thus the first order
conditions of the agent imply $c_{n}=0$, which contradicts interiority
of the implemented action. Hence, by Theorem \ref{thm: MultiTask},
the welfare-maximizing rating is an upper- or a lower-censorship rating.
By Lemma \ref{lem: P1P2}, $P_{2}=\left(\mu_{1}c_{2}-\mu_{2}c_{1}\right)/\Psi$
at the implemented action. When (\ref{eq: impact-1}) holds, $P_{2}\leq0$
and thus by Lemma \ref{lem: signP2} the optimal rating cannot be
lower censorship; it is therefore upper censorship, degenerating to
full disclosure when (\ref{eq: impact-1}) holds with equality. When
(\ref{eq: impact-1}) fails, $P_{2}>0$ and the optimal rating cannot
be upper censorship; it is therefore lower censorship. This establishes
the claim.
\end{proof}

\section{Validity of the First Order Approach\label{sec:Validity-of-the}}

In this section, we describe conditions that make the first order
approach valid. To do so, we use an approach similar to that of \citet{jewitt1988justifying}
and more recently \citet{chade2020no}. More specifically, it is sufficient
to show that given our optimal ratings (lower- or upper-censorship),
the payoff of the agent is quasi--concave in her effort. To show
quasi--concavity of the payoff, it is sufficient to show that $U'\left(a\right)=0$
implies $U''\left(a\right)<0$ if $U$ is twice continuously differentiable.
This would imply that $U$ cannot have more than one local maximum,
i.e., it is single peaked, and is thus quasi--concave.

To see this, suppose that there are two points $a_{1}<a_{2}$ such
that $U'\left(a_{1}\right)=U'\left(a_{2}\right)=0$. Since $U''\left(a_{1}\right)<0$
it must be that there is an interval of values above $a_{1}$ for
which $U'\left(a\right)<0$. Now, without loss of generality, let
us assume that $a_{2}=\inf_{a'>a_{1},U'\left(a'\right)=0}a'$. Since
$U$ is assumed to be twice continuously differentiable, $U'$ is
continuous and thus $U'\left(a_{2}\right)=0$ and since $U'\left(a\right)<0$
for an interval around $a_{1}$, $a_{1}<a_{2}$. Moreover, we must
have that for all $a\in\left(a_{1},a_{2}\right)$, $U'\left(a\right)<0$.
This implies that $U'\left(a\right)<U'\left(a_{2}\right)=0$ for values
of $a$ below $a_{2}$. Since $U$ is assumed to be twice continuously
differentiable, we must have that $U''\left(a_{2}\right)\geq0$ which
is a contradiction. This implies that $U$ is single peaked.

\subsection{A Single Task\label{subsec:FOA-single}}

The ratings that arise in Sections \ref{sec:Application-1:-Rating}
and \ref{subsec:Redistributive-Test-Design} -- full disclosure,
upper- and lower-censorship, and mid-censorship -- are all deterministic
monotone partitions. The following lemma provides a condition for
the validity of FOA that applies to all of them at once:
\begin{lem}
\label{lem: FOAgeneral} Let $p$ be a non-decreasing interim price
schedule and suppose that for all $a,\hat{a}\in A$, 
\[
\int p\left(y;\hat{a}\right)\frac{\partial^{2}g\left(y|a\right)}{\partial a^{2}}dy<\frac{c''\left(a\right)}{c'\left(a\right)}\int p\left(y;\hat{a}\right)\frac{\partial g\left(y|a\right)}{\partial a}dy
\]
 Then the payoff of the agent is quasi-concave in effort and the FOA
is valid under $p$.
\end{lem}
\begin{proof}
The payoff of the agent is $U\left(a\right)=\int p\left(y;\hat{a}\right)g\left(y|a\right)dy-c\left(a\right)$.
Suppose that $U'\left(a\right)=0$ at some effort level $a$. Then
$\int p\,g_{a}dy=c'\left(a\right)>0$ and 
\[
U''\left(a\right)=\int p\,g_{aa}dy-c''\left(a\right)=\int p\,g_{aa}dy-\frac{c''\left(a\right)}{c'\left(a\right)}\int p\,g_{a}dy<0
\]
 by the assumption of the lemma. Given our argument above, $U$ is
single-peaked and thus quasi-concave.
\end{proof}

For upper- and lower-censorship ratings, integration by parts turns
the condition of Lemma \ref{lem: FOAgeneral} into a condition on
the distribution function directly. This gives the following:
\begin{lem}
\label{lem: Lemma4}Suppose that the family of distributions $\left\{ G\left(y|a\right)\right\} _{a\in A}$
is twice continuously differentiable and $A$ is a convex subset of
$\mathbb{R}$. Suppose further that $G$ satisfies the following properties
\begin{align*}
0 & >\int^{\hat{y}}_{-\infty}\left(\overline{v}\left(y,\hat{a}\right)-\mathbb{E}\left[\overline{v}|y\geq\hat{y}\right]\right)\frac{\partial}{\partial a}\frac{g_{a}\left(y|a\right)}{c'\left(a\right)}dy,\forall a,\hat{a}\in A,\hat{y}\in\mathbb{R},\\
0 & >\int^{\infty}_{\hat{y}}\left(\overline{v}\left(y,\hat{a}\right)-\mathbb{E}\left[\overline{v}|y\leq\hat{y}\right]\right)\frac{\partial}{\partial a}\frac{g_{a}\left(y|a\right)}{c'\left(a\right)}dy,\forall a,\hat{a}\in A,\hat{y}\in\mathbb{R}.
\end{align*}

Then the FOA is valid under lower- and upper-censorship policies.
\end{lem}
\begin{proof}
Consider an upper-censorship policy that pools realizations of $y$
above $\hat{y}$. If the market believes the agent chooses effort
$\hat{a}$, then the payoff of the agent is given by
\[
U\left(a\right)=\int^{\hat{y}}_{-\infty}\overline{v}\left(y;\hat{a}\right)g\left(y|a\right)dy+\frac{\int^{\infty}_{\hat{y}}\overline{v}\left(y;\hat{a}\right)g\left(y|\hat{a}\right)dy}{1-G\left(\hat{y}|\hat{a}\right)}\left(1-G\left(\hat{y}|a\right)\right)-c\left(a\right)
\]
If we let $p_{H}=\mathbb{E}\left[\overline{v}\left(y;\hat{a}\right)|y\geq\hat{y}\right]>\overline{v}\left(\hat{y};\hat{a}\right)$,
we can write
\begin{align*}
U'\left(a\right) & =\int^{\hat{y}}_{-\infty}\overline{v}\left(y;\hat{a}\right)g_{a}\left(y|a\right)dy-p_{H}G_{a}\left(\hat{y}|a\right)-c'\left(a\right)\\
 & =\int^{\hat{y}}_{-\infty}\left(\overline{v}\left(y;\hat{a}\right)-p_{H}\right)g_{a}\left(y|a\right)dy-c'\left(a\right)
\end{align*}
Now, suppose that $U'\left(a_{1}\right)=0$ at some effort level $a_{1}$.
Then,
\begin{align*}
U''\left(a_{1}\right)= & \int^{\hat{y}}_{-\infty}\left(\overline{v}\left(y;\hat{a}\right)-p_{H}\right)g_{aa}\left(y|a_{1}\right)dy-c''\left(a_{1}\right)\\
= & \int^{\hat{y}}_{-\infty}\left(\overline{v}\left(y;\hat{a}\right)-p_{H}\right)g_{aa}\left(y|a_{1}\right)dy\\
 & -\frac{c''\left(a_{1}\right)}{c'\left(a_{1}\right)}c'\left(a_{1}\right)\\
= & \int^{\hat{y}}_{-\infty}\left(\overline{v}\left(y;\hat{a}\right)-p_{H}\right)g_{aa}\left(y|a_{1}\right)dy\\
 & -\frac{c''\left(a_{1}\right)}{c'\left(a_{1}\right)}\int^{\hat{y}}_{-\infty}\left(\overline{v}\left(y;\hat{a}\right)-p_{H}\right)g_{a}\left(y|a_{1}\right)dy\\
= & \int^{\hat{y}}_{-\infty}\left(\overline{v}\left(y;\hat{a}\right)-p_{H}\right)\left[g_{aa}\left(y|a_{1}\right)-\frac{c''\left(a_{1}\right)}{c'\left(a_{1}\right)}g_{a}\left(y|a_{1}\right)\right]dy\\
= & \int^{\hat{y}}_{-\infty}\left(\overline{v}\left(y;\hat{a}\right)-p_{H}\right)c'\left(a_{1}\right)\frac{\partial}{\partial a}\frac{g_{a}\left(y|a_{1}\right)}{c'\left(a_{1}\right)}dy
\end{align*}
Since $c'\left(a_{1}\right)>0$, the assumption on $G$ in the statement
of the lemma guarantees that $U''\left(a_{1}\right)<0$. Given our
argument above, this implies that $U$ is quasi--concave and thus
FOA is valid. The argument for lower-censorship ratings is the mirror
of this argument.
\end{proof}

The essence of the conditions in Lemma \ref{lem: Lemma4} is that
they put a restriction on how convex the cost is, captured by $c''\left(a\right)/c'\left(a\right)$
relative to that of expected interim price. Indeed, the conditions
can be rewritten as
\[
\frac{\int^{\infty}_{-\infty}p\left(y;\hat{y},\hat{a}\right)g_{aa}\left(y|a\right)dy}{\int^{\infty}_{-\infty}p\left(y;\hat{y},\hat{a}\right)g_{a}\left(y|a\right)dy}<\frac{c''\left(a\right)}{c'\left(a\right)},\forall a\in A
\]
where in the case of lower and upper censorship respectively, interim
prices are
\begin{align*}
p\left(y;\hat{y},\hat{a}\right) & =\begin{cases}
\overline{v}\left(y;\hat{a}\right) & y\geq\hat{y}\\
\mathbb{E}_{\hat{a}}\left[v|y\leq\hat{y}\right] & y\leq\hat{y}
\end{cases}\\
p\left(y;\hat{y},\hat{a}\right) & =\begin{cases}
\mathbb{E}_{\hat{a}}\left[v|y\geq\hat{y}\right] & y\geq\hat{y}\\
\overline{v}\left(y;\hat{a}\right) & y\leq\hat{y}
\end{cases}
\end{align*}

In the special case where $\frac{\partial}{\partial y}v\left(y,a\right)=\beta\left(a\right)$
and $y=a+\varepsilon$ with $\varepsilon\sim H\left(\varepsilon\right)$
and density $h\left(\varepsilon\right)=H'\left(\varepsilon\right)$,
these become
\begin{align*}
\frac{\int^{\hat{y}}_{-\infty}\left(y-\mathbb{E}_{\hat{a}}\left[y|y\geq\hat{y}\right]\right)g_{aa}\left(y|a\right)dy}{\int^{\hat{y}}_{-\infty}\left(y-\mathbb{E}_{\hat{a}}\left[y|y\geq\hat{y}\right]\right)g_{a}\left(y|a\right)dy} & <\frac{c''\left(a\right)}{c'\left(a\right)}\\
\frac{\int^{\infty}_{\hat{y}}\left(y-\mathbb{E}_{\hat{a}}\left[y|y\leq\hat{y}\right]\right)g_{aa}\left(y|a\right)dy}{\int^{\infty}_{\hat{y}}\left(y-\mathbb{E}_{\hat{a}}\left[y|y\leq\hat{y}\right]\right)g_{a}\left(y|a\right)dy} & <\frac{c''\left(a\right)}{c'\left(a\right)}
\end{align*}
 and we have
\begin{align*}
\int^{\hat{y}}_{-\infty}\left(y-\mathbb{E}_{\hat{a}}\left[y|y\geq\hat{y}\right]\right)g_{a}\left(y|a\right)dy & =\\
\int^{\hat{y}}_{-\infty}\left(y-\hat{y}-\mathbb{E}_{\hat{a}}\left[y-\hat{y}|y\geq\hat{y}\right]\right)dG_{a}\left(y|a\right) & =\\
-\int^{\hat{y}}_{-\infty}G_{a}\left(y|a\right)dy-\mathbb{E}_{\hat{a}}\left[y-\hat{y}|y\geq\hat{y}\right]G_{a}\left(\hat{y}|a\right) & =\\
-\int^{\hat{y}}_{-\infty}\frac{\partial}{\partial a}H\left(y-a\right)dy-\mathbb{E}_{\hat{a}}\left[y-\hat{y}|y\geq\hat{y}\right]\frac{\partial}{\partial a}H\left(\hat{y}-a\right) & =\\
H\left(\hat{y}-a\right)+\mathbb{E}_{\hat{a}}\left[y-\hat{y}|y\geq\hat{y}\right]h\left(\hat{y}-a\right)\\
\int^{\hat{y}}_{-\infty}\left(y-\mathbb{E}_{\hat{a}}\left[y|y\geq\hat{y}\right]\right)g_{aa}\left(y|a\right)dy & =\\
-h\left(\hat{y}-a\right)-\mathbb{E}_{\hat{a}}\left[y-\hat{y}|y\geq\hat{y}\right]h'\left(\hat{y}-a\right)
\end{align*}
Let us make the following assumption on $H$:
\begin{assumption}
\label{assu:The-cummulative-distribution}The cumulative distribution
function $H\left(\varepsilon\right)$ satisfies:
\begin{enumerate}
\item $\log\left(1-H\left(\varepsilon\right)\right)$ is concave in $\varepsilon$,
\item $\log H\left(\varepsilon\right)$ is concave in $\varepsilon$,
\item $\forall d>0$, $\frac{h\left(x\right)}{1-H\left(x\right)}-\frac{h\left(x-d\right)}{1-H\left(x-d\right)}\leq\kappa_{1}\left(d\right),\frac{h\left(x-d\right)}{H\left(x-d\right)}-\frac{h\left(x\right)}{H\left(x\right)}\leq\kappa_{2}\left(d\right)$
\end{enumerate}
\end{assumption}
The first two conditions are standard log--concavity conditions while
the last condition implies that the variations in the derivative of
$\log\left(1-H\left(x\right)\right)$ are uniformly bounded above.
A sufficient condition for the latter is that the expressions $h'\left(x\right)/\left(1-H\left(x\right)\right)+h\left(x\right)^{2}/\left(1-H\left(x\right)\right)^{2}$
and $h'\left(x\right)/H\left(x\right)-h\left(x\right)^{2}/\left(H\left(x\right)\right)^{2}$
are bounded above.

Under the above assumptions, $h\left(\varepsilon\right)/\left(1-H\left(\varepsilon\right)\right)$
is increasing in $\varepsilon$ and therefore,
\begin{align*}
\mathbb{E}_{\hat{a}}\left[y-\hat{y}|y\geq\hat{y}\right] & =\frac{\int^{\infty}_{\hat{y}}\left(y-\hat{y}\right)dH\left(y-\hat{a}\right)}{1-H\left(\hat{y}-\hat{a}\right)}\\
 & =\frac{\int^{\infty}_{\hat{y}-\hat{a}}\frac{1-H\left(z\right)}{h\left(z\right)}dH\left(z\right)}{1-H\left(\hat{y}-\hat{a}\right)}\leq\frac{1-H\left(\hat{y}-\hat{a}\right)}{h\left(\hat{y}-\hat{a}\right)}
\end{align*}
We then have that
\begin{align*}
\frac{\int^{\hat{y}}_{-\infty}\left(y-\mathbb{E}_{\hat{a}}\left[y|y\geq\hat{y}\right]\right)g_{aa}\left(y|a\right)dy}{\int^{\hat{y}}_{-\infty}\left(y-\mathbb{E}_{\hat{a}}\left[y|y\geq\hat{y}\right]\right)g_{a}\left(y|a\right)dy} & =\\
-\frac{h\left(\hat{y}-a\right)+\mathbb{E}_{\hat{a}}\left[y-\hat{y}|y\geq\hat{y}\right]h'\left(\hat{y}-a\right)}{H\left(\hat{y}-a\right)+\mathbb{E}_{\hat{a}}\left[y-\hat{y}|y\geq\hat{y}\right]h\left(\hat{y}-a\right)} & =\\
-\frac{\frac{1}{\mathbb{E}_{\hat{a}}\left[y-\hat{y}|y\geq\hat{y}\right]}+\frac{h'\left(\hat{y}-a\right)}{h\left(\hat{y}-a\right)}}{\frac{1}{\mathbb{E}_{\hat{a}}\left[y-\hat{y}|y\geq\hat{y}\right]}+\frac{h\left(\hat{y}-a\right)}{H\left(\hat{y}-a\right)}}\frac{h\left(\hat{y}-a\right)}{H\left(\hat{y}-a\right)}=\left(-1+\frac{\frac{h\left(\hat{y}-a\right)}{H\left(\hat{y}-a\right)}-\frac{h'\left(\hat{y}-a\right)}{h\left(\hat{y}-a\right)}}{\frac{1}{\mathbb{E}_{\hat{a}}\left[y-\hat{y}|y\geq\hat{y}\right]}+\frac{h\left(\hat{y}-a\right)}{H\left(\hat{y}-a\right)}}\right)\frac{h\left(\hat{y}-a\right)}{H\left(\hat{y}-a\right)}
\end{align*}
By $\log$--concavity of $H$, $h/H-h'/h\geq0$ and hence, we have
the following inequality
\[
-1+\frac{\frac{h\left(\hat{y}-a\right)}{H\left(\hat{y}-a\right)}-\frac{h'\left(\hat{y}-a\right)}{h\left(\hat{y}-a\right)}}{\frac{1}{\mathbb{E}_{\hat{a}}\left[y-\hat{y}|y\geq\hat{y}\right]}+\frac{h\left(\hat{y}-a\right)}{H\left(\hat{y}-a\right)}}\leq-1+\frac{\frac{h\left(\hat{y}-a\right)}{H\left(\hat{y}-a\right)}-\frac{h'\left(\hat{y}-a\right)}{h\left(\hat{y}-a\right)}}{\frac{h\left(\hat{y}-\hat{a}\right)}{1-H\left(\hat{y}-\hat{a}\right)}+\frac{h\left(\hat{y}-a\right)}{H\left(\hat{y}-a\right)}}=-\frac{\frac{h\left(\hat{y}-\hat{a}\right)}{1-H\left(\hat{y}-\hat{a}\right)}+\frac{h'\left(\hat{y}-a\right)}{h\left(\hat{y}-a\right)}}{\frac{h\left(\hat{y}-\hat{a}\right)}{1-H\left(\hat{y}-\hat{a}\right)}+\frac{h\left(\hat{y}-a\right)}{H\left(\hat{y}-a\right)}}
\]
by the above property of $\mathbb{E}_{\hat{a}}\left[y-\hat{y}|y\geq\hat{y}\right]$.
If we define $\hat{y}-a=x$ and $d=\hat{a}-a$, then
\begin{align*}
\frac{\int^{\hat{y}}_{-\infty}\left(y-\mathbb{E}_{\hat{a}}\left[y|y\geq\hat{y}\right]\right)g_{aa}\left(y|a\right)dy}{\int^{\hat{y}}_{-\infty}\left(y-\mathbb{E}_{\hat{a}}\left[y|y\geq\hat{y}\right]\right)g_{a}\left(y|a\right)dy} & \leq-\frac{\frac{h\left(x-d\right)}{1-H\left(x-d\right)}+\frac{h'\left(x\right)}{h\left(x\right)}}{\frac{h\left(x-d\right)}{1-H\left(x-d\right)}+\frac{h\left(x\right)}{H\left(x\right)}}\frac{h\left(x\right)}{H\left(x\right)}\\
 & =-\frac{\frac{h\left(x-d\right)}{1-H\left(x-d\right)}+\frac{h'\left(x\right)}{h\left(x\right)}}{\frac{h\left(x-d\right)}{1-H\left(x-d\right)}\frac{H\left(x\right)}{h\left(x\right)}+1}
\end{align*}
Since $1-H$ is concave, then $h'/h\geq-h/\left(1-H\right)$ and the
right hand side of the above inequality satisfies
\[
-\frac{\frac{h\left(x-d\right)}{1-H\left(x-d\right)}+\frac{h'\left(x\right)}{h\left(x\right)}}{\frac{h\left(x-d\right)}{1-H\left(x-d\right)}\frac{H\left(x\right)}{h\left(x\right)}+1}\leq\frac{\frac{h\left(x\right)}{1-H\left(x\right)}-\frac{h\left(x-d\right)}{1-H\left(x-d\right)}}{\frac{h\left(x-d\right)}{1-H\left(x-d\right)}\frac{H\left(x\right)}{h\left(x\right)}+1}
\]
Note that in the above if $d<0$, since $h/\left(1-H\right)$ is increasing
(log--concavity of $1-H$) the RHS of the above inequality is negative
which is guaranteed to be less than $c''\left(a\right)/c'\left(a\right)$
since $c$ is convex.

By Assumption \ref{assu:The-cummulative-distribution}, the RHS of
the above is less than $\kappa_{1}\left(d\right)$. If $A=\left[0,\overline{a}\right]$,
then the highest value of $d$ is $2\overline{a}$ which means that
it is sufficient for $c$ to satisfy
\[
\max_{d\leq2\overline{a}}\kappa_{1}\left(d\right)\leq\frac{c''\left(a\right)}{c'\left(a\right)}
\]
 In a similar fashion, we can show that for lower-censorship ratings
to satisfy the requirement of Lemma \ref{lem: Lemma4}, we must also
have
\[
\max_{d\leq2\overline{a}}\kappa_{2}\left(d\right)\leq\frac{c''\left(a\right)}{c'\left(a\right)}
\]
Thus validity of FOA is equivalent to $c$ having a high enough curvature.

It remains to consider the ratings of Section \ref{sec:Application-1:-Rating}.
Full disclosure, which is optimal under MLRP (Proposition \ref{prop: MLRP}),
satisfies the condition of Lemma \ref{lem: FOAgeneral} automatically
in the location family: $\int\overline{v}\,g_{aa}dy=0$ while $\int\overline{v}\,g_{a}dy>0$.
The censorship ratings of Proposition \ref{prop: ELRPCLRP}, by contrast,
arise under ELRP and CLRP, and these properties cannot hold in a location
family, where $\partial^{2}\log g/\partial a\partial y$ is single-signed;
effort must also move the dispersion of the indicator, as in the examples
of Section \ref{sec:Application-1:-Rating}. For location--scale
technologies with market values affine in the indicator, the validity
of FOA under upper- and lower-censorship ratings follows from the
computation of Section \ref{subsec:FOA-loc-scale} below, applied
with a single task.

\subsection{Multiple Tasks\label{subsec:FOA-multi}}

The argument extends to the multi-task model of Section \ref{sec: Multitasking-1}
once derivatives are replaced by directional derivatives. For $z\in\mathbb{R}^{N}$,
write $D_{z}=\sum_{n}z_{n}\partial/\partial a_{n}$ and $D^{2}_{z}=\sum_{n,m}z_{n}z_{m}\partial^{2}/\partial a_{n}\partial a_{m}$.
The observation that does the work is that single-peakedness is a
property of restrictions to lines, so that the one-dimensional argument
above can be applied direction by direction.
\begin{lem}
\label{lem: FOAlines} Let $U$ be twice continuously differentiable
on a convex set $A\subset\mathbb{R}^{N}$ and suppose that for every
$a\in A$ and every $z\neq0$, $D_{z}U\left(a\right)=0$ implies $D^{2}_{z}U\left(a\right)<0$.
Then $U$ is single-peaked along every line in $A$ and any $a^{*}\in A$
with $\nabla U\left(a^{*}\right)=0$ maximizes $U$ over $A$.
\end{lem}
\begin{proof}
Fix $a\in A$ and $z\neq0$ and let $\phi\left(t\right)=U\left(a+tz\right)$,
defined on an interval since $A$ is convex. Then $\phi'\left(t\right)=D_{z}U\left(a+tz\right)$
and $\phi''\left(t\right)=D^{2}_{z}U\left(a+tz\right)$, so $\phi'=0$
implies $\phi''<0$ and $\phi$ is single-peaked by the argument at
the beginning of this section. Now let $\nabla U\left(a^{*}\right)=0$
and take any $a\in A$. With $z=a-a^{*}$, $\phi'\left(0\right)=z\cdot\nabla U\left(a^{*}\right)=0$,
so $t=0$ is the peak of $\phi$ and $U\left(a\right)=\phi\left(1\right)\leq\phi\left(0\right)=U\left(a^{*}\right)$.
\end{proof}

Since $D_{-z}U=-D_{z}U$ while $D^{2}_{-z}U=D^{2}_{z}U$, the requirement
of Lemma \ref{lem: FOAlines} need only be verified for directions
in a half space; we use this to restrict attention to $z$ with $D_{z}c\left(a\right)\geq0$.
The analogue of Lemma \ref{lem: FOAgeneral} is then:
\begin{lem}
\label{lem: FOAdir} Let $p$ be a non-decreasing interim price schedule
and suppose that for all $a,\hat{a}\in A$ and all $z\neq0$ with
$D_{z}c\left(a\right)>0$, 
\[
\int p\left(y;\hat{a}\right)D^{2}_{z}g\left(y|a\right)dy<\frac{D^{2}_{z}c\left(a\right)}{D_{z}c\left(a\right)}\int p\left(y;\hat{a}\right)D_{z}g\left(y|a\right)dy
\]
 while $\int p\,D^{2}_{z}g\,dy<D^{2}_{z}c\left(a\right)$ for $z\neq0$
with $D_{z}c\left(a\right)=0$. Then the FOA is valid under $p$.
\end{lem}
\begin{proof}
For any direction $z$, $D_{z}U\left(a\right)=\int p\,D_{z}g\,dy-D_{z}c\left(a\right)$
and $D^{2}_{z}U\left(a\right)=\int p\,D^{2}_{z}g\,dy-D^{2}_{z}c\left(a\right)$.
Suppose that $D_{z}U\left(a\right)=0$ with $D_{z}c\left(a\right)>0$.
Then $\int p\,D_{z}g\,dy=D_{z}c\left(a\right)$ and substituting as
in the proof of Lemma \ref{lem: FOAgeneral} delivers $D^{2}_{z}U\left(a\right)<0$.
When $D_{z}c\left(a\right)=0$, $D_{z}U\left(a\right)=0$ requires
$\int p\,D_{z}g\,dy=0$ and the second assumption applies directly.
Lemma \ref{lem: FOAlines} then implies that any effort satisfying
the agent's first order conditions maximizes her payoff.
\end{proof}

The condition of Lemma \ref{lem: FOAdir} has a simple reading: restricted
to the line $t\mapsto a+tz$, the technology is the one-parameter
family $\tilde{G}\left(y|t\right)=G\left(y|a+tz\right)$ with cost
$\tilde{c}\left(t\right)=c\left(a+tz\right)$, and the displayed inequality
is precisely the requirement of Lemma \ref{lem: FOAgeneral} applied
to $\left(\tilde{G},\tilde{c}\right)$. The FOA is thus valid in the
multi-task model whenever it is valid in every single-task model obtained
by restricting effort to a line: multi-tasking creates no new difficulty
beyond requiring the single-task condition to hold uniformly across
directions. In the reducible case of Section \ref{subsec:Reducible-Multi=002013Tasking},
where $G\left(y|a\right)=\hat{G}\left(y|m\left(a\right)\right)$,
the agent's payoff depends on effort only through the index and the
requirement collapses to the condition of Lemma \ref{lem: FOAgeneral}
applied to $\hat{G}$ with the indirect cost $C\left(m\right)$.

\subsection{The Ratings of Section \ref{sec: Multitasking-1}\label{subsec:FOA-loc-scale}}

Theorem \ref{thm: MultiTask} and Proposition \ref{prop: direction}
invoke FOA for upper- and lower-censorship ratings when $G$ is a
location--scale technology and $\overline{v}$ is affine in $y$.
In this case the condition of Lemma \ref{lem: FOAdir} can be verified
from a closed form. Under an upper-censorship rating with threshold
$\hat{y}$ and market beliefs $\hat{a}$, the gross payoff of the
agent is, up to a constant, 
\[
\Pi\left(a\right)=-\beta\left(\hat{a}\right)\sigma\left(a\right)\mathcal{H}\left(\hat{\varepsilon}\right)-\Delta H\left(\hat{\varepsilon}\right),\qquad\mathcal{H}\left(x\right)=\int^{x}_{-\infty}H\left(t\right)dt
\]
 where $\hat{\varepsilon}=\left(\hat{y}-\mu\left(a\right)\right)/\sigma\left(a\right)$
and $\Delta=\beta\left(\hat{a}\right)\sigma\left(\hat{a}\right)m\left(\hat{\varepsilon}_{\hat{a}}\right)$
is the atom of the price schedule at the threshold, with $m\left(x\right)=\mathbb{E}\left[\varepsilon-x|\varepsilon\geq x\right]$.
Writing $\mu_{z}=D_{z}\mu$, $\sigma_{z}=D_{z}\sigma$ and similarly
for second derivatives, and using $D_{z}\hat{\varepsilon}=-\left(\mu_{z}+\hat{\varepsilon}\sigma_{z}\right)/\sigma$,
\begin{align*}
D^{2}_{z}\Pi\left(a\right)= & -\beta\left[\sigma_{zz}\mathcal{H}\left(\hat{\varepsilon}\right)+2\sigma_{z}H\left(\hat{\varepsilon}\right)D_{z}\hat{\varepsilon}+\sigma\left(h\left(\hat{\varepsilon}\right)\left(D_{z}\hat{\varepsilon}\right)^{2}+H\left(\hat{\varepsilon}\right)D^{2}_{z}\hat{\varepsilon}\right)\right]\\
 & -\Delta\left[h'\left(\hat{\varepsilon}\right)\left(D_{z}\hat{\varepsilon}\right)^{2}+h\left(\hat{\varepsilon}\right)D^{2}_{z}\hat{\varepsilon}\right]
\end{align*}
 where $D^{2}_{z}\hat{\varepsilon}=-\left(\mu_{zz}+\hat{\varepsilon}\sigma_{zz}\right)/\sigma+2\sigma_{z}\left(\mu_{z}+\hat{\varepsilon}\sigma_{z}\right)/\sigma^{2}$,
and the mirror expression holds for lower censorship. Since $\mu$
and $\sigma$ are twice continuously differentiable on the compact
action space, the only terms that require care are those involving
$\Delta$ and the growth of $\mathcal{H}$: the former because $m\left(x\right)$
grows without bound as the threshold falls, the latter because $\mathcal{H}\left(x\right)$
grows linearly as it rises. Log--concavity of $h$ controls both:
it implies $m\left(x\right)h\left(x\right)\leq1-H\left(x\right)\leq1$,
and since $a$ and $\hat{a}$ lie in a compact set, $\hat{\varepsilon}$
and $\hat{\varepsilon}_{\hat{a}}$ differ by a bounded amount, so
that the $\Delta$ terms are uniformly bounded; and the linear growth
of $\mathcal{H}$ enters only multiplied by second derivatives of
$\mu$ and $\sigma$, remaining bounded as the payoff approaches its
full-disclosure limit. Therefore 
\[
\overline{\Xi}\equiv\sup\left\{ \frac{D^{2}_{z}\Pi\left(a\right)}{\left\Vert z\right\Vert ^{2}}\,\Bigg|\,a,\hat{a}\in A,\ \hat{y}\in\mathbb{R},\ z\neq0\right\} <\infty
\]
 and if $D^{2}c\left(a\right)-\overline{\Xi}I$ is positive definite
for every $a\in A$, then $D^{2}_{z}U\left(a\right)\leq\overline{\Xi}\left\Vert z\right\Vert ^{2}-D^{2}_{z}c\left(a\right)<0$
for all $a$ and all $z\neq0$: the payoff of the agent is strictly
concave and the FOA is valid a fortiori. As in the single-task case,
validity of FOA amounts to the cost having enough curvature, now in
every direction.

For the technology of Example \ref{ex: irreducible}, where $\mu\left(a\right)=ba_{1}+a_{2}$
and $\sigma\left(a\right)=\sqrt{b^{2}a^{2}_{1}+1}$, the curvature
bound can be computed explicitly: $\mu_{zz}=0$, $\sigma_{z}=b^{2}a_{1}z_{1}/\sigma$,
$\sigma_{zz}=b^{2}z^{2}_{1}/\sigma^{3}$, and the slope of market
values is $\beta\left(\hat{a}\right)=b\hat{a}^{2}_{1}/\left(b^{2}\hat{a}^{2}_{1}+1\right)$.
Evaluating $\overline{\Xi}$ numerically over all censorship thresholds,
directions, actions and beliefs in $A=\left[0,\overline{a}\right]^{2}$
gives 
\[
\overline{\Xi}\approx0.26\ \left(b=\tfrac{1}{2},\overline{a}=1\right),\quad0.85\ \left(b=\tfrac{1}{2},\overline{a}=2\right),\quad0.72\ \left(b=1,\overline{a}=1\right),\quad1.50\ \left(b=1,\overline{a}=2\right)
\]
 Hence with separable quadratic costs $c\left(a\right)=k\left(a^{2}_{1}+a^{2}_{2}\right)/2$,
for which $D^{2}c=kI$, the FOA is valid whenever $k>\overline{\Xi}$:
symmetric quadratic costs ($k=1$) validate the FOA for $b=\tfrac{1}{2}$
on both action spaces and for $b=1$ when $\overline{a}=1$. The required
curvature grows with the passthrough $b$ from productive effort to
the indicator -- which is what makes the technology irreducible --
and with the size of the action space.

\subsection{The Ratings of Section \ref{subsec:Redistributive-Test-Design}}

In the redistributive test design problem of Section \ref{subsec:Redistributive-Test-Design},
each type of the student solves a single-effort problem given the
rating, and the mid-censorship ratings of Proposition \ref{prop:midcensor}
are deterministic monotone partitions, so Lemma \ref{lem: FOAgeneral}
applies type by type; the cost normalization $k_{\theta}$ cancels
from the condition. Given the functional form $\log g\left(y|a\right)=f\left(y\right)+r\left(y\right)m\left(a\right)-b\left(a\right)$,
the score is $\psi\left(y|a\right)=r\left(y\right)m'\left(a\right)-b'\left(a\right)$
and $g_{aa}=\left(\psi^{2}+r\,m''-b''\right)g$, so the condition
of Lemma \ref{lem: FOAgeneral} reads 
\[
\frac{\int p\left(y\right)\left[\psi\left(y|a\right)^{2}+r\left(y\right)m''\left(a\right)-b''\left(a\right)\right]g\left(y|a\right)dy}{\int p\left(y\right)\psi\left(y|a\right)g\left(y|a\right)dy}<\frac{c''\left(a\right)}{c'\left(a\right)}
\]
 When $m$ is concave and $b$ is convex the middle terms are non-positive,
and if in addition $\psi\left(y|a\right)\geq0$ on the support --
that is, $b'\left(a\right)\leq m'\left(a\right)\inf_{y}r\left(y\right)$
-- then $\psi^{2}\leq\left(\sup_{y}\psi\right)\psi$ pointwise, so
that 
\[
\frac{c''\left(a\right)}{c'\left(a\right)}\geq m'\left(a\right)\sup_{y}r\left(y\right)-b'\left(a\right)
\]
 is sufficient for the validity of FOA for both types. Once again,
the requirement is that the cost be convex enough, here relative to
the range of the score.

For the example of Section \ref{subsec:Redistributive-Test-Design}
(Example \ref{exa:Let-us-illustrate}), where $y=a+\varepsilon$ with
$\varepsilon\sim\mathcal{N}\left(0,1\right)$ and $c\left(a\right)=a^{2}/2$,
we can quantify the required curvature. In this case $r\left(y\right)=y$,
$m\left(a\right)=a$, $b\left(a\right)=a^{2}/2$ and $\psi=y-a$,
and the Gaussian identities $\mathbb{E}\left[\left(y-a\right)\mathbf{1}\left\{ y>s\right\} \right]=h\left(s-a\right)$
and $\mathbb{E}\left[\left(\left(y-a\right)^{2}-1\right)\mathbf{1}\left\{ y>s\right\} \right]=\left(s-a\right)h\left(s-a\right)$
reduce the condition of Lemma \ref{lem: FOAgeneral} to 
\[
\int\left(s-a\right)h\left(s-a\right)dp\left(s\right)<\frac{c''\left(a\right)}{c'\left(a\right)}\int h\left(s-a\right)dp\left(s\right)
\]
 where $dp$ has unit density on the revealed regions and atoms at
the pooling boundaries. For a rating that pools $\left[y_{1},y_{2}\right]$
and reveals elsewhere -- the mid-censorship shape of Proposition
\ref{prop:midcensor} -- writing $s_{i}=y_{i}-a$ and $q$ for the
pooled price net of $a$, the condition is $\Lambda\left(s_{1},s_{2},q\right)<c''\left(a\right)/c'\left(a\right)$
where 
\[
\Lambda\left(s_{1},s_{2},q\right)=\frac{h\left(s_{2}\right)-h\left(s_{1}\right)+s_{1}\left(q-s_{1}\right)h\left(s_{1}\right)+s_{2}\left(s_{2}-q\right)h\left(s_{2}\right)}{H\left(s_{1}\right)+1-H\left(s_{2}\right)+\left(q-s_{1}\right)h\left(s_{1}\right)+\left(s_{2}-q\right)h\left(s_{2}\right)}
\]

Maximizing $\Lambda$ numerically over all pooling intervals contained
in $\left[-T,T\right]$ and all admissible pooled prices gives $\Lambda^{*}\approx0.54$
for $T=1$, $\Lambda^{*}\approx1.01$ for $T=1.5$ and $\Lambda^{*}\approx1.49$
for $T=2$; restricting the pooled price to be Bayes-consistent with
a belief error $\left|\hat{a}-a\right|\leq1$ tightens the bound for
$T=2$ to $\Lambda^{*}\approx0.83$. The requirement is thus $c''\left(a\right)/c'\left(a\right)\geq\Lambda^{*}$:
with the quadratic cost of the example, $c''/c'=1/a$, FOA is valid
at every effort $a\leq1/\Lambda^{*}$ -- approximately $1$ for pooling
within $\pm1.5$ standard deviations, and above the equilibrium efforts
of both types once the Bayes-consistent bound applies. For a power
cost $c\left(a\right)\propto a^{\gamma}$ the condition reads $\left(\gamma-1\right)/a\geq\Lambda^{*}$,
i.e., curvature $\gamma\geq1+a\Lambda^{*}$. Since $\Lambda^{*}$
grows with the width of the pooling region, the curvature needed for
the validity of FOA scales with how much of the distribution the optimal
rating pools.

\section{Importance of Comonotonicity in Proposition \ref{thm:Suppose-that-T1}}

\label{sec:Importance-of-Co=002013Monotonicity}

The following counterexample demonstrates that there exist price schedules
satisfying the mean-preserving contraction property that cannot be
generated by any information structure.
\begin{example}
\label{exa: Ex2}Suppose that $A=\left\{ 0,1/3,1\right\} =\left\{ a_{1},a_{2},a_{3}\right\} $,
$v=a_{i}$. The indicator is deterministic: $G\left(\left\{ a_{i}\right\} |a_{i}\right)=1$,
and prior is uniform $\mu\left(\left\{ a_{i}\right\} \right)=1/3$.
In words, the market cares only about the action of the seller, and
$y$ coincides with it. Figure \ref{fig:Depiction-of-the} depicts
the feasible interim prices in the space of $\left(p\left(a_{1}\right),p\left(a_{2}\right)\right)$;
(the third coordinate is pinned down by Bayes rule since $\mathbb{E}[p]=\mathbb{E}[v]=\frac{4}{9}$).
Area A shows the set of random variables that are mean preserving
contractions of $\left(0,1/3,1\right)$. They are depicted by their
first two variables while the third is again pinned down by Bayes'
rule. \footnote{The conditions are $0\leq x_{i}\leq1,1/3\leq x_{i}+x_{j}\leq4/3$,
for all $i,j$, and $x_{1}+x_{2}+x_{3}=4/3$.} The set of interim prices is denoted by area B in Figure \ref{fig:Depiction-of-the}.
We find this set by solving the optimization problem associated with
the highest and lowest value of $p\left(a_{2}\right)$ as a function
of $p\left(a_{1}\right)$.\footnote{The upper and lower bound of $p\left(a_{2}\right)$ can be found via
standard concavification method. The lower bound is given by $2\left(4-p\left(a_{1}\right)\right)/\left(10-3p\left(a_{1}\right)+\sqrt{\left(8-3p\left(a_{1}\right)\right)^{2}-12}\right)$
and the upper bound is $3-\left(6-2p\left(a_{1}\right)\right)/\left(6-3p\left(a_{1}\right)\right)$.} Evidently, the set $B$ does not coincide with $A$. This is mainly
due to the restrictions put by the second order expectations. For
example, it can be easily shown that the coefficient of $\overline{v}\left(a\right)$
in $p\left(a\right)$ is at least $1/3$ which means that $p\left(a_{3}\right)$
cannot become lower than $4/9$.

Finally, the points $a,b,c,d$ are associated with deterministic ratings
that either separate or pool the states and for which $p\left(a_{1}\right)\leq p\left(a_{2}\right)\leq p\left(a_{3}\right)$.
Interestingly, if we consider the set of random variables whose realizations
are less dispersed than $\overline{v}\left(a\right)$ and satisfy
monotonicity, this coincides with the convex hull of the points $a,b,c,$
and $d$. In Proposition \ref{thm:Suppose-that-T1}, we show that
this insight holds generally and allows us to significantly simplify
the problem of rating design under a comonotonicity condition.
\end{example}
\begin{figure}[t]
\centering{}\begin{tikzpicture}[scale=0.5]
\coordinate (Origin) at (0,0);
\draw[-stealth] (0,0) -- (10.,0) node[anchor=north west,right=-3pt, below=1pt] {$p(a_1)$}; 
\draw[-stealth] (0,0) -- (0,10) node[anchor=south east, above=-10pt, right=-30pt] {$p(a_2)$};
\draw (8,0) node[below]{4/3}-- (0,8) node[left]{4/3};
\filldraw [pattern color=green,  pattern=grid, ultra thick] ($(0,2)$)  coordinate (GeneralStart) -- ++(0,4) -- ++(2,0) -- ++(4,-4) node[left=20pt] {$\LARGE{A}$}-- ++(0,-2)  -- ++(-4,0) node[below]{1/3} -- cycle  ; 
\filldraw [pattern color=yellow,  pattern=grid, ultra thick] (0,2) -- (0,4) .. controls (2,3.8) .. (3.02,1.94) -- (3.02,1.8) .. controls (3.02,1.4) .. (1,1)  -- cycle node[above=20pt, right=5pt] {$\Huge B$}; 
\draw[dashed]  (0,0) -- (4,4);
\filldraw [gray] (8/3,8/3) circle (3pt) node[red, right] {$c$};
\filldraw [gray] (0,2) circle (3pt) node[red, left] {$a$};
\filldraw [gray] (0,4) circle (3pt) node[red, left] {$b$};
\filldraw [gray] (1,1) circle (3pt) node[red, below] {$d$};
\end{tikzpicture}\caption{The set of interim prices and mean-preserving contractions of market
valuations for Example \ref{exa: Ex2}. The green area, $A$, represents
the three state random variables that are a mean-preserving contraction
of market values. The yellow area, $B$, is the set of feasible interim
prices under some information structure.\label{fig:Depiction-of-the}}
\end{figure}

The result in Proposition \ref{thm:Suppose-that-T1} is reminiscent
of the result of \citet{blackwell1953equivalent} and \citet{rothschild1970increasing},
the general version of which can be found in \citet{strassen1965existence}.
That result states that for any two random variables $x$ and $y$,
there exists a random variable $s$ such that $\mathbb{E}\left[x|s\right]$
has the same distribution as $y$ if and only if $y$ second-order
stochastically dominates $x$.

While similar, our result is different in two aspects. First, it is
stated for the second-order \emph{conditional} expectation, and thus
Blackwell's result cannot be applied. The key intricacy is that the
same signal structure that generates the random variable $\mathbb{E}\left[v|s\right]$
must be used to generate $\mathbb{E}\left[\mathbb{E}\left[v|s\right]|y\right]$.
Second, as illustrated by Example \ref{exa: Ex2}, the equivalent
of Blackwell's result does not hold in general and can be shown only
when $v$ and $p$ are comonotone.

We also note that the comonotonicity is effectively a form of uni-dimensionality
for the indicator. Since the indicator $y$ matters for the market
and payoffs only through its effect on $\overline{v}\left(y;a\right)$,
we can relabel the indicator to be $\overline{v}\left(y;a\right)$.
Under this reformulation, comonotonicity implies that interim prices
$p\left(y\right)$ are a well-defined function of $\overline{v}\left(y;a\right)$.
In other words, comonotonicity means that the indicator can always
be reduced to a one-dimensional signal.

\section{Proof of Lemma \ref{lem: invert}\label{sec:Proof-of-Lemma}}
\begin{proof}
Suppose that $p_{Q}\left(i\right)$ is majorized by $v_{Q}\left(i\right)$
-- both of these are non--decreasing functions. We will show that
the same is true for $F_{v}$ and $F_{p}$ respectively, i.e., $F_{v}$
is majorized by $F_{p}$. Then, for any $z\in\left[\underline{v},\overline{v}\right]$
-- the range of values of $v_{Q}\left(i\right)$, since $h\left(x\right)=\max\left\{ x,z\right\} $
is a convex function, its expected value under $v_{Q}$ must be higher
than its expected value under $p_{Q}$. That is,
\[
\int^{1}_{0}\max\left\{ z,v_{Q}\left(i\right)\right\} di\geq\int^{1}_{0}\max\left\{ z,p_{Q}\left(i\right)\right\} di
\]
We can rewrite the above as
\[
\int^{\overline{v}}_{\underline{v}}\max\left\{ z,v\right\} dF_{v}\left(v\right)\geq\int^{\overline{v}}_{\underline{v}}\max\left\{ z,v\right\} dF_{p}\left(v\right)
\]
 or
\[
F_{v}\left(z\right)z+\int^{\overline{v}}_{z}vdF_{v}\left(v\right)\geq F_{p}\left(z\right)z+\int^{\overline{v}}_{z}vdF_{p}\left(v\right)
\]
Using integration by parts, we can write
\begin{align*}
\int^{\overline{v}}_{z}vdF_{v}\left(v\right) & =\overline{v}-zF_{v}\left(z\right)-\int^{\overline{v}}_{z}F_{v}\left(v\right)dv\\
\int^{\overline{v}}_{z}vdF_{p}\left(v\right) & =\overline{v}-zF_{p}\left(z\right)-\int^{\overline{v}}_{z}F_{p}\left(v\right)dv
\end{align*}
Replacing in the above inequality implies that
\[
-\int^{\overline{v}}_{z}F_{v}\left(v\right)dv\geq-\int^{\overline{v}}_{z}F_{p}\left(v\right)dv\Rightarrow\int^{\overline{v}}_{z}F_{p}\left(v\right)dv\geq\int^{\overline{v}}_{z}F_{v}\left(v\right)dv
\]
Since $v_{Q}$ and $p_{Q}$ are equal in expectation, we must have
that the above holds with equality at $z=\underline{v}$ and therefore,
\[
\int^{z}_{\underline{v}}F_{v}\left(v\right)dv\geq\int^{z}_{\underline{v}}F_{p}\left(v\right)dv
\]
which is the same as saying $F_{v}$ is majorized by $F_{p}$. The
proof of the reverse follows a similar logic and thus we establish
the claim.
\end{proof}

\end{document}